\documentclass[letterpaper]{article}

\usepackage{jcappub}
\usepackage{amssymb,amsmath,graphicx,times,epsfig}
\usepackage{fullpage}
\usepackage{hyperref}
\usepackage{deluxetable}
\usepackage{epstopdf}
\usepackage{aas_macros} 
\usepackage{multirow}

\title{Optimization Study for the Experimental Configuration of CMB-S4}

\author[a]{Darcy~Barron,}
\author[a,b]{Yuji~Chinone,}
\author[c,*]{Akito~Kusaka,}
\author[d,e]{Julian~Borril,}
\author[f]{Josquin~Errard,}
\author[g]{Stephen~Feeney,}
\author[h]{Simone~Ferraro,}
\author[d,e]{Reijo~Keskitalo,}
\author[a,c]{Adrian~T.~Lee,}
\author[c]{Natalie~A.~Roe,}
\author[c]{Blake~D.~Sherwin}
\author[i]{and Aritoki~Suzuki}

\note[*]{Electronic mail: akusaka@lbl.gov}
\affiliation[a]{Department of Physics, University of California, Berkeley, CA 94720, USA}
\affiliation[b]{Kavli IPMU (WPI), UTIAS, The University of Tokyo, Kashiwa, Chiba 277-8583, Japan}
\affiliation[c]{Physics Division, Lawrence Berkeley National Laboratory, Berkeley, CA 94720, USA}
\affiliation[d]{Computational Cosmology Center, Lawrence Berkeley National Laboratory, Berkeley, CA 94720, USA}
\affiliation[e]{Space Sciences Laboratory, University of California, Berkeley, CA 94720, USA}
\affiliation[f]{AstroParticule et Cosmologie, Univ Paris Diderot, CNRS/IN2P3,CEA/Irfu, Obs de Paris, Sorbonne Paris Cit\'e, France}
\affiliation[g]{Center for Computational Astrophysics, Flatiron Institute, 162 5th Ave, New York, NY 10010, USA}
\affiliation[h]{Department of Astronomy and Miller Institute, University of California, Berkeley, CA 94720, USA}
\affiliation[i]{Radio Astronomy Laboratory, University of California, Berkeley, CA 94720, USA}

\usepackage{color}

\begin{document}

\abstract{The CMB Stage 4 (CMB-S4) experiment is a next-generation,
ground-based experiment that will measure the cosmic microwave background (CMB) polarization to
unprecedented accuracy, probing the signature of inflation, the
nature of cosmic neutrinos, relativistic thermal relics in the early
universe, and the evolution of the universe.
CMB-S4 will consist of $O(500{,}000)$ photon-noise-limited detectors
that cover a wide range of angular scales  in
order to probe the cosmological signatures from both the early and late
universe.
It will measure a wide range of microwave frequencies
to cleanly separate the CMB signals from galactic and extra-galactic
foregrounds.

To advance the progress towards designing the instrument for CMB-S4, we have
established a framework to optimize the instrumental configuration to
maximize its scientific output.  
The framework combines cost and instrumental models 
with a cosmology forecasting tool, and evaluates the scientific sensitivity
as a function of various instrumental parameters.
The cost model also
allows us to perform the analysis under a fixed-cost constraint,
optimizing for the scientific output of the experiment given finite resources.

In this paper, we report our first results from this framework, using
simplified instrumental and cost models.
We have primarily studied two classes of
instrumental configurations:  arrays
of large-aperture telescopes with diameters ranging from 2--10\,m,
and hybrid arrays that combine small-aperture telescopes (0.5-m diameter)
with large-aperture telescopes.  We explore performance as a function of telescope aperture size, distribution of the
detectors into different microwave frequencies, 
survey strategy and survey area, 
low-frequency noise performance, 
and  balance between small and large aperture telescopes for
hybrid configurations.
Both types of configurations must cover  both large
($\sim \mathrm{degree}$) and small ($\sim \mathrm{arcmin}$) angular
scales, and the performance depends on assumptions for performance vs. angular scale.

The  configurations with large-aperture telescopes have a shallow optimum 
around 4--6\,m in aperture diameter, assuming that large telescopes can achieve good performance
for low-frequency noise.  We explore some of the
uncertainties of the instrumental model and cost parameters,
and we find that the optimum has a weak dependence on these parameters.
The hybrid configuration shows an even broader optimum, 
spanning a range of 4--10\,m in aperture for the large telescopes.
We also present two strawperson configurations as an outcome of this
optimization study, and we discuss some ideas for improving 
our simple cost and instrumental models used here.

There are several areas of this analysis that deserve further improvement.
In our forecasting framework, we adopt a simple two-component foreground
model with spatially varying power-law spectral indices.
We estimate de-lensing performance statistically
and ignore
non-idealities such as
anisotropic mode coverage, boundary effect, and possible
foreground residual.
Instrumental systematics, which is not accounted for in our analyses, 
may also influence the conceptual design.
Further study of the instrumental and cost models
will be one of the main areas of study by the entire CMB-S4 community.
We hope that our framework will be useful for estimating the influence
of these improvements in the future, and we will incorporate them in order to
further improve the optimization.}

\maketitle

\section{Introduction}
\label{sec:intro}

The Particle Physics Project Prioritization Panel (P5), a subpanel of the High Energy Physics Advisory Panel (HEPAP), submitted a report in 2014 that laid out a roadmap for the next ten years of research in particle physics and cosmology.  The P5 report recommended that DOE and NSF support a future CMB Stage 4 (CMB-S4) experiment, a next-generation, ground-based CMB polarization experiment.  
This experiment will probe the signatures of cosmic inflation,
a rapid expansion of the universe during its first $10^{-36}$ seconds,
and elusive dark elements of the universe, such as neutrinos,
dark radiation, dark matter, and early time behavior of dark energy.

CMB-S4 is expected to field 250{,}000 -- 1{,}000{,}000
photon-noise-limited detectors covering more than $ 50\%$ of the sky,
over the frequency range $\sim 20$--$280$
GHz~\cite{2015APh....63...55A,2015APh....63...66A,2016arXiv161002743A}.
Over a 5-year survey, it should reach a sensitivity on the
 tensor-to-scalar ratio $r$ of $\sigma(r) \approx 0.0005 \sim 0.001$.  In addition,
CMB-S4 will be sensitive to the sum of neutrino masses due to
gravitational lensing effects.  In combination with the Stage IV DESI
BAO experiment, the sensitivity to $\Sigma m_{\nu}$ is expected to reach
of order 0.02 eV, which is sufficient to detect the lowest allowed value in the Standard Model at $3 \sigma$.   CMB-S4 will also measure the effective number of light relativistic species  $N_{\rm eff}$ and 
the spectral index of the primordial scalar perturbation $n_s$, another
important parameter to constrain inflationary models, and constrain dark energy by measuring the kinetic Sunyaev Zeldovich effect, among other scientific goals.

In designing the optimal configuration for CMB-S4, many experimental choices must be made, including the number and diameter of the telescopes; the telescope optical design; the type and number of detectors and their allocation by frequency; the detector readout system; baffling and polarization modulation to reduce systematic errors; etc.  There are also choices that involve the survey strategy, for example, the fraction of time spent surveying deep, narrow fields (to study the degree scale signature of inflation) vs. wider, shallower fields (to study arc-minute signatures of lensing, clusters, kSZ effect, etc). The location of the experiment is also important, for both site characteristics and the size and region of accessible sky, including overlap with other surveys that will cover the same area.  The optimal experimental configuration and survey strategy will depend on how one prioritizes the scientific objectives.  In addition, some assumptions must be made about the limiting systematic errors on various techniques as well as the properties of galactic foregrounds and how well they can be measured and subtracted (either by CMB-S4 itself, or by other planned experiments that are likely to proceed on the same time scale).  In all of these experimental choices,  cost is a very important consideration that will determine the possible scope of CMB-S4 as well as schedule considerations, such as how long it will take to get CMB-S4 approved, built, and operating.   

In this paper, we present a framework to optimize the design of the CMB-S4 experiment to maximize the scientific productivity as a function of construction cost, where only hardware components are explicitly considered (an algorithm can be used to roughly translate hardware costs to total cost including engineering, technical, and management costs).  The framework we have developed is based on the Fisher matrix forecasting code of Errard et al.~\cite{Errard2016}, together with a  parametric model for the construction cost based on telescope size and the number of detectors, readout channels, and receivers. We have  prioritized the scientific goals to focus on topics that can uniquely be addressed with CMB-S4.  The analysis includes the effects of  foregrounds and lensing, but we have not attempted a detailed analysis of foreground model uncertainties and residual systematic effects.   

The work presented in this paper is not intended to be a detailed cost exercise.
Detailed cost modeling is an active area of study and discussion
in the entire CMB-S4 community.  Our work is intended to be complementary
to such efforts by providing a framework and methodology for optimization
together with initial results based on a simplified cost model.  We present
global trends of the optimization and discuss their sensitivity to
the assumptions of the cost model.  We find that some of these trends are robust
against possible variations of the cost model, while others
show significant dependence on the cost model assumptions.  This,
in turn, informs us where improvements in cost models are most crucial.
We expect that community-wide efforts toward improved  cost estimates
will feed into the optimization framework, providing a path towards an optimized
conceptual design for CMB-S4.

Another active area of community-wide development
is the forecasting and foreground modeling.
In our study, 
we assume simple two-component (dust and synchrotron) foregrounds with spatially varying
power-law spectral indices.
We estimate de-lensing performance statistically;
non-idealities such as
anisotropic mode coverage, boundary effect, and possible bias due to
residual foregrounds are not accounted for in our forecast and may
degrade the performance.
Instrumental systematics, which are also not accounted for in our analyses, 
may influence the conceptual design.
We hope the community-wide effort to address these aspects will make
forecasting more realistic and accurate, and
we will improve our optimization further by incorporating these
developments.

This paper is organized as follows:  in Section~\ref{sec:science},  we discuss the key scientific goals of CMB-S4 and motivate our choice of scientific parameters for the optimization exercise.  In Section~\ref{sec:methodology}, the optimization methodology is introduced, including the instrumental performance parameters, prior and external data sets, and the Fisher matrix forecasting framework including the treatment of foregrounds, de-lensing, and noise.  The instrument configuration and cost modeling is described in Section~\ref{sec:technical}.  In Section~\ref{sec:results}, we provide our optimization results, beginning with some general trends for two types of configurations, those involving large aperture telescopes only, and hybrid arrays with a mix of large and small apertures. We study the limit of diminishing returns, 
variations according to the uncertainties in the cost model used, and
the dependence on the survey strategy chosen.   In Section~\ref{sec:strawperson}, we present two detailed strawperson models to illustrate the results of the study, including some limitations and areas for future study.  Our conclusions are presented in Section~\ref{sec:conclusion}.

\section{Key Science Goals}
\label{sec:science}

Among the four science goals discussed here, we use the
tensor-to-scalar ratio $r$ and the number of relativistic species
$N_\mathrm{eff}$ to define the figure of merit for the CMB-S4 instrumental
configurations.  We choose not to
assess the {\it importance} of each science goal.
This choice is based on the following two reasons:   first, we chose $r$
and $N_\mathrm{eff}$ because they encompass the parameter space of the
instrument, e.g., angular scales and frequency coverage.  For example,
an instrumental configuration optimized for $N_\mathrm{eff}$, which
requires arcminute resolution, is nearly
optimal for measuring neutrino mass and kSZ as well.  We will discuss
this in the optimization section.  Second, $r$ and $N_\mathrm{eff}$ are
the observables that are unique to CMB polarization, and no other
cosmological probes, such as optical surveys, are competitive with CMB-S4.
More details about the science goals can be found in the CMB-S4 Science Book \cite{2016arXiv161002743A}.

\subsection{Inflation through Primordial B-modes}
Inflation, a phase of accelerating expansion in the very early universe, is currently the most promising mechanism to explain both the presence of small initial density fluctuations and the large-scale homogeneity and flatness of the universe \cite{1981PhRvD..23..347G, Linde:2005ht, 2016arXiv161002743A}. While the inflationary framework has been verified via the predictions it makes for the properties of the scalar density fluctuations (e.g., Gaussianity, isotropy, super-horizon correlations, near-scale invariance with a red spectral tilt, adiabaticity), a more specific prediction of many inflationary models is the production of a stochastic background of gravitational waves \cite{Rubakov:1982df, Fabbri:1983us, Abbott:1984fp}. The detection of this background of inflationary gravitational waves would not only provide confirmation of the inflationary framework, but by measuring the strength of this gravitational wave background -- parametrized by the tensor-scalar-ratio $r$ -- the energy scale of inflation can be determined (see for example \cite{2016arXiv161002743A}, Chapter 2). This measurement would thus probe physics at the GUT scale, far beyond the reach of even futuristic particle colliders. Even improved non-detection upper limits are extremely valuable: increasing the strength of the constraints on $r$ by two orders of magnitude would rule out broad classes of large-field inflation models.

The most promising method for detecting inflationary gravitational waves is through the measurement of the characteristic large-scale B-mode polarization it produces. The B-mode polarization channel is unique as it is not limited by cosmic variance from scalar fluctuations (at leading order), so that even small values of $r$ can be probed \cite{Seljak:1996ti, Kamionkowski:1996zd, Seljak:1996gy}. The measurement of inflationary B-mode polarization at low levels suffers from three main challenges. First, the instrumental requirements on measuring or constraining small B-mode polarization signals are extremely stringent. Second, galactic foregrounds such as galactic dust and synchrotron can produce B-modes as well, which can be confused with inflationary signals. These foreground signals must be removed or accounted for in inflationary searches; the most promising method for this is to separate primordial signals from foreground emission using multifrequency data. Third, by remapping polarization anisotropies, gravitational lensing by large-scale structure converts some of the primordial E-mode polarization into B-mode polarization \cite{Zaldarriaga:1998ar}. This lensing B-mode polarization acts as a source of noise that can obscure any primordial inflationary B-mode signal. Fortunately, CMB-S4 will be able to reconstruct the CMB lensing signal so well that de-lensing methods can be applied: from the reconstructed lensing, we can infer the lensing B-mode and subtract it from the measured B-mode map, thereby greatly reducing the lensing B-mode noise and potentially revealing any underlying inflationary signal.

\subsection{Extra Relativistic Species}
Many extensions to the standard model of particle physics predict the
presence of new light particles.  While these particles may interact too
weakly to be produced in terrestrial experiments, the early universe is
so hot and dense that they could be created in thermal equilibrium. As
the universe cools, these ``relic'' particles may persist.  Their energy
density, while small, can affect cosmology and, in turn, the properties
of the CMB (see Chapter 4 of \cite{2016arXiv161002743A} for a review).

The presence of these light particles manifests itself in the CMB
through two main effects. First, the early expansion rate is modified
due to the presence of additional energy density; this decreases the
amount of Silk damping in the power spectra when the acoustic
scale is held fixed. Second, the presence of free streaming particles changes
the propagation of acoustic oscillations in the primordial plasma,
leading to a small phase shift in the positions of the CMB acoustic
peaks \cite{Bashinsky:2003tk, Baumann:2015rya} . By measuring these effects, CMB-S4 can provide an extremely
precise measurement of the energy density of light, weakly coupled
particles.

The magnitude of the effects depend on the energy in these light
particles and hence when they froze out: a particle that falls out of
thermal equilibrium very early does not gain energy from subsequent
phase transitions, where the known particles annihilate and deposit
their energy into the thermally coupled phases. Particles that freeze out
extremely early, before the QCD phase transition, give a contribution
equivalent to $\Delta N_\mathrm{eff}>0.027$, where $N_\mathrm{eff}$ is an effective
number of
neutrino-like species, and $\Delta N_\mathrm{eff}$ is a deviation from the
standard
model without new light particles.  For particles that freeze out later,
$\Delta N_\mathrm{eff}$ is larger.
CMB-S4 approaches the sensitivity needed to explore
 $\Delta N_\mathrm{eff} \sim 0.03$,
which is comparable to this lower bound \cite{2016arXiv161002743A}.

\subsection{Neutrino Mass through Gravitational Lensing}
Though neutrinos comprise three of the twelve elementary fermions, the absolute scale of their masses is not well known, in contrast to the other nine fermions; only the two mass splittings among the three neutrino species have been well measured, setting a lower bound on the sum of the neutrino masses of $\approx 0.06$ eV \cite{Agashe:2014kda}.  Measuring the sum of neutrino masses thus probes a fundamental unknown scale in physics and could also determine the neutrino mass hierarchy. A cosmological measurement of the neutrino mass scale, complemented by terrestrial particle physics experiments, will hence form an important part of a program of understanding the neutrino sector and might even give insight into the origin of the remarkably small masses of these particles.

The mass scale of neutrinos can be probed in cosmology because the masses of neutrinos suppress the growth of cosmic structure. Measurements of the gravitational lensing of the CMB is a direct probe of this large-scale structure: by measuring new mode correlations that lensing induces into the CMB, the gravitational lensing field can be mapped \cite{Lewis:2006fu}. This lensing field directly probes the density of mass and dark matter, projected out to high redshifts (with the largest contribution arising from the redshift range $z=0.5-3$). By reconstructing the lensing maps and statistically characterizing them with the lensing power spectrum, we can probe any physics -- such as neutrino mass -- that affects the growth of the large-scale structure or geometry of the universe. Measurements of the lensing power spectrum have already made rapid progress; however, with its high sensitivity and angular resolution, CMB-S4 will provide measurements of the CMB lensing power spectrum with unprecedented precision, allowing definitive measurements of the neutrino mass when combined with baryon acoustic oscillation (BAO) measurements from the planned DESI experiment (see Chapter 3 of \cite{2016arXiv161002743A}).

\subsection{Galaxy Clusters and Astrophysics}

Galaxy clusters are the largest gravitationally bound objects in the Universe, and many physical processes related to their formation and evolution are still poorly understood. The interaction of CMB photons with clusters leaves an imprint on the observed anisotropy, making high-resolution observations of the CMB a powerful tool to study these objects and potentially a very powerful probe of cosmology.  There are a number of effects that are relevant, as summarized below.

Galaxy clusters host large quantities of hot, ionized gas with typical electron temperature $T_e \sim 10^8$ K.  A CMB photon propagating through this hot medium can inverse-Compton-scatter off the cluster electrons and, on average, gain energy.  This effect is known as the thermal Sunyaev-Zel'dovich effect \cite{Sunyaev:1980vz,Sunyaev:1972eq} (tSZ). This produces a spectral distortion of the CMB and is easily identifiable by combining measurements at different frequencies. The net effect on the CMB anisotropy is of order $\tau_{\rm cluster} T_e / m_e \propto n_e T_e$ and is proportional to the thermal pressure of the gas. Being a probe of the thermal pressure, it helps to characterize the amount of energy injection in the cluster and quantify the amount of non-thermal pressure. Recent studies have found evidence of feedback from the central supermassive black hole in stacked tSZ maps \cite{Crichton:2015joa,Greco:2014vwa,Ruan:2015vca}.

Moreover, the tSZ effect is one of the most effective tools to find high-redshift ($z \gtrsim 1$) clusters, since the magnitude of the signal is redshift independent\footnote{However, the angular size does depend on redshift.}. Cluster number counts are a very powerful probe of cosmology, since they are very sensitive to the amplitude of the perturbations and neutrino masses \cite{Ade:2015fva,deHaan:2016qvy,Sehgal:2010ca}. If we allow deviations from General Relativity, cluster abundance is also one of the most informative tests of gravity \cite{Ferraro:2010gh,Schmidt:2009am}.

The bulk motion of a cluster also produces a signature in the observed CMB, known as the kinematic Sunyaev-Zel'dovich effect (kSZ) \cite{Ostriker:1986fua,Sunyaev:1980vz,Sunyaev:1972eq}.  The size of the temperature shift (essentially a Doppler effect) for a cluster with radial velocity $v_r$ is $\tau_{\rm cluster} v_r \propto n_e v_r$. It is thus a probe of the total electron abundance associated with the halo as well as of the gas profile. Recent work has shown large differences between the gas and dark matter profiles, indicating powerful physical processes at play \cite{Ade:2015lza,Schaan:2015uaa}. Precision measurement of the gas profile through the kSZ effect will inform us about cluster physics and provide an important tool to help calibrate weak lensing surveys, since baryons account for $\sim 20$ \% of the total mass.

Cluster properties are expected to depend both on mass and redshift of the host halo and could depend on other properties, such as star formation rate, color, presence of an Active Galactic Nucleus (AGN), etc. The large sky coverage of CMB-S4, together with better characterization of several galaxy properties (compared to a photometric survey),  will  shed light on the effect of feedback and star formation on the gas. When combined with tSZ measurements, the temperature of the IGM as well as the amount of energy injection can be constrained. If the optical depth of the cluster can be obtained (for example, through tSZ or X-ray observations), the kSZ signal measures the statistics of the radial velocities, which are proportional to the rate of growth of structure and which provide competitive constraints on the theory of gravity as well as neutrino masses \cite{Kosowsky:2009nc,Mueller:2014dba}.

Galaxy clusters, due to their large mass, also lens the primary CMB, creating a typical signature in temperature and polarization \cite{Hu:2007bt,Seljak:1999zn}. This can be used to accurately measure cluster masses, which is one of the main uncertainties when extracting cosmological parameters from cluster counts.

Lastly, the kSZ signal can also be used to explore the epoch of reionization. High-resolution CMB observations will accurately measure the duration and time of reionization, which in turn will place tight constraints on the physics of the universe at an intermediate redshift \cite{Calabrese:2014gwa,smithferraro2016}.

\section{Optimization Methodology}
\label{sec:methodology}

\begin{figure}[htbp]
  \begin{center}
   \includegraphics[width=0.95\textwidth]{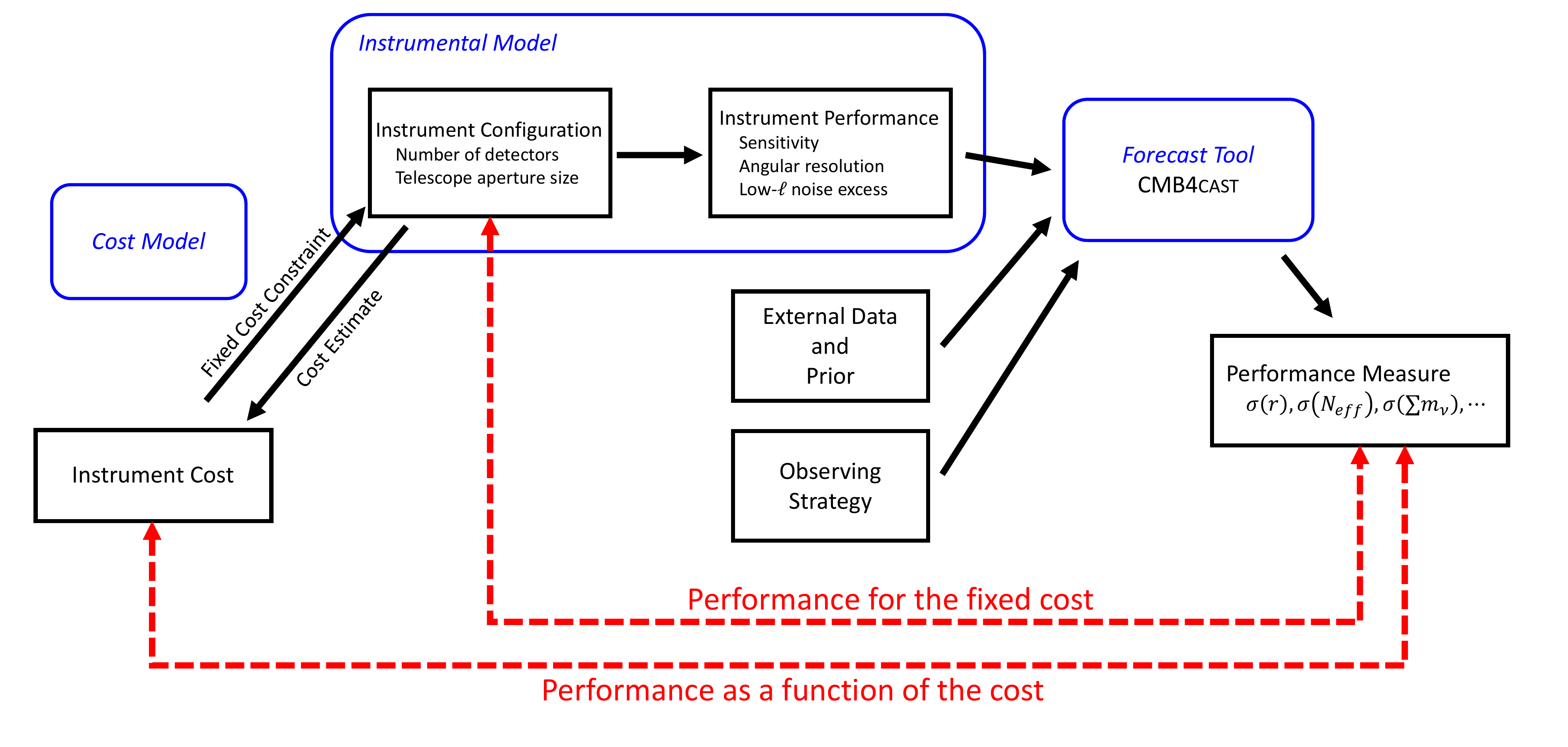}
  \caption{\label{fig:methodology_framework}
  Schematic figure showing the optimization framework.  The framework
  consists of the Cost Model, Instrumental Model, and the Forecast
  Tool.
  }
  \end{center}
\end{figure}

Our goal is to optimize the science output of the CMB-S4 instrument for a
given fixed cost.  For this optimization, we establish a framework that 
combines a forecasting tool with an instrumental model and a cost model
(Fig.~\ref{fig:methodology_framework}).
  Our goal is to explore the following dependencies through this framework:
  \begin{enumerate}
\item The relationship between the instrumental configuration and the
  performance metric given a cost constraint.  For example, we compare different
  telescope array configurations under a fixed cost assumption and 
  compare their relative effects on the
      error on $r$.
  \item The relationship between the cost and the performance metric
  for a given instrumental configuration.  In this case, as we vary the
  cost, we simply scale
  the instrument (numbers of telescopes, detectors, readout, and cryostats)
  for specific configurations and see how the metric improves for additional cost.
  \end{enumerate}
  
In this section,  we describe 
 the forecasting tool we have adopted, \textsc{CMB4cast}~\cite{Errard2016},
including its treatment of foregrounds, lensing, and noise.
The details of the instrumental model and the cost model will be discussed
in the next section.

\subsection{Instrumental Input to Forecast}
Based on the instrumental model described in Sec.~\ref{sec:technical}, we generate
the input to the forecast.  As shown in
Fig.~\ref{fig:methodology_framework},
the instrumental inputs to the forecast model are the sensitivity
and angular resolution for each
frequency band as well as the low angular-frequency noise excess.

The experimental sensitivity is calculated according to the instrumental
model, the observing time, and the observed sky area.
We account for possible degradation of the white noise level due to
non-idealities such as data selection efficiency
(Sec.~\ref{sec:realistic_sensit}).
The aperture size and the wavelength determine 
the angular resolution for each frequency band.
CMB experiments suffer from low-frequency noise, 
or so-called $1/f$ noise, leading to excess noise in the low-$\ell$ region.
\textsc{CMB4cast} incorporates this noise excess using a
parameterization discussed in Sec.~\ref{sec:low_ell_noise}.
The degree of excess depends on various aspects of the instrument;
further discussion can be found in Sec.~\ref{sec:results}.

The relative number of detectors within each
frequency band is determined based on an overall optimization
(Sec.~\ref{sec:freq_combination_and_aperture}).
The map depths calculated for each frequency band are then combined to separate out the foreground
components from the CMB signal and to estimate the noise variance in the reconstructed CMB map,
$\sigma_{\rm CMB}$, as described in Sec.~\ref{ssec:fgs}
and~\cite{Errard2016}.

\subsection{Prior and External Data}
The external priors required to measure $r$ from CMB-S4
are the scalar amplitude and index:  $A_S$ and $n_S$. These priors are expected to be provided by 
Planck and WMAP data.  For simplicity, in this
study, we use only the CMB-S4 data; we do not combine with Planck or WMAP, and  we do not marginalize over $A_S$ or $n_S$.  We have
confirmed that this treatment  differs negligibly from the case where CMB-S4 data is combined
with Planck or WMAP data in order to constrain $A_S$ and $n_S$.

For measuring $N_\mathrm{eff}$ and $M_\nu$ ($\equiv \sum m_\nu$), we assume a prior from the DESI galaxy redshift survey.
We also include the Planck dataset, where we incorporate a naive
white noise model in the map, as specified in Ref.~\cite{Errard2016}, Table~4,
corresponding to an error on the optical depth $\tau$ of 
$\simeq 0.004$.  While the current constraint by Planck is about two times worse than 
this~\cite{2016arXiv160502985P}, 
we expect that future experiments (satellite, balloon, or even ground-based such as CLASS) will improve the constraint on $\tau$.  We consider this assumption to be appropriate for forecasting the performance of CMB-S4, since we wish to explore other limiting factors, but it is important to keep this in mind.

\subsection{Forecasting Framework}
We describe the \textsc{CMB4cast}~\cite{Errard2016} tool, which is an implementation within a consistent framework of a parametric component separation algorithm, a de-lensing of $B$-modes and an estimation of constraints on cosmological parameters.
There are differences in methodology and
assumptions when
comparing multiple forecasting codes.  Some of the differences are
pointed out in Sec.~8.10.1.1 of Ref.~\cite{2016arXiv161002743A},
where \textsc{CMB4cast} is compared to another Fischer code.
We note, however, that the assumptions we adopt differ from those used
in Ref.~\cite{2016arXiv161002743A} for \textsc{CMB4cast}.  Our
assumptions are described
below.  We also note that the frequency band definition of the detectors and
 per-detector sensitivity differ between our
 study (Table~\ref{tab:NETSummary2}) and Ref.~\cite{2016arXiv161002743A}.

\subsubsection{Foregrounds}
\label{ssec:fgs}
We use the parametric maximum-likelihood approach as introduced in, e.g., \cite{Brandt1994,Eriksen2006,Stompor2009}. For a given sky pixel $p$, the measured amplitudes at all frequencies are concatenated in a data vector $d$, such that
\begin{eqnarray}
	\centering
		d_p = \mathbf{A}_p\,s_p + n_p
	\label{eq:data_modeling}
\end{eqnarray}
where 
\begin{itemize}
	\item $\mathbf{A}$ is the so-called \emph{mixing matrix}, which contains the frequency scaling laws of all sky components (CMB, foregrounds). Under the parametric formalism, we assume that the mixing matrix $\mathbf{A}$ can be parametrized by a set of \emph{spectral parameters} $\beta$:
	\begin{eqnarray}
		\centering
			\mathbf{A} \equiv \mathbf{A}(\beta).
	\end{eqnarray}
	\item $s_p$ contains the amplitudes of each sky component;
	\item $n_p$ is the instrumental noise, assumed white in our analysis.
\end{itemize}

Given Eq.~\ref{eq:data_modeling}, the component separation is performed in two steps:
\begin{itemize}
	\item the estimation of the mixing matrix or, equivalently, the
	      estimation of the spectral parameters. This is achieved
	      through the optimization of a spectral likelihood,
	      $\mathcal{L}_{\rm spectral}(\beta)$, as detailed
	      in~\cite{Stompor2009}. In \textsc{CMB4cast}, following the
	      formalism developed in~\cite{Errard2011}, we do not
	      optimize the spectral likelihood itself, but instead we assume that a given instrumental setup is able to recover the true spectral parameters, with some uncertainties related to the finite sensitivity (or limited number of frequency channels) of the instrument. The error bars on the spectral parameters, $\sigma(\beta)$, are derived from the curvature of the spectral likelihood at its peak, averaged over noise realizations, i.e. 
	\begin{eqnarray}
		\centering
			\sigma (\beta_i ) = \sqrt{\left[\mathbf{\Sigma}\right]_{ii}}\ \ \ \ {\rm with}\ \ \ \ \mathbf{\Sigma}^{-1} \equiv \left\langle \frac{\partial^2\mathcal{L}_{\rm spectral}}{\partial \beta_i \partial \beta_j} \right\rangle_{\rm noise}
		\label{eq:spectral_indices_uncertainties}
	\end{eqnarray}
	Ref.~\cite{Errard2011} proposes a semi-analytical expression for $\mathbf{\Sigma}$, hence providing a computationally efficient framework to evaluate the performance of a given observational configuration. This approach assumes that the ``true'' scaling laws are recovered with some error bars, which leads to the presence of ``statistical'' foreground residuals in the cleaned CMB map. By reducing the analysis to $\mathbf{\Sigma}$, the curvature of the spectral likelihood, we do not account for possible bias in the estimation of spectral parameters, which could generate ``systematic'' foregrounds residuals, and could bias the estimation of cosmological parameters\footnote{An extension of the \textsc{CMB4cast} framework, called \textsc{xForecast} --- estimating the possible bias on spectral and cosmological parameters, has recently been proposed in \cite{Stompor2016}}.
	\item the ``inversion'' of Eq.~\ref{eq:data_modeling} with the estimated $\mathbf{A}$, in order to disentangle sky components and obtain estimates of the sky signals $\tilde{s}$, given by:
	\begin{eqnarray}
		\centering
			\tilde{s} = \left(\mathbf{A}^T\mathbf{N^{-1}}\mathbf{A}\right)^{-1}\mathbf{A}^T\mathbf{N^{-1}}d
		\label{eq:normal_equation}
	\end{eqnarray}
	From Eq.~\ref{eq:normal_equation}, one can see that the noise variance, $\sigma_{\rm CMB}$, associated with the recovered CMB map is given by
	\begin{eqnarray}
		\centering
			\sigma_{\rm CMB} \equiv \sqrt{ \left[  \left(\mathbf{A}^T\mathbf{N^{-1}}\mathbf{A}\right)^{-1} \right]_{\rm CMB \times CMB}}
		\label{eq:noise_variance_CMB_map}
	\end{eqnarray}
\end{itemize}
Furthermore, the statistical residual foregrounds left in the CMB map after component separation can be derived using the error bars $\mathbf{\Sigma}$ from Eq.~\ref{eq:spectral_indices_uncertainties}; their power spectrum is given by
\begin{equation}
	\centering 
		C_\ell^{\rm res}  \equiv \sum_{k,k'}\sum_{j,j'}\,\mathbf{\Sigma}_{kk'}\,{\kappa}^{jj'}_{kk'}  \, C^{jj'}_{\ell}, 
		\label{eq:Clres_def}
\end{equation}
where $C^{jj'}_{\ell}$ are the input foreground spectra with $j,\,j'\ \in\, \{$cmb, dust, synchrotron$\}$. The element ${\kappa}^{jj'}_{kk'}$ is as defined in~\cite{Errard2011}:
\begin{eqnarray}
	\centering
		{\kappa}^{jj'}_{kk'} &\equiv& \alpha_k^{0j}\alpha_{k'}^{0j'}\\
		{\rm with}\  \alpha_k^{0j} &\equiv& -\left[\left(\mathbf{A}^T\mathbf{N^{-1}}\mathbf{A}\right)^{-1} \mathbf{A}^T\mathbf{N^{-1}}  \frac{\partial \mathbf{A}}{\partial \beta_{k}} \right]_{0j}.
	\label{eq:kappa_def}
\end{eqnarray}
The residual foregrounds can ultimately bias the estimation of CMB power spectra and therefore the estimation of cosmological parameters. 
\textsc{CMB4cast} parameterizes this residual foreground power as a
power law in $\ell$ space, with an amplitude $A_{\rm res}$ and tilt $b_{\rm res}$:
\begin{eqnarray}
	\centering
	   C_\ell^{\rm res} = A_{\rm res}\times \left( \frac{\ell}{\ell_0} \right)^{b_{\rm res}}.
	\label{eq:Clres_param}
\end{eqnarray}
While  \textsc{CMB4cast} allows us to marginalize over $A_{\rm res}$ and
$b_{\rm res}$, we do not perform this marginalization in our study for
two reasons.  First, the expectation value of $C_\ell^{\rm res}$ is
small, and this bias term is non-negligible only when $A_{\rm res}$
is $O(100)$ larger than the nominal value.
Second, turning on this marginalization corresponds to distinguishing the
cosmological signal from the foreground residual merely from the power
spectrum shape.  This is particularly challenging for primordial
gravitational waves and may not be the most efficient way to
achieve redundancy in foreground removal.

In this study, we consider the two main diffuse polarized astrophysical
foregrounds: dust and synchrotron.  They are assumed to follow,
respectively, a gray-body and power-law spectra.  The power-law spectrum
for synchrotron is
\begin{eqnarray}
	\centering
		A^{\rm raw}_{\rm sync}(\nu, \nu_{\rm ref}) \equiv \left(\frac{\nu}{\nu_{\rm ref}}\right)^{\beta_s},
	\label{eq:As_def}
\end{eqnarray}
where the reference frequency $\nu_{\rm ref}=150$\;GHz. We consider a modified grey-body emission law for the dust
\begin{eqnarray}
	\centering
		A^{\rm raw}_{\rm dust}(\nu, \nu_{\rm ref}) \equiv \left( \frac{\nu}{\nu_{\rm ref}}\right)^{\beta_d+1}\frac{e^{\frac{h\nu_{\rm ref}}{k\,T_d}}-1}{e^{\frac{h\nu}{kT_d}} -1 }.
	\label{eq:Ad_def}
\end{eqnarray}
The present study follows the ``$n_p$-approach'' described in~\cite{Errard2016}, which assumes that dust and synchrotron spectral indices vary on angular scales larger than $15\,\deg$ (healpix resolution with $n_{\rm side}=4$).
Foregrounds due to point sources, whether galactic or extra-galactic, are not considered in this study.

\subsubsection{De-lensing}

Removing the CMB lensing contaminant through de-lensing requires a measurement of the lensing potential, which can be used to estimate the lensed CMB $B$ modes for subtraction from the total observed signal. \textsc{CMB4cast} follows  the approach in~\cite{smith2012}, which provides the following analytical expression for the estimated lensing $B$ modes:
\begin{equation}
	\label{eq:smith_delensing} 
		C_\ell^{BB,\,\rm estimated} = \frac{1}{2\ell + 1}\sum_{\ell_1, \ell_2} \left|f_{\ell \ell_1 \ell_2}^{EB}\right|^2 \times \frac{(C_{\ell_1}^{EE})^2}{C_{\ell_1}^{EE}+N_{\ell_1}^{EE}} \frac{(C_{\ell_2}^{\phi \phi})^2}{C_{\ell_2}^{\phi \phi}+N_{\ell_2}^{\phi \phi} } \, , 
\end{equation}
where $f_{\ell \ell_1 \ell_2}^{EB}$ is a geometric coupling factor. The de-lensed $B$ mode is then given by
\begin{equation}
		C_\ell^{BB,\ \rm delensed} \equiv C_\ell^{BB,\,\rm fiducial,\, lensed} - C_\ell^{BB,\,\rm estimated}.
	\label{eq:delensing_def}
\end{equation}
The presence of noise in Eq.~\eqref{eq:smith_delensing} always guarantees that $C_\ell^{BB,\,\rm fiducial,\, lensed} \geq C_\ell^{BB,\,\rm estimated}$.

\textsc{CMB4cast} proposes three sources for the lensing potential
estimate: the CMB polarization itself (``CMB$\times$CMB'' de-lensing),
the cross-correlation of the CMB and the cosmic \emph{infrared}
background (``CMB$\times$CIB''), and measurements of the large-scale
structure using, for example, cosmic shear or 21cm radiation
(``CMB$\times$LSS''). In the CMB$\times$CMB case, the noise on this
estimate is given as the following~\cite{Hirata2003}:
\begin{equation}
	\label{eq:eb_estimator}
		N_\ell^{\phi\phi} = \left[ \frac{1}{2\ell+1} \sum_{\ell_1 \ell_2} |f_{\ell_1 \ell_2 \ell}^{EB}|^2 \left( \frac{1}{C_{\ell_1}^{BB} + N_{\ell_1}^{BB}} \right) \times \left( \frac{(C_{\ell_2}^{EE})^2}{C_{\ell_2}^{EE} + N_{\ell_2}^{EE}} \right) \right]^{-1}.
\end{equation}
Iterating over this estimator can significantly improve the ability of a given instrument to delense the CMB -- for realistic instrumental configurations, this process converges after a few steps once the convergence criterion is satisfied:
\begin{equation}
		\left| \sum_\ell \frac{ N_{\ell}^{\phi \phi, i} - N_{\ell}^{\phi \phi, i-1} }{ N_{\ell}^{\phi \phi,i} } \right|\,\leq\,1\%.
	\label{eq:iterative_delensing_def}
\end{equation}

Our forecasts for de-lensing may be complicated in real data by multiple
issues. First, some modes of the CMB E-polarization may remain very
noisy -- and hence effectively unobserved -- if the instrument scans the map
only from a restricted range of directions (for example, modes along the
Fourier-y-axis). The B-modes sourced by these unobserved E-modes cannot 
be de-lensed, which results in a reduced efficiency for lensing
B-mode removal. The extent to which this is problematic depends, of
course, on how much of the E-mode Fourier plane is unobserved.  A
second, related caveat is that of boundary effects. For small maps, the
lensing B-modes in the map may be sourced by E-mode polarization and
lensing features located outside the map region. The de-lensing would
then be incomplete near the boundaries, leaving some level of residual
B-modes in the map. Finally, there are caveats regarding foregrounds:
dust, synchrotron, and other foreground residuals may induce biases in
the lensing map and could also have non-trivial correlations with
large-scale dust residuals. The extent to which realistic levels of
foreground residuals can degrade the de-lensing efficiency or bias the
de-lensing procedure is currently a topic of active research.

\subsubsection{Fisher estimate for constraints on cosmological parameters}

\textsc{CMB4cast} adopts a Fisher matrix approach to estimate the scientific performance of a given configuration. Following, e.g., \cite{Smith2009}, the Fisher matrix element $F_{ij}$ for CMB spectra is written as
\begin{equation}
	\centering
		F_{ij} = \sum_{\ell=\ell_{\rm min}}^{\ell_{\rm max}} \frac{2\ell+1}{2} f_{\rm sky} {\rm tr} \left(  \boldsymbol{C}^{-1}_\ell \frac{\partial \boldsymbol{C}_\ell}{\partial p_i} \boldsymbol{C}^{-1}_\ell \frac{\partial \boldsymbol{C}_\ell}{\partial p_j}  \right) \, ,
	\label{eq:fisher_definition}
 \end{equation}
where $p_i$ and $p_j$ are two cosmological parameters, and the covariance matrix $\boldsymbol{C}_\ell$ is defined as
\begin{equation}
	\label{eq:covariance_definition} 
		 \boldsymbol{C}_\ell \equiv 
			 \begin{bmatrix}
			\bar{C}_\ell^{TT} + N_\ell^{TT} & \bar{C}_\ell^{TE} & 0 & C_\ell^{Td} \\
			\bar{C}_\ell^{TE} & \bar{C}_\ell^{EE} + N_\ell^{EE} & 0 & C_\ell^{Ed} \\
			0 & 0 & \bar{C}_\ell^{BB} + N_\ell^{BB} & 0  \\
			C_\ell^{Td}  & C_\ell^{Ed} & 0 & C_\ell^{dd} + N_\ell^{dd}
			\end{bmatrix},
\end{equation}
where $C_\ell$ are the various auto- and cross-power spectra of the CMB
temperature ($T$), polarization ($E, B$), and deflection ($d$)
components. In order to not double-count the lensing information encapsulated in the deflection field, we use only unlensed $T$, $E$, and $B$ information, as denoted by barred $C_\ell$s~\cite{Lesgourgues}. More details on the construction of the Fisher matrix are given in~\cite{Errard2016}.
In Eq.~\ref{eq:covariance_definition}, the diagonal elements of the
covariance matrix contain all of the Gaussian noise terms $N^{XX}_\ell$. For the components $X = \{T,E,B\}$, this noise power spectrum accounts for the effects of instrumental noise, imperfect foreground removal and, in the case $X = B$, de-lensing:
 \begin{equation}
 	\centering
		N^{BB}_\ell = N^{BB,\,{\rm  inst}}_\ell + C_\ell^{\rm res} + C_\ell^{BB,\ \rm delensed};
	\label{eq:ellnoise}
\end{equation}
$C_\ell^{\rm res}$ is parameterized as in Eq.~\eqref{eq:Clres_param} and
$C_\ell^{BB,\ \rm delensed}$ in Eq.~\eqref{eq:delensing_def}. As
mentioned in paragraph~\ref{ssec:fgs}, \textsc{CMB4cast} can derive all
of the Fisher constraints on cosmological parameters after marginalizing over $A_{\rm res}$ and $b_{\rm res}$. 
The instrumental noise power spectra,  $N^{XX,\,{\rm  inst}}_\ell$, are given by~\cite{Knox1995}:
\begin{eqnarray}
	\centering
		N^{XX,\,{\rm  inst}}_\ell  &=& \left[ \sum_\nu{  N^{XX,\,\nu}_\ell } \right]^{-1},\\
					{\rm with}\ \ \ \	 N^{XX,\,\nu}_\ell &\equiv& w_{X,\nu} \exp \left[- \ell(\ell+1) \frac{\theta^{\ 2}_{\textsc{fwhm},\nu}}{8\log2}\right]
	\label{eq:beamnoise}
\end{eqnarray}
where $w_{X,\nu}^{-1/2}$ is the instrumental white noise level of a given frequency channel $\nu$ in $\mu$K$_{\rm CMB}$-rad (see Eq.~\ref{eq:mapdepth}), and $\theta_{\textsc{fwhm},\nu}$ is the full-width at half-maximum beam size in radians. We assume fully polarized detectors, such that $w_E^{-1/2} = w_B^{-1/2} = \sqrt{2} w_T^{-1/2}$. Eq.~\eqref{eq:beamnoise} is only valid in its given format in the case of no component separation. For the realistic cases in which component separation is performed, we use the noise variance after component separation, as given in Eq.~\ref{eq:noise_variance_CMB_map}:
\begin{eqnarray}
	\centering
		N^{XX,\,{\text{post-comp-sep}}}_\ell  =   \left[  \left(\mathbf{A}^T\mathbf{N_\ell^{-1}}\mathbf{A}\right)^{-1} \right]_{\rm CMB \times CMB}
\end{eqnarray}
where the diagonal elements of $\mathbf{N_\ell}$ are given by $N^{XX,\,\nu}_\ell$ from Eq.~\ref{eq:beamnoise}.

The Fisher formalism allows forecasting of uncertainties that are either
conditional on the other parameters that take their fiducial values or
marginalized over the parameters that take any value. Conditional errors are given simply by the inverse of individual entries in the Fisher matrix, $1/\sqrt{F_{ij}}$; marginal errors, which we employ throughout, are given by inverting the Fisher matrix:
\begin{equation}
\sigma_i \equiv \sigma (p_i) = \sqrt{[\mathbf{F}^{-1}]_{ii}}.
\end{equation}

\subsubsection{Noise Modeling and Low-frequency Noise Excess}
\label{sec:low_ell_noise}
\textsc{CMB4cast} uses a generalized version of Eq.~\ref{eq:beamnoise} to include low-$\ell$ noise:
\begin{equation}
	\centering
		N^{XX,\,{\rm  inst}}_\ell  \rightarrow N^{XX,\,{\rm  inst}}_\ell \times \left[ 1 + \left( \frac{\ell_{\rm knee}}{\ell}\right)^{\alpha_{\rm knee}} \right]
	\label{eq:low_freq_excess}
\end{equation}
The actual parameters $\ell_{\rm knee}$ and $\alpha_{\rm knee}$ depend
on a variety of instrumental and environmental conditions: 
the aperture size; the field of view;
the observing site; scan strategy; polarization modulators; and temperature
stability of cryogenic stages, warm electronics, and optical elements.
In Sec.~\ref{sec:results}, we discuss the parameters we use for each configuration.

\section{Instrument and Cost Modeling}
\label{sec:technical}
In this section, we discuss the instrumental and cost models.
We strive to model the instrument as abstractly as possible in order to be agnostic to
the technical instrumental design choices that will come later.  
While we use the performance of existing instruments  to determine realistic choices for the model 
parameters, we do not favor any specific instrumental approaches.
The cost model defined here is simple and will need refinement in
future studies.  The cost estimate only includes major hardware components
and does not include labor costs for design, test,  and assembly.  The implicit assumption
is that the total cost will scale as a function of the underlying hardware costs.
We use the cost estimate as a metric for optimization,
which does not strive for absolute accuracy but can serve as a benchmark
that provides insight about how the cost optimization drives the instrumental
configuration.  For this reason, we use an abstract
unit,  the {\it Parametric Cost Unit (PCU)}, throughout this
paper.  One PCU is the equivalent of \$1M in raw hardware costs.
Further discussion about this unit can be found in
Sec.~\ref{sec:cost_model}.

\subsection{Detector Assumptions}
\label{sec:detector_assumptions}
\begin{figure}[hb]
\begin{center}
\includegraphics[height=1.5in,keepaspectratio]{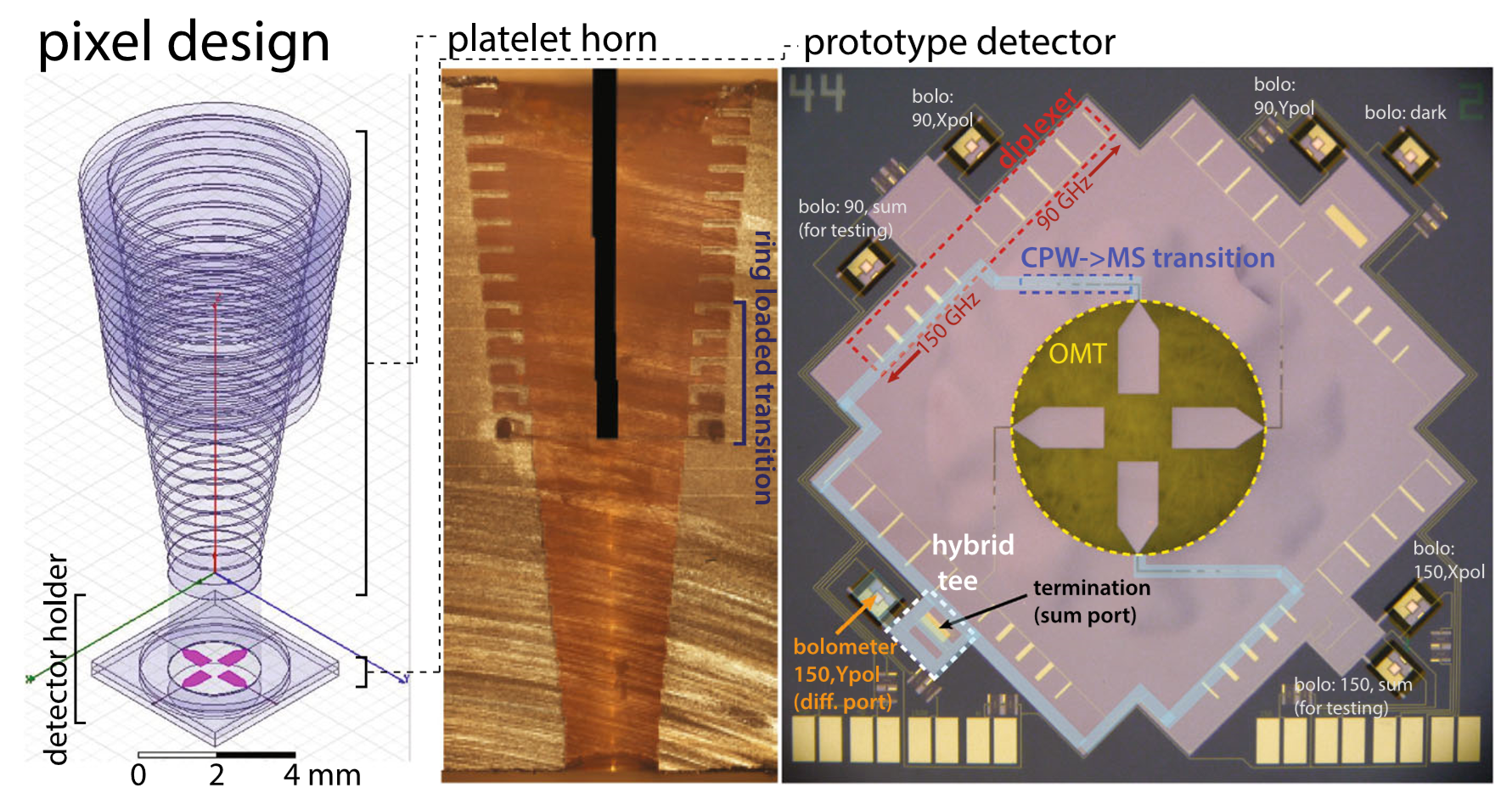}
\includegraphics[height=1.5in,keepaspectratio]{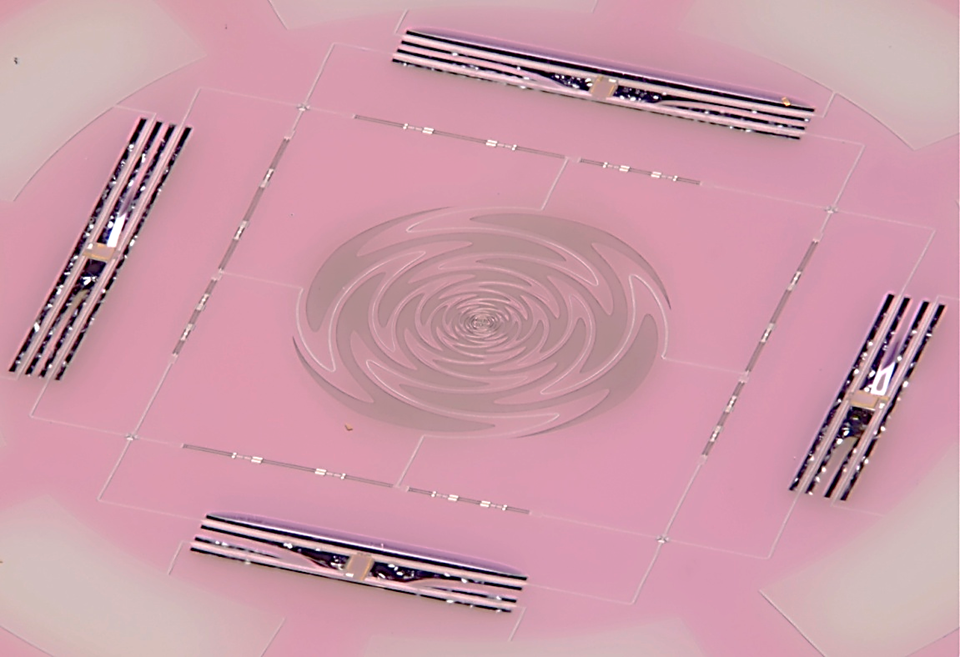}
\end{center}
\caption{Example of a multi-chroic pixel. (Left) Broadband horn coupled to broadband OMT. (Right) Broadband antenna at the center (sinuous antenna) captures wide range of frequency band. (Both) On-chip band-pass filter at T-junction partitions the broadband signal into different frequency bands. The signal is detected by the TES bolometers (dark rectangular object). \cite{2016arXiv160506569T, Suzuki_LTD15}}
\label{fig:sinuous}
\end{figure}

For this study, we adopted a model for the CMB-S4 experimental configuration that provides a realistic estimate of the detector performance for a given hardware cost.  For the detectors, we assumed the frequency bands and noise performance summarized  in Table~\ref{tab:NETSummary2}. 
We assume instruments are split into three groups of frequency bands:
low-frequency (LF), mid-frequency (MF), and high-frequency (HF) instruments.
Each group covers multiple frequency bands
 with one pixel
 (see Fig.~\ref{fig:sinuous} for examples of such technologies);
by measuring two orthogonal linear polarizations for each frequency band,
a single pixel in a LF, MF, and HF instrument
  is assumed to comprise 6, 4, and 4 detector channels, respectively.

In calculating the noise performance, or noise-equivalent temperature
(NET), we studied two receiver configurations.
The first configuration (Conf1) is for a small-aperture instrument and
assumes a fully cryogenic optics system.
The second configuration (Conf2) has two warm mirrors with multiple
cryogenically cooled lenses in the receiver; this configuration is
assumed for a large-aperture instrument.
For the atmospheric conditions, we assume a 1-mm
precipitible water vapor
(pwv) at 60\,degrees elevation at a site with an altitude of
$\sim 5000$\,meters.
More description on possible
observing sites can be found in Sec.~\ref{sec:site}.  Although the
environmental
conditions assumed above are closer to those at the Atacama desert
in Chile than that of the South Pole, the impact of the differences on
the detector sensitivities in Table~\ref{tab:NETSummary2} is small and
does not significantly change our optimization results.

We followed standard methods to calculate the photon noise, detector noise, and readout noise~\cite{Richards, Griffin:02}. Further details on the assumptions for NET calculation
  are given in Appendix~\ref{sec:detsens}.

In addition to the model presented in Table~\ref{tab:NETSummary2}, we
also looked at a ``staggered'' frequency band configuration that has two
different frequency schedules shifted by one-half of the bandwidth to
provide more spectral information
 (see, e.g., Ref.~\cite{2016JLTP..184..824M}).
In order to assess the merits of the different frequency configurations, it is necessary to implement foreground complexity beyond the simple power-law synchrotron and gray-body dust models.
This is an active area of research. For this note, we assumed the foreground model described in Section~\ref{ssec:fgs}.

\begin{table}[ht]
\centering
\caption{Summary of NET per pixel for 1 mm pwv at 60 deg.
 elevation. Conf1 refers to a receiver with fully cryogenic
 optics. Conf2 refers to a configuration with two warm mirrors and
 a three cryogenic lens system. \label{tab:NETSummary2}}
	\begin{tabular}{c c c c c}
	\hline
	\hline
	Pixel Type & Frequency & Frac BW & $ \mathrm{NET}_{\mathrm{CMB,Bolo}}^{\mathrm{Conf1,100mK}}$ & $ \mathrm{NET}_{\mathrm{CMB,Bolo}}^{\mathrm{Conf2,100mK}}$ \\
	                 & [GHz] & [\%] & [$\mu\mathrm{K}\cdot\sqrt{\rm s}$] & [$\mu\mathrm{K}\cdot\sqrt{\rm s}$]  \\
	\hline
        LF1 & 21 & 25 & 311 & 371 \\
        LF2 & 29 & 25 & 216 & 269 \\
        LF3 & 40 & 25 & 225 & 270 \\
        \hline  
        MF1 & 95 & 30 & 243 & 296 \\
        MF2 & 150 & 25 & 267 & 331 \\
        \hline  
        HF1 & 220 & 20 & 728 & 909 \\
        HF2 & 270 & 20 & 1237 & 1509 \\
	\hline
	\end{tabular}
\end{table} 

\subsection{Telescope Assumptions}
\label{sec:telescope_assumption}
Broadly speaking, there are three types of optical architectures that
are widely used in
the field of CMB polarimetry: offset-Gregorian~\cite{2012SPIE.8452E..3EM,2016arXiv160506569T,2014SPIE.9153E..1PB,2003ApJ...585..566P,2010A&A...520A...2T},
cross-Dragone~\cite{2013ApJ...768....9B,2010arXiv1008.3915E,2012SPIE.8444E..2YR,2012SPIE.8452E..1MT,doi:10.1117/12.2232008}, and cryogenic
fully-refractive~\cite{2014SPIE.9153E..1NA,2010SPIE.7741E..1OR} optics.
Offset-Gregorian designs are commonly adopted by large-aperture ($>1\,$m)
systems with warm reflectors.  Cross-Dragone designs are used both as 
large-aperture systems with warm reflectors or small-aperture systems
with cryogenic reflectors; they offer a more compact physical profile than
an offset-Gregorian system.
For large-aperture systems with warm reflectors, both 
offset-Gregorian and cross-Dragone designs may employ a cryogenic
corrector re-imaging lenses.
Cryogenic, fully refractive designs are commonly used for
small-aperture applications.
There are also possibilities other than those enumerated above; examples
include three-mirror anastigmat (TMA) optical designs.

We take a general approach to modeling the telescope without assuming a
specific architecture.  The telescope instrument is simply characterized
by its effective aperture size, $D_\mathrm{tel}$ (meters), and the number of pixels
it can accommodate, $N_\mathrm{pix}$.  For simplicity, we assume the
following:
\begin{itemize}
 \item Throughput scaling with wavelength and aperture: we assume the
       following relation because of the scale invariance of the
       electromagnetism in the optics design.  If a telescope with
       aperture $D_\mathrm{tel,1}$ can accommodate $N_\mathrm{pix}$
       pixels at a frequency $\nu_1$, or wavelength $\lambda_1 (= c/\nu_1)$, a telescope with aperture
       $D_\mathrm{tel,2} = ( \nu_1/ \nu_2 ) \cdot D_\mathrm{tel,1} = ( \lambda_2/ \lambda_1 ) \cdot D_\mathrm{tel,1}$
       accommodates the same $N_\mathrm{pix}$ pixels at a frequency of $\nu_2 (=c/\lambda_2)$.
 \item The full-width half maximum (FWHM) beam size,
       $\theta_\mathrm{beam}\,$ in arcmin, is related to the aperture size in m and the
       frequency $\nu$\ in GHz by
       $\theta_\mathrm{beam} = 3.5 \cdot (150/\nu) \cdot (2.5/D_\mathrm{tel})$.
 \item Each telescope is dedicated to either low-frequency
       (LF), mid-frequency (MF), or high-frequency (HF) pixels.
\end{itemize}
As for the last point, in principle it is possible to let LF, MF and HF
pixels coexist on a single focal plane, though we do not include this for simplicity.
However, we note that such a variation would simply result in a
reduction of the total telescope cost\footnote{
For example, a HF telescope with a fully populated focal
plane can accommodate some additional MF pixels around the edges 
of the HF region.
Using the telescope throughput model discussed below with $\alpha_1 = 0.5$,
the number of MF detectors around the HF pixel region
corresponds to $\sim 65\%$ of the detector count on
a dedicated MF telescope.}.
We investigate how such a change in cost could affect the optimization
results in later sections.
We also note that mixing LF, MF, and/or HF pixels may not necessarily be
optimum since some of the requirements on the telescopes, for example the
mirror surface roughness, will depend on frequency and
the cost advantage may be somewhat less than the 
naive savings calculated from a reduction
in the total number of telescopes.

The telescope throughput $N_\mathrm{pix}$ is modeled for MF pixels as
\begin{equation}
 N^\mathrm{MF}_\mathrm{pix} = C_\mathrm{pix} \left( \frac{D_\mathrm{tel}}{2.5} \right)^{\alpha_1}
  \label{equ:npix_telescope_power_law}
\end{equation}
assuming a power-law scaling.  According to the assumptions above, this can be generalized
for an arbitrary frequency $\nu$ as
\begin{equation}
  N^{\nu}_\mathrm{pix} = C_\mathrm{pix} \left( \frac{\nu}{125} \cdot \frac{D_\mathrm{tel}}{2.5} \right)^{\alpha_1}
  \:.
\end{equation}
Thus, the models for LF and HF pixels are
\begin{equation}
 N^\mathrm{LF}_\mathrm{pix} = C_\mathrm{pix} \left( \frac{29}{125} \cdot \frac{D_\mathrm{tel}}{2.5} \right)^{\alpha_1}
  \quad \mathrm{and} \quad
 N^\mathrm{HF}_\mathrm{pix} = C_\mathrm{pix} \left( \frac{250}{125} \cdot \frac{D_\mathrm{tel}}{2.5} \right)^{\alpha_1}
 \:,
 \label{equ:npix_telescope_power_law_lf_hf}
\end{equation}
respectively.

\begin{figure}[htbp]
 \begin{center}
  \includegraphics[width=0.6\textwidth]{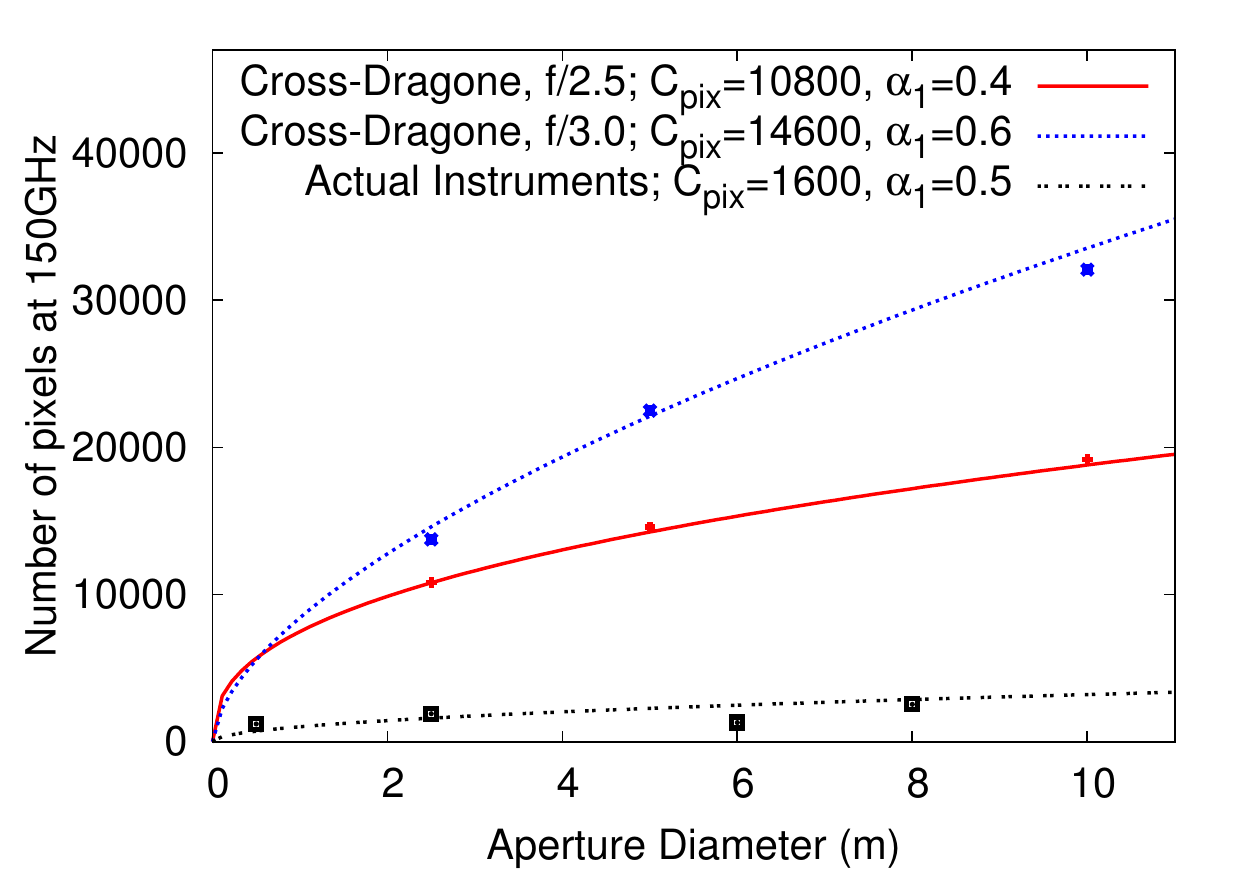}
  \caption{\label{fig:throughput_vs_aperture} 
  Relation between $N^\mathrm{MF}_\mathrm{pix}$ and the telescope
  aperture $D_\mathrm{tel}$ with some example data points.
  For cross-Dragone systems, two curves are shown based on
  Ref.~\cite{2016ApOpt..55.1686N}.
  Also shown are the instruments deployed or under construction.
}
 \end{center}
\end{figure}

A typical value for $\alpha_1$ is $0.4 \sim 0.6$.  The value of
$C_\mathrm{pix}$, on the other hand, can vary from $\sim 2000$ for
currently fielded offset-Gregorian systems to $\sim 15000$ for an
ambitious proposal adopting cross-Dragone optics~\cite{2016ApOpt..55.1686N}.
Figure~\ref{fig:throughput_vs_aperture} summarizes the relation between
$N_\mathrm{pix}$ and the aperture size for some examples.  
We will assume $\alpha = 0.5$ and $C_\mathrm{pix} = 5000$ as fiducial values.

The power-law scaling of the telescope throughput as a function of the
aperture size and the frequencies,
Eqs.~(\ref{equ:npix_telescope_power_law}) and (\ref{equ:npix_telescope_power_law_lf_hf}),
involves an
implicit assumption that the throughput is primarily limited by image
quality, or Strehl ratio, across the focal plane.
However, there are other throughput-limiting factors than the image
quality.
For example, 
geometric constraints may limit the
throughput for a dual-reflector optics.
For fully refractive optics,
a large-throughput configuration at small-aperture/low-frequency limit
may be achieved
from geometric optics and an aberration perspective.
However, such a configuration would involve 
a large range of incident angles and may result in inadequate performance
with standard anti-reflection coatings.
These factors come into play particularly at the small-aperture/low-frequency corner
of the parameter space, and thus the power-law scaling breaks down there.

In our study, a small-aperture ($D_{\rm tel} = 0.5$\,m) LF
instrument corresponds to this corner of parameter space, where
Eq.~(\ref{equ:npix_telescope_power_law_lf_hf})
with the fiducial values for $C_{\rm pix}$ and $\alpha_1$ yields $\sim 1000$ pixels,
corresponding to a focal plane diameter of approximately 1.2\,m.
To avoid this breakdown, we impose an additional throughput constraint
applied only to the small-aperture LF instrument: $N_{\rm pix}^{\rm LF} \leq 140$, or
$\lesssim 7$ wafers per  small-aperture telescope.
This will significantly affect the cost of the small-aperture LF instrument.
We will discuss the difference in the optimization results with
and without this additional throughput constraint in
Sec.~\ref{sec:with_and_without_lf_throughput_constraint}.
As we discuss in Sec.~\ref{sec:strawperson_2}, this configuration 
is likely to be suboptimal, and this is an area that requires
further study.

\subsection{Receiver Cryostat}
\label{sec:cryostat_model}
The receiver cryostat consists of a focal plane and cryogenic optics;
the latter can be either re-imaging optics or a cryogenic reflective
or refractive telescope.  The standard configuration of the cryogenics
is to combine pulse-tube cooler(s) and a sub-K refrigerator, where the
latter is typically a ${}^3\mathrm{He}$/${}^4\mathrm{He}$ sorption
refrigerator or a ${}^3\mathrm{He}$/${}^4\mathrm{He}$ dilution
refrigerator.
The two differ in the achievable temperature, cooling capacity (and thus the
number of pixels per unit), and cost.

Table~\ref{tab:cryogenics_options} shows some typical parameters of these 
refrigerator and cryostat  systems.  We list two entries for the
${}^3\mathrm{He}$/${}^4\mathrm{He}$ sorption refrigerator option
that correspond to different numbers of refrigerators per cryostat.
As can be seen in this table, the cost is similar for the
dilution-refrigerator and sorption-refrigerator options.  The
slightly higher cost of the dilution refrigerator is offset by the
reduction in detector noise when operating at the lower temperature.
  There are also other possibilities such as continuous
adiabatic demagnetization refrigerators, yet we expect no significant
differences in their per-cost capacity.

For the purpose of the optimization study, we only require
sensible assumptions regarding the capacity and cost of the cryostat and
cryogenic systems.  We select the dilution-based refrigerator system and 
adopt its capacity as listed in Table~\ref{tab:cryogenics_options} as the default
assumption.
As noted above, there is no significant difference between 
the refrigerator systems, and thus our optimization results are
approximately agnostic regarding this choice.
In practice, we expect the choice will be made not merely based on the cost
and capacity of the cryostat and cryogenics but will also be driven by the
ease of the detector fabrication requirements and cryogenic engineering.

While our basic assumption is one refrigerator per cryostat,
our model is also a good approximation for a configuration where
one cryostat is equipped with multiple refrigerators.
Large-aperture telescopes might adopt a large cryostat with multiple
refrigerators
that accommodate a  large number of detector
pixels~\cite{2016ApOpt..55.1686N}.  Since we will assume a dilution-based system,
the cost scaling will not depend strongly on
whether the system consists of one large
cryostat with $N$ refrigerators or $N$ cryostats with one refrigerator each.

\begin{table}[htbp]
 \begin{center}
  \caption{\label{tab:cryogenics_options} 
    Cryogenics parameters assumed in our optimization.  Note that the
  cooling capacity shown here is only for the coldest stage, and there
  are other factors that affect the number of pixels ($N_{\rm pix}$) that
  can be supported by a refrigerator.
  As noted in the main text, the number of pixels can be increased or
  decreased by varying the capacity of the sub-K fridge systems.
  Here, we show a couple of examples for the
  ${}^3\mathrm{He}$/${}^4\mathrm{He}$ sorption
  refrigerator.
  }
  \begin{tabular}{cccccc}
   \hline \hline 
   Type & Temperature & Capacity & Duty & Number of Pixels ($N_{\rm pix}$) & Cryostat Cost \\
   \hline 
   ${}^3\mathrm{He}$/${}^4\mathrm{He}$ dilution based & 100\,mK &
	   100\,$\mu \mathrm{W}$ & 100\% & 8,000 & \$1.0M \\
   ${}^3\mathrm{He}$/${}^4\mathrm{He}$ sorption based & 250\,mK &
	   10\,$\mu \mathrm{W}$ & 80\% & 2,000 & \$0.5M \\
    &                                   &
	   40\,$\mu \mathrm{W}$ & 80\% & 8,000 & \$0.7M \\
   \hline 
  \end{tabular}
 \end{center}
\end{table}

\subsection{Site and Observing Strategy}
\label{sec:site}
In our optimization study, we do not assume a specific site.  However, some aspects of
the study assume that a large fraction of the sky area is available, which would
require at least one mid-latitude site.

\begin{figure}[htbp]
 \centering
 \includegraphics[width=1.0\textwidth]{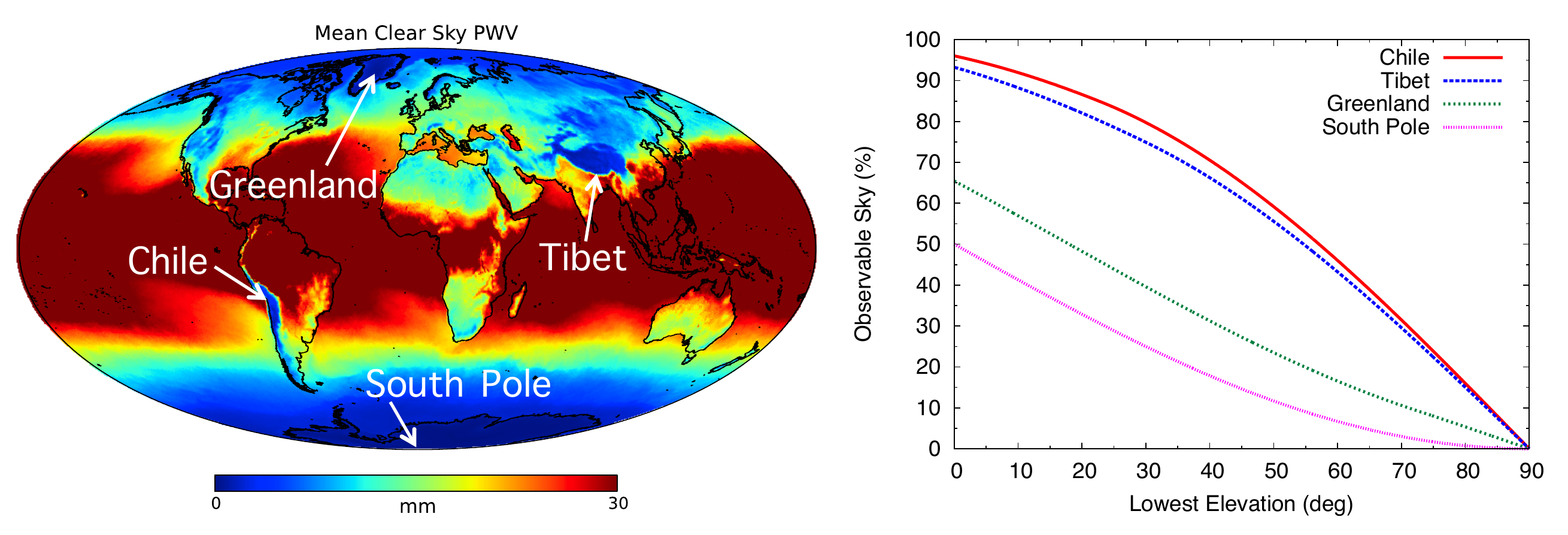}
 \caption{Left: Global distribution of mean clear-sky PWV. The original data are from \cite{6710211}, and the plot uses the Mollweide projection.
 Right: Observable sky as a function of the lowest elevation from a given site.
 The curves from top to bottom correspond to Chile~($23^\circ$S), Tibet~($30^\circ$N), Greenland~($72^\circ$N), and South Pole~($90^\circ$S).}
 \label{fig:pwv_and_observable}
\end{figure}

The two strongest candidates for the CMB-S4 site are the South Pole and the Atacama desert in Chile.
There is significant infrastructure and a well characterized site for CMB observations at the South Pole,  which has hosted a series of successful CMB polarization experiments, including
DASI, QuaD, BICEP / Keck Array, and SPT.  The weather condition is very dry,
stable, and consistent, and there is low atmospheric noise and low loading from precipitable water vapor~(Figure~\ref{fig:pwv_and_observable}),
which can reduce atmospheric noise due to the absorption and emission of water in observation frequencies.  These site characteristics are very important because the sensitivity of current and future experiments will be limited by photon noise.
Typically, the ``day-time season'' data at the South Pole are not
used for CMB observations.

The Atacama Desert in Chile is another excellent site for ground-based millimeter-wave observations;
there have been many successful experiments performed there, including ACT, ALMA, APEX, ASTE, CBI,
NANTEN, POLARBEAR, QUIET, and Simons Observatory.
The Atacama Desert also has very stable weather except for the ``Bolivian Winter'' from the end of December to early April.
Therefore the majority of the data are taken under very low atmospheric noise and low loading.
The mid-latitude location would have the advantage
 of being able to access a large fraction of the sky for observations up to 80\%~(Figure~\ref{fig:pwv_and_observable}).
A large-scale structure map of 80\% of the sky from CMB lensing would have the potential to map out most of the matter in the universe.

A survey from either Chile or the South Pole would overlap with premier optical surveys (e.g., 
DES, HSC, PFS, and LSST) and could provide a rich set of cross-correlation
science.

\subsection{Estimating integrated experimental sensitivity (map noise)}
\label{sec:realistic_sensit}

The integrated sensitivity of an experiment is given by the map noise achieved at the end of observations. This depends on the combined sensitivity of the detector arrays, ${\rm NET}_{\rm array}$, the length of observations, $T_{\rm obs}$, and the fraction of the sky observed, $f_{\rm sky}$.  To accurately predict the sensitivity of a potential instrument, we add estimates of the degradation in map depth based on the published achievements of ground-based CMB polarization experiments and realistic expected improvements. The achieved polarization map depth $\sigma$\footnote{
The polarization map depth $\sigma$, or the white-noise level of Q or U
polarizations, is worse than the temperature map depth $\sigma_I$ by a
factor of $\sqrt{2}$, i.e., $\sigma= \sqrt{2} \sigma_I$.
} at a single frequency band is given by Eq.~\ref{eq:mapdepth}, where $\varepsilon$ is the overall observing efficiency,  $\beta$ is the degradation to ${\rm NET}_{\rm det}$, and $Y$ is detector yield:

\begin{equation}\label{eq:mapdepth}
	\sigma ({\rm freq}) [{\rm \mu K.arcmin}] \equiv  3.07 \times
		  \sqrt{\frac{f_{\rm sky} }{T_{\rm obs} [{\rm yr}] \times \varepsilon }} \times
	\frac{\beta \times {\rm NET}_{\rm det}({\rm freq}) [{\rm \mu K . \sqrt{{\rm s}}}]}{\sqrt{N_{\rm det}({\rm freq}) \times Y}} ,
\end{equation}

For these forecasts, $T_{\rm obs}$ is 5 years for the total survey (see Section \ref{sec:survey} for a discussion of survey strategy). 
The observing efficiency, $\varepsilon$, is estimated to be 25\% based on the performance of Stage-2 CMB experiments, comparing published map depth to the achieved median ${\rm NET}_{\rm array}$.
This factor includes seasonal downtime (e.g., Bolivian winter, austral summer), other poor observing weather throughout the year, telescope maintenance and downtime, and data quality cuts.
The degradation in ${\rm NET}_{\rm det}$, $\beta$, is an estimate of the difference in achieved median ${\rm NET}_{\rm det}$ compared to the nominal ${\rm NET}_{\rm det}$ given in Table \ref{tab:NETSummary2}, which is calculated at an elevation of 60 degrees with 1\,mm of precipitable water vapor. 
There can be many sources of excess noise that will increase the achieved median ${\rm NET}_{\rm det}$, including the actual observing conditions and scan elevations, and achieved readout noise levels. We use a value of 1.15 for all frequency bands.
We assume $\beta = 1.15$ in our study.
We also include a factor corresponding the end-to-end yield of deployed detectors that send data into final maps, $Y$.
For Stage-2 CMB experiments, this yield of detectors in science results was approximately 50\% \cite{bicep2_instrument2014,PB_ClBB_2014}.
In this study, we estimated the yield to be 85\%, which would be a significant improvement over current achievements.
The yield of deployable detector wafers is included in our cost estimation, since we assume that wafers will be screened before deployment (see Section \ref{sec:cost_model}).
 More aggressive screening of wafers to ensure high on-sky yield is considered part of detector costs.
  Lower on-sky yield $Y$ than assumed here would lead to higher overall costs, either due to more required instruments (e.g., telescopes, cryostats) than assumed here or due to longer survey time needed to achieve the same final sensitivity.

With these combined degradation factors, the map depth is a factor of 2.5 higher than an ideal experiment.

\subsection{Cost Modeling}
\label{sec:cost_model}
We estimate the costs of the overall instrument by parameterizing and combining the cost of detectors and readout, telescopes, and cryostats.
We note that the cost model presented here is by no means mature or
established.  Our intention is to present an example  that
can be used to run the optimization process.  We anticipate the
community will work to establish more sophisticated cost models to
finalize the design of CMB-S4.
Our estimate only includes raw hardware cost and does not include labor
cost for component testing and integration.
Empirically, the actual
cost including labor is likely to be $2\sim 3$ times higher than
the raw hardware cost.  

In order to signify the fact that our cost model is simplistic and includes the raw
hardware cost only, we introduce the
 {\it Parametric Cost Model Unit (PCU)}.
One PCU is equal to one million dollars in our cost model.  Thus, when
labor is included, one PCU would roughly correspond to $2\sim 3$
million dollars.

\subsubsection{Detector costs}
The cost to fabricate a detector array with $O(500{,}000)$  detectors was estimated with the following assumptions:
\begin{enumerate}
\item We assumed a fabrication yield at the wafer level of 50\%, that is, two wafers must be fabricated 
to yield one science-grade wafer.
\item We conservatively estimated that one 150-mm wafer will hold 1{,}000 detectors when averaged over all frequency ranges; thus, 500 wafers are needed.  (Note that a multi-chroic pixel measuring two polarization modes at two frequencies will have four detectors.)   
\item We calculated the detector fabrication cost, including the capital investment, facility maintenance cost, support for fabrication engineers, support for equipment engineers, support for scientists, and supply cost, based on the detector fabrication experience from the current Stage-3 experiments.
\end{enumerate}

These assumptions lead to an estimate of approximately \$30\,M over 4 years to produce 1{,}000 wafers, yielding 500 science grade wafers.  Thus, the approximate cost per deployed wafer is $\sim$\$60K.
Assuming a focal-plane f/\# of $1.5\sim 2.0$, the wafer would have $\sim 20$, $\sim 300$, and $\sim 1200$ pixels for LF, MF, and
HF, resulting in a per-detector
cost of \$500, \$50, and \$12.5 for LF, MF, and HF, respectively.
Table~\ref{tab:wafer_cost_to_detector_cost} summarizes these assumptions.
It is important to note that this cost estimate does not include assembly, inspection, and testing costs.

\begin{table}[htbp]
  \begin{center}
   \caption{\label{tab:wafer_cost_to_detector_cost} Assumptions and
   per-detector cost for LF, MF, and HF detectors.
   We assume common per-wafer costs, yields, and focal-plane f/\# 
   for all  frequencies.}
    \begin{tabular}{ccccccc}
     \hline \hline
     Frequency & Per-wafer cost & Yield & f/\# & $N_{\rm pix}$ per wafer &
     $N_{\rm det}$ per pixel & per-detector cost \\
     \hline
     LF & 30k & 50\% & $1.5\sim 2.0$ & $\sim 20$   & 6 & $\sim$ \$500 \\
     MF &     &      &       & $\sim 300$  & 4 & $\sim$ \$50 \\
     HF &     &      &       & $\sim 1200$ & 4 & $\sim$ \$12.5 \\
     \hline
    \end{tabular}
  \end{center}
\end{table}

\subsubsection{Readout costs}
Readout systems for CMB detectors have been driven to high levels of multiplexing in order to reduce thermal loading on the cryogenic stages, as well as cost and complexity.  The cost for readout of the detectors is partly a linear function of the total number of detector channels, and some fixed costs are associated with shared multiplexing components like FPGAs and SQUID amplifiers.  
The current generation of frequency domain multiplexing used on several CMB experiments has multiplexing factors of $40\times$ to $68\times$. 
The readout costs for this system are approximately \$30-50 per channel for room temperature readout components and approximately \$30-50 per channel for cryogenic readout components, including all hybridization and interconnect costs.  
Increasing the multiplexing factor by a factor of two to three, which is possible with modest development efforts, would reduce total readout costs per channel by a similar factor.  
For this cost model, we estimate the readout costs at \$20 per channel (i.e., a factor of four improvement from current costs) based on these anticipated improvements in multiplexing as well as cost benefits from scaled up production of readout components.
These estimated costs include only the manufacturing costs for readout
hardware and exclude development cost, the labor required for
integration, and characterization necessary for the readout system.

\subsubsection{Telescope costs}
\label{sec:telescopecost}
The telescope cost includes the warm optics system as well as the
telescope mount system.
We model the baseline cost of a telescope by a power law using an index
$\alpha_{tc}$:
\begin{equation}
 M_{\rm tel}^0 = C_{\rm tel} \left( \frac{D_{\rm tel}}{2.5} \right)^{\alpha_{tc}} \:.
\end{equation}
This model breaks down at small apertures.
For a small-aperture system where the optics are fully
cryogenic, the only cost associated in this category is the drive
system, which we estimate to be $\sim $\$200k each.  On the other
hand, a 0.5-m telescope costs only \$40k with the above parameters.
To amend this breakdown, we define the telescope cost as follows:
\begin{equation}
M_{\rm tel}^0 = C_{\rm tel} \left( \frac{D_{\rm tel}}{2.5} \right)^{\alpha_{tc}} + \mathrm{\$200k} \:.
\end{equation}
Note that the cryogenic optics cost is commonly included in the cryostat cost
for both large aperture systems with warm mirrors and small
aperture systems with fully cryogenic optics.  Thus, we assume that the
``telescope cost'' of the small aperture system is dominated by the
drive system.

\begin{figure}[htbp]
\begin{center}
 \includegraphics[width=0.65\textwidth]{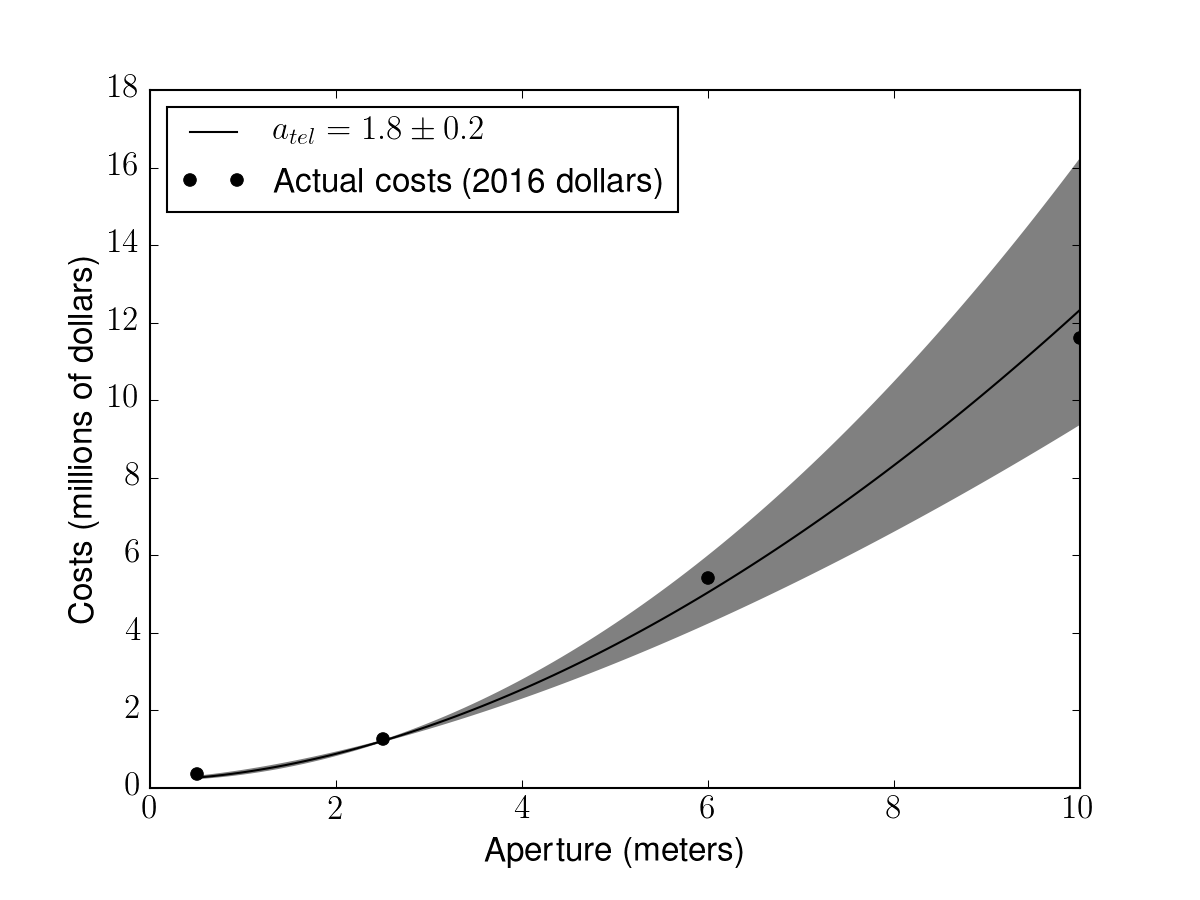}
 \caption{\label{fig:telescopecosts_actual} 
 Comparison of telescope cost model (curve) with historical CMB
 telescope costs (data points).
 The central curve shows a power-law model with
 $\alpha_{tc} = 1.8$ and $C_{\rm tel} = $\$1M,
 and the gray shaded area corresponds to $\alpha_{tc} = 1.8 \pm 0.2$.
 The points are based on the experience
 in the field and in 2016 dollars with a correction for nominal inflation.
 }
\end{center}  
\end{figure}

It is empirically known that $C_{\rm tel} \sim $\$1M and $\alpha_{tc}$ is
$1.5\sim 2.0$.  In this model, a telescope with an effective aperture of
2.5\,m (6\,m) costs \$1M (\$4M $\sim$ \$6M).
In our study, we set $C_{\rm tel} = $\$1M and
$\alpha_{tc} = 1.8$.  
As shown in Fig.~\ref{fig:telescopecosts_actual}, this roughly reflects
the
experience in the field~\cite{BICEPcost,PBcost,ACTcost,SPTcost},
where we corrected for inflation factor\footnote{Costs in 2016 dollars calculated using the U.S. Bureau of Labor Statistics CPI inflation calculator (www.data.bls.gov)}.
We will explore the possible impact of the error in these parameters
on the optimization results.  The power law index
$\alpha_{tc}$ is varied by $\pm 0.2$.  We also vary the telescope
throughput parameter $C_{\rm pix}$,
 a scale factor for the number of pixels per telescope
 in Eq.~\ref{equ:npix_telescope_power_law},
from the nominal value of 5000 to 2000 and
14000.  This is equivalent to varying the overall telescope cost $C_{\rm tel}$
by a factor of $\sim 2.5$.

We note that there are two power-law indices involved in the telescope
modeling: the throughput scaling index $\alpha_1$ in
Eq.~(\ref{equ:npix_telescope_power_law}) and
the cost scaling index $\alpha_{tc}$.  
These two parameters are degenerate.  The important parameter is
the power-law index of the telescope cost per pixel:
$\alpha_{tc} - \alpha_1$.  The uncertainty in $\alpha_1$ is relatively
minor since $\alpha_{tc}$ has a larger uncertainty.

In practice, we expect a cost break (both as the cost itself and its
derivative) at around $D_{\rm tel}$ of $\sim 6$\,m due to a transition from
a monolithic mirror to a segmented mirror.  
The transition would correspond to a physical size of $\sim 7$\,m;
$D_{\rm tel}$ is the illuminated and effective aperture size, and
the corresponding physical mirror diameter
is larger for offset systems typically employed for CMB telescopes.
A mirror of composite material (e.g., carbon fiber) is likely to
follow a different cost model.  Further study in this area is needed.

\subsubsection{Cryostat costs}
\label{sec:cryostat_cost}
In the cryostat cost,  we include all mechanical and cryogenic components that support the focal plane, cold optics, and cryocoolers. The cost of the cryostats is also roughly a function of the number of detectors, but there are also fixed costs associated with each individual cryostat and its cryocoolers as well as limitations in the number of pixels and detectors that can be supported by each cryostat. We assume no major improvements in technology but only optimization of existing technologies. We parameterize the cost $M_{\rm cryo}$ in Equation \ref{eq:cryocost} with $N_{\rm cryo}$, the total number of cryostats, and $C_{\rm cryo}$, the fixed cost per cryostat. The number of cryostats needed for each telescope is determined by the number of pixels illuminated by the telescope design, $N_{\rm pix}$, and the number of pixels that can be accommodated by a single independent cryostat, $N_{\rm pix}^{\rm max}$, as given in Equation \ref{eq:n_cryo}. 
	\begin{equation}\label{eq:cryocost}
		M_{\rm cryo} = C_{\rm cryo} \times N_{\rm cryo}^{\rm tot}
	\end{equation}
	\begin{equation}\label{eq:n_cryo}
	    N_{\rm cryo} = \frac{N_{\rm pix}}{N_{\rm pix}^{\rm max}}
	\end{equation}
In this study, we assume the use of a dilution refrigerator with  maximum
number of pixels per cryostat $N_{\rm pix}^{\rm max} = 8{,}000$,
as discussed in Sec.~\ref{sec:cryostat_model}.
We also impose a constraint that the number of pixels per cryostat is not greater than the number of pixels illuminated by the telescope design; that is, each telescope has at least one cryostat. 
This reduces $N_{\rm pix}^{\rm max}$ for some of the configurations.
For most of the configurations in our model, the optical throughput is well matched to the cryostat capacity, and this has a small effect.
For smaller apertures at low-frequency, where the telescope throughput is limited to less than 7 wafers (as described in Section \ref{sec:telescope_assumption}), and where the telescope cost is smaller than the cryostat cost, this constraint would lead to cryostat costs dominating the overall cost.
For the results in Section \ref{sec:hybridresults}, and the configuration described in Section \ref{sec:strawperson_2}, we removed this constraint specifically for the 0.5 meter LF instrument (i.e., the small aperture LF instrument can have one cryostat with many telescopes). 
We explore the effect of changes in the cost modeling of the small aperture on forecast results, including this constraint, in Section \ref{sec:with_and_without_lf_throughput_constraint}.

\subsection{Cost per Mapping Speed and Aperture Scaling}
The instrument and cost modeling approaches described above already have some
implications regarding the instrumental configurations.
These allow us to narrow down the parameter space that we will explore  in the next section,
specifically with regard to the telescope aperture scaling as a function of the
frequency.

\begin{figure}[htbp]
 \begin{center}
  \includegraphics[width=0.3\textwidth]{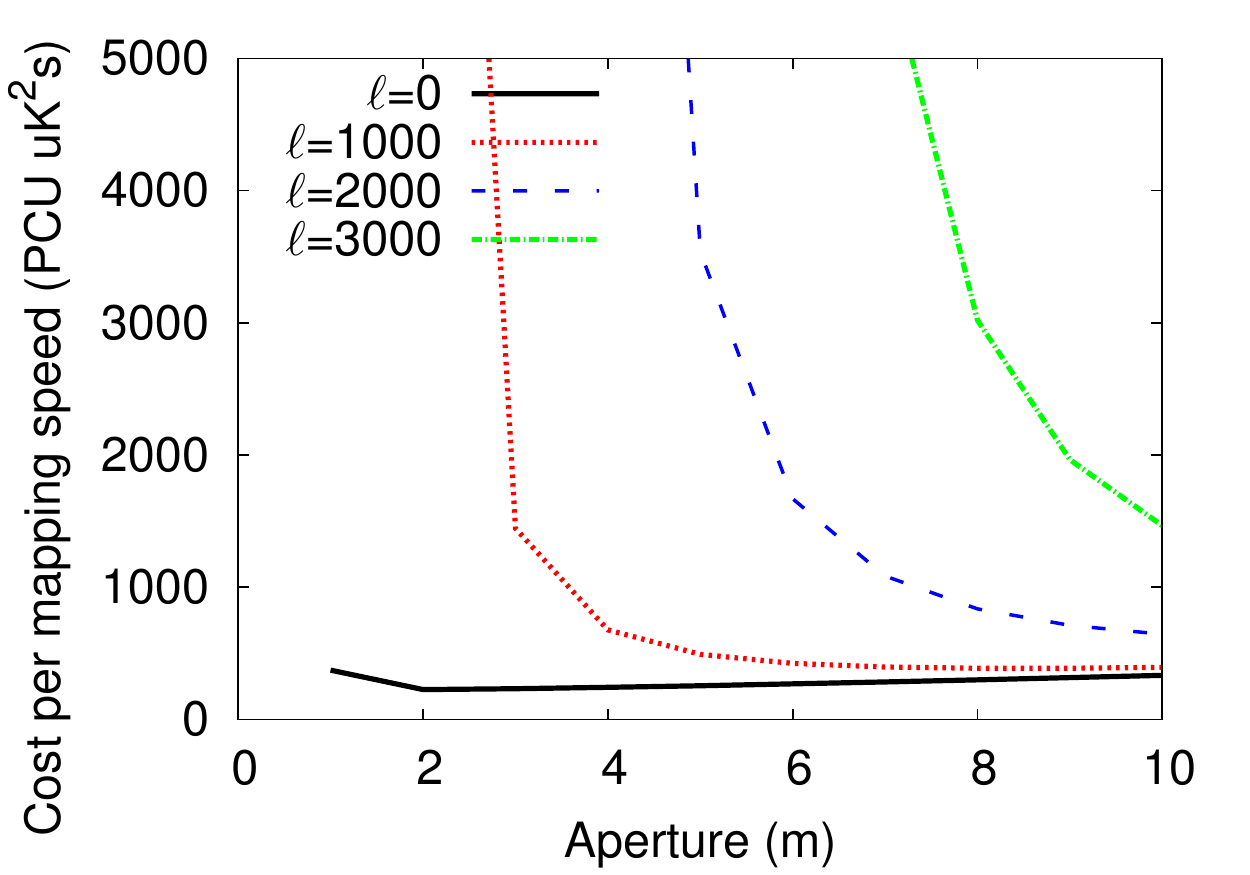}
  \hspace{0.02\textwidth}
  \includegraphics[width=0.3\textwidth]{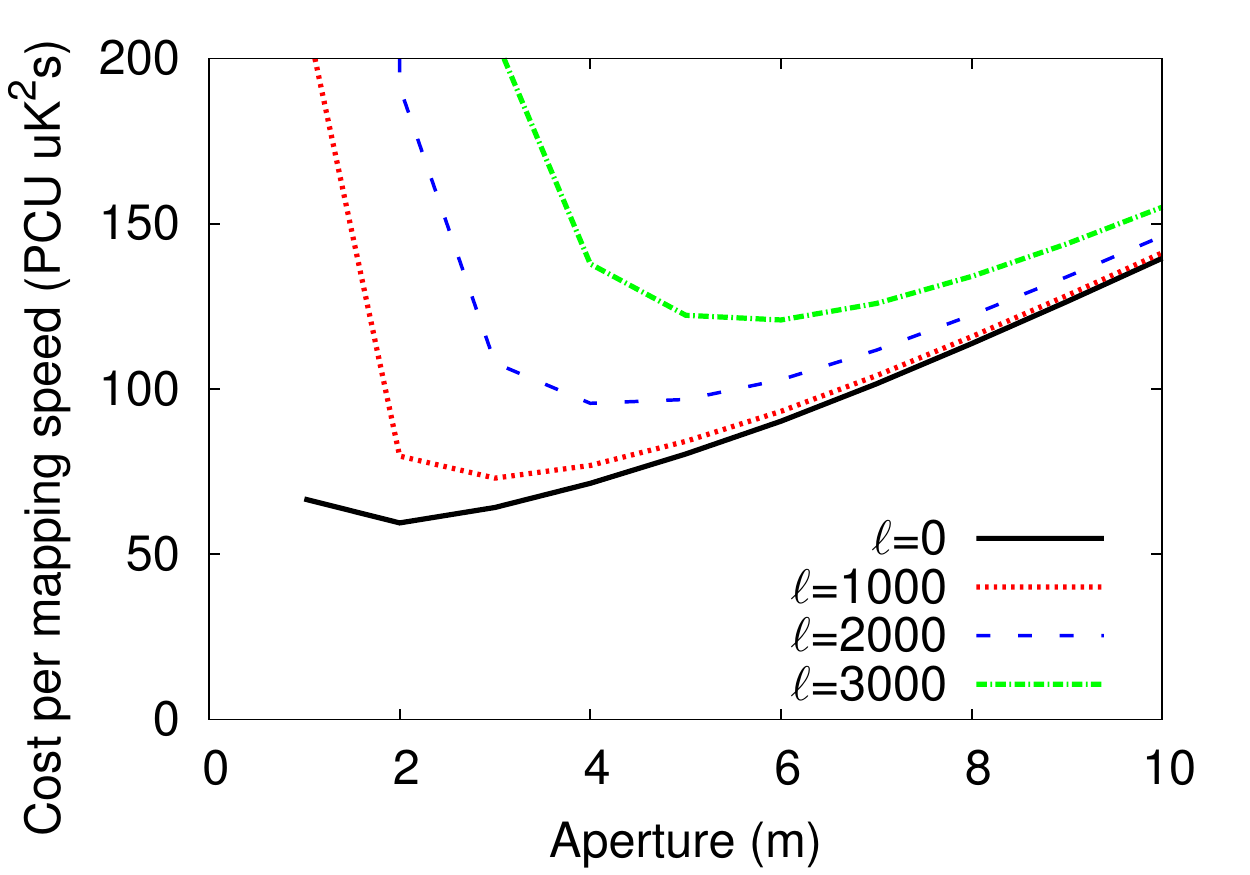}
  \hspace{0.02\textwidth}
  \includegraphics[width=0.3\textwidth]{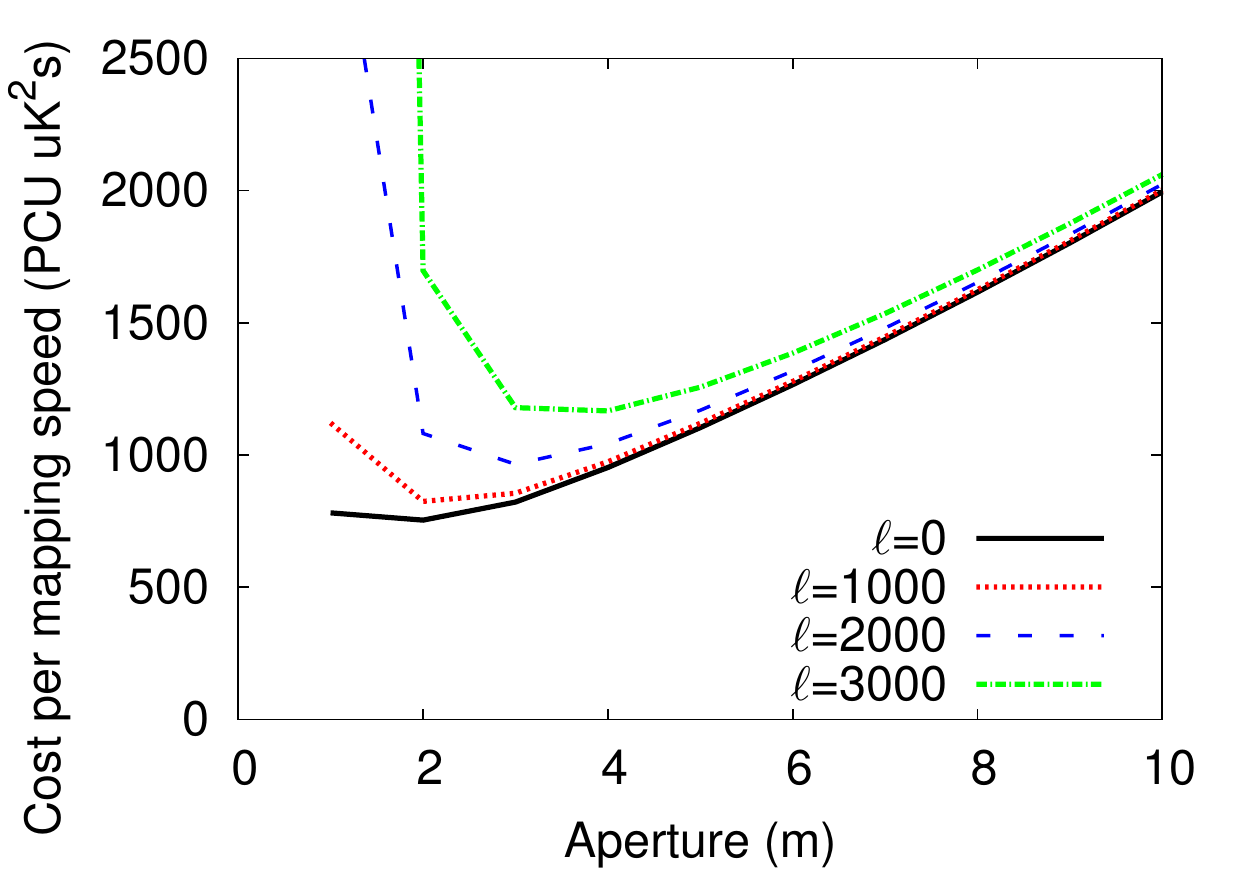}
  \caption{
  \label{fig:cost_per_ms_and_beam}
  Cost per mapping speed for 40\,GHz (left), 150\,GHz (middle),
  and 270\,GHz (right)
  for a configuration with a large aperture telescope with two warm
  mirrors
  and a dilution refrigerator.
  Different curves correspond to different $\ell$.
  }
 \end{center}
\end{figure}

Figure~\ref{fig:cost_per_ms_and_beam}
shows the total cost (PCU) per mapping speed ($1/\mu\mathrm{K}^2 \cdot s$),
or the sensitivity squared,  as a
function of $D_{\rm tel}$ for the
 dilution-based systems.
We show 
the mapping speeds over the full range of $\ell$, by applying a beam
window function $\exp \left[ -2b \ell (\ell + 1) \right]$
with $b = \mathrm{fwhm}^2 / (16 \log 2)$; the fwhm is in radians.
In Fig.~\ref{fig:cost_per_ms_and_beam}, we present some examples for different frequencies 
as well as possible cost variations.
No $1/f$ noise or low-frequency noise excess is accounted for in these figures. 
From these plots, we see that the optimal aperture shifts towards larger apertures
at lower frequencies and at higher values of $\ell$ due to the beam.
The optimal aperture does not follow na\"{i}ve scaling by
wavelength due to the increase in telescope cost with aperture.
This is in particular the case for the LF telescope,
where the telescope size tends to be large, and the cost increase tends
to be steep.
For example, the optimal aperture sizes for $\ell=1000$
are $\sim$8\,m, $\sim$3\,m, and $\sim$2\,m for LF, MF, and HF, respectively.
Based on this trend, we study the following two configurations in the next section:
\begin{enumerate}
 \item Fixed aperture size, $D_{\rm tel}$, for all LF, MF, and HF
       telescopes.
 \item Aperture sizes scaled by factors of two: $2 D_{\rm tel}$, $D_{\rm tel}$,
       and $D_{\rm tel}/2$ for LF, MF, and HF, respectively.
\end{enumerate}
As we will see later, these choices lead to only minor
differences in the optimization results.

\section{Optimization Results}
\label{sec:results}
In this section, we present results from a variety of optimization exercises in which
we use the modeling approach described in Section~\ref{sec:methodology}, combined
with the technical and cost framework described in Section~\ref{sec:technical}, to 
determine how to optimize the CMB-S4 experimental configuration to maximize scientific 
performance with a fixed cost constraint.
This will necessarily be an iterative process, given the large number of
experimental parameters and technical issues to explore.  We will provide
some examples, study various trends, and point out areas for future study.

The CMB-S4 experiment will consist of an array of telescopes covering a wide
range in frequency bands in order to provide sufficient characterization of foregrounds.  
The performance will depend on the instrument configuration and on the 
survey strategy, which will include both deep coverage over small fields (to optimize
the inflation sensitivity) and wide but shallower coverage (to study large scale structure phenomena).  

In the following sections, we will generally assume that the instrument spends 2.5 years
on small-sky observations ($f_{\rm sky}=$5\%) and 2.5 years on wide-sky
observations ($f_{\rm sky}=$50\%).  We will later discuss varying these fractions;
the optimum has a broad minimum that is generally consistent with this assumption.

\subsection{Types of Configurations}
\begin{figure}[htbp]
 \begin{center}
  \includegraphics[width=0.95\textwidth]{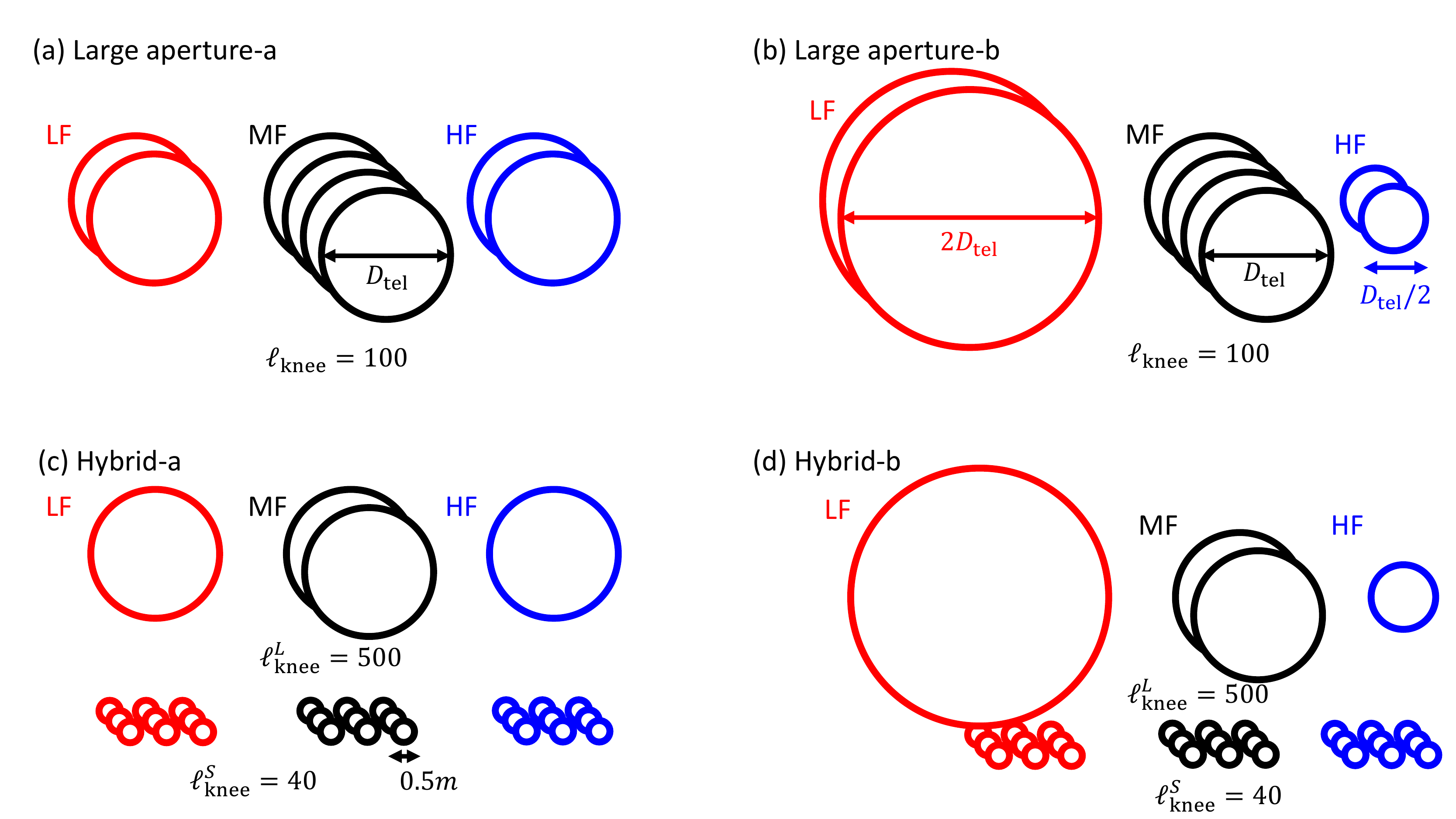}
  \caption{\label{fig:configuration_types}
  Schematic figure symbolically showing four types of configurations that we consider
  in the optimization study.  Each circle symbolizes telescopes and
  their aperture size.
  (a) {\it Large aperture-a}: a homogeneous telescope array with same aperture
  size ($D_{\rm tel}$) across all the frequency bands.
  We set the nominal knee frequency to be $\ell_{\rm knee} = 100$.
  (b) {\it Large aperture-b}: a  telescope array with scaled aperture
  sizes ($D_{\rm tel}$ for MF) over the frequency bands and with
  $\ell_{\rm knee} = 100$.
  (c) {\it Hybrid-a}: a hybrid telescope array that mixes large-aperture and
  small-aperture telescopes.  The large telescopes have
  the same aperture size ($D_{\rm tel}$) across all the frequency bands,
  while the small telescopes have an aperture size of 0.5\,m.
  The knee frequencies ($\ell_{\rm knee}$) are set to 500 and 40
  for large- and small-aperture telescopes, respectively.
  Nominally, the cost is split between the large and small telescopes
  in equal parts (i.e., half and half), resulting in a 50\% number of large-aperture
  telescopes/detectors compared to {\it Large aperture-a}.
  (d) {\it Hybrid-b}: a hybrid telescope array, where the large 
  telescopes have a scaled aperture ($D_{\rm tel}$ for MF) over the
  frequency bands and the small telescopes have an aperture size of
  0.5\,m.
  The knee frequencies ($\ell_{\rm knee}$) are set to 500 and 40
  for large- and small-aperture telescopes, respectively.
  Nominally, the cost is split equally between the large and small telescopes.
  }
 \end{center}
\end{figure}

In the following optimization study, we study four types of
instrument configuration (Fig~\ref{fig:configuration_types}).
The configurations are broadly categorized into large aperture arrays (a)
and (b), and hybrid arrays (c) and (d).  The large aperture arrays
measure the entire angular scale, or $\ell$ range of approximately
30--4000, with apertures of diameter 2 - 10 m, assigning a single size telescope to each frequency band.  On the
other hand, the hybrid arrays split the angular scales into two  regions,
$30\lesssim\ell\lesssim 400$ and $400\lesssim \ell \lesssim 4000$,
which are measured by the small  (0.5 m) and large (2--10 m) telescopes, respectively.

The collective experience of the CMB community suggests that
small telescope apertures perform better at larger scales, in particular at
the degree angular scales where the primordial gravitational signal should be present.
This trend is characterized by a smaller value of $\ell_{\rm knee}$, as defined in
Eq.~\ref{eq:low_freq_excess}.
However, this relation is not simple nor proven,
as  $\ell_{\rm knee}$ depends on a variety of instrumental and
environmental conditions in addition to the aperture size.
These factors include
the field of view (typically correlated to the aperture size);
observing site; scan strategy; use of polarization modulators; and the temperature
stability of the cryogenic stages, warm electronics, and optical elements.

It is beyond the scope of this paper to analyze this issue in detail.
Therefore, we take an empirical approach and investigate both large aperture arrays and
hybrid arrays, covering a large parameter space in the 
possible $\ell_{\rm knee}$ dependence for the instrumental 
configurations.  
Eq.~\ref{eq:low_freq_excess} also defines a power law index $\alpha_\mathrm{knee}$, which we fix at $\alpha_\mathrm{knee}=3.0$ for this study.
 For hybrid arrays, we assume 
$\ell_\mathrm{knee}^\mathrm{S} = 40$ and
 $\ell_\mathrm{knee}^\mathrm{L} = 500$
for small and large aperture telescopes, respectively.
These are roughly consistent with values that have already been achieved by 
existing CMB instruments.\footnote{
For small aperture, $\ell_\mathrm{knee}^\mathrm{S} = 40$
approximates the $\ell$ dependence of the error bars on
$C_\ell$ achieved by BICEP2 and Keck Array~\cite{2016PhRvL.116c1302B}.
The error bars on $C_\ell$ are used to determine
$\ell_\mathrm{knee}^\mathrm{S}$, this $\ell_\mathrm{knee}^\mathrm{S}$
empirically includes both of the effects from the noise increase and
mode decrease due to filtering.
For large aperture, $\ell_\mathrm{knee}^\mathrm{L} = 500$
is a conservatively large number.  We use 500 so that
the large aperture telescope only contributes to the high-$\ell$
observation, such as those for delensing in the {\it Hybrid}
configurations.}
For large-aperture configurations, we use
 $\ell_\mathrm{knee} = 100$ as a fiducial value.
 We will also explore variations in $\ell_\mathrm{knee}$
 and study how the results depend on it
 in Sec.~\ref{sec:lknee_vary}.
This analysis shows that $\ell_\mathrm{knee} < 100$ is required for a large
 aperture array to be competitive with a hybrid array of the same cost.
 While $\ell_\mathrm{knee} = 100$ with a large aperture telescope has
 not yet been demonstrated,
 we find that this is a good target for this type of array.

\subsection{Large Aperture Telescope Array Configurations}
We consider two types of large aperture arrays:
{\it Large Aperture-a} (Fig.~\ref{fig:configuration_types}a)
with the same (or fixed) aperture size across all the frequency bands,
and {\it Large Aperture-b} (Fig.~\ref{fig:configuration_types}b)
with the scaled aperture sizes over frequency bands: $2D_{\rm tel}$,
$D_{\rm tel}$, and $D_{\rm tel}/2$ for LF, MF, and HF, respectively.  
As described above, we use $\ell_{\rm knee} = 100$ as a fiducial value,
and later explore variations of $\ell_{\rm knee}$.

\subsubsection{Frequency Combination and Aperture Scaling}
\label{sec:freq_combination_and_aperture}

Here, we optimize for the weighting among LF (20--40 GHz), MF (95--150
GHz), and HF (220--270 GHz) instruments
for a fixed cost of 50\,Parametric Cost Units (PCU).
These bands are defined in Table~\ref{tab:NETSummary2}.  
We assume an aperture size
of $D_{\rm tel} = 6\,\mathrm{m}$, which is sufficiently near the optimum,
as we will later show.
We compare the errors on $r$ and $N_{\rm eff}$ as
a function of the ratio of the number of detectors in the three
frequency bands.

\begin{figure}[htbp]
 \begin{center}
  \includegraphics[width=0.45\textwidth]{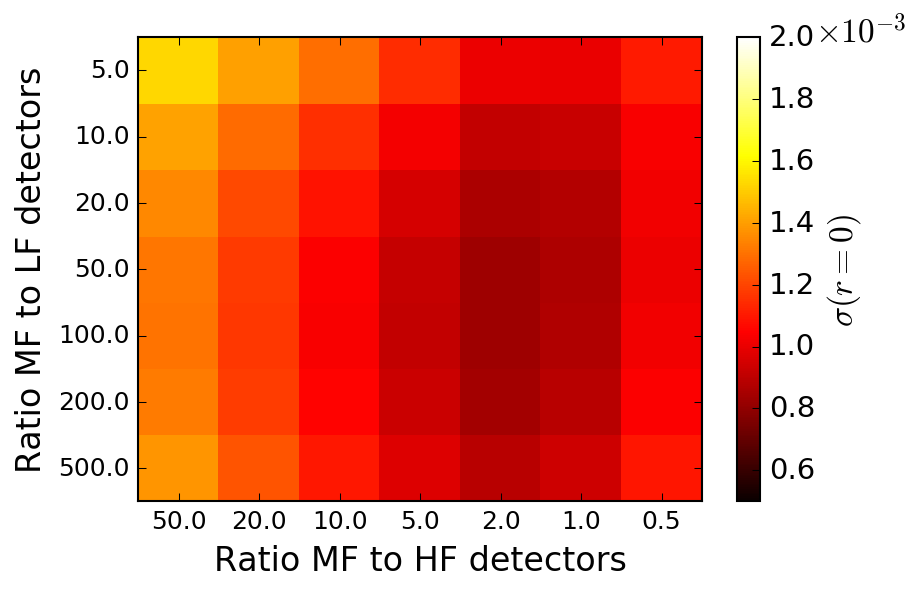}
  \includegraphics[width=0.45\textwidth]{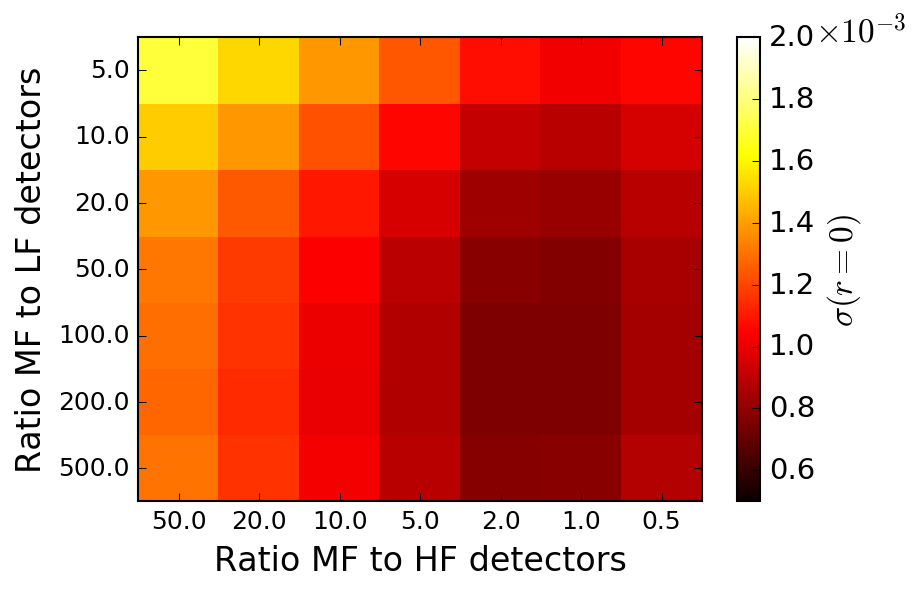}
  \\
  \vspace{0.5cm}
  \includegraphics[width=0.45\textwidth]{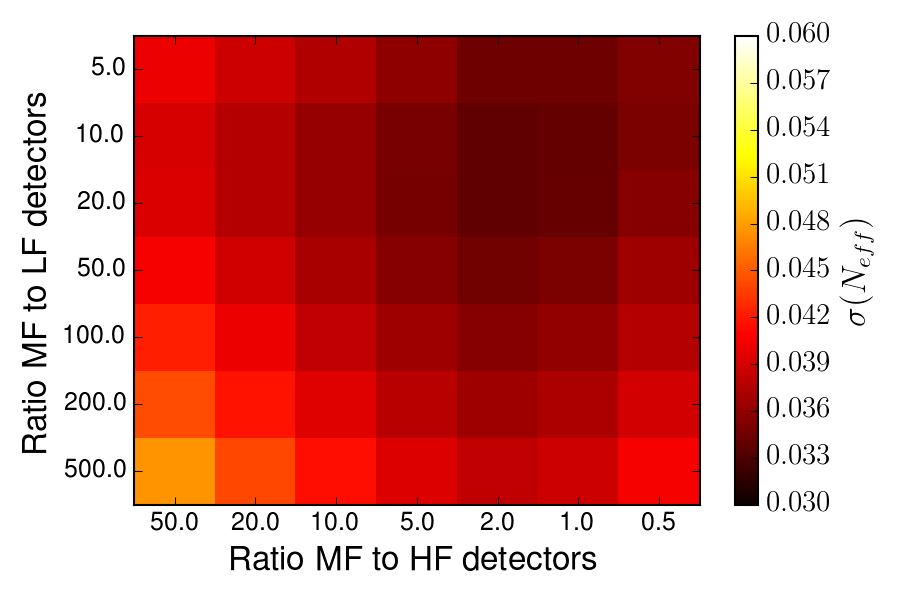}
  \includegraphics[width=0.45\textwidth]{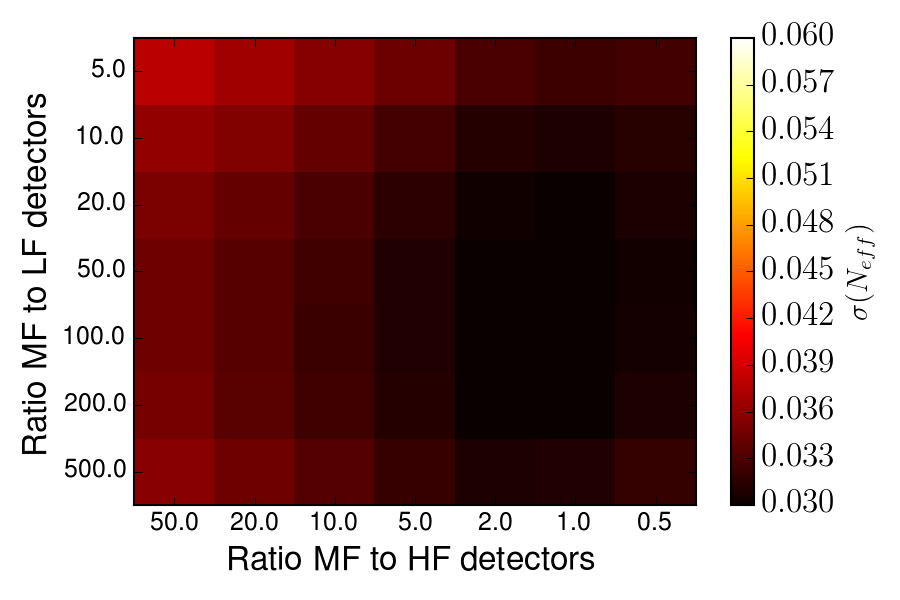}
  \caption{\label{fig:frequency_optimization} 
  Optimization over the frequency weighting;
  $\sigma(r)$ (top)
  and $\sigma(N_{\rm eff})$ (bottom)
  as a function of the ratios of MF/LF and MF/HF.
  A 5\% (50\%) sky coverage is assumed in optimizing for $r$
  ($N_{\rm eff}$).
  The left panels show the case where the aperture size of all the
  frequencies are fixed to 6\,m ({\it Large aperture-a} with
  $D_{\rm tel}=6$\,m).
  The right panels show the case
  where the aperture sizes are scaled as 12\,m, 6\,m, and 3\,m for LF,
  MF, and HF, respectively ({\it Large aperture-b} with $D_{\rm tel}=6$\,m).
  The aperture scaling used in the latter is near the optimum; we explored
  varying levels of the
  scaling and did not find a strong improvement beyond this level.
  For both, we find that MF/LF and MF/HF of 20 and 2 are near the optimum.
  }
 \end{center}
\end{figure}

Figure~\ref{fig:frequency_optimization} shows the expected error on $r$ and
$N_\mathrm{eff}$ as a function of the ratio of MF/LF and
MF/HF.\footnote{
Note that the sub-band ratio within LF, MF, and HF (i.e., the
ratio of LF1:LF2:LF3 etc.) is kept at unity.}
Both figures have
shallow minima around MF/LF= 10 -- 200 and MF/HF= 1 -- 5
for both the fixed aperture size ({\it Large aperture-a}) and
the scaled aperture size ({\it Large aperture-b}).  
We choose MF/LF=20 and MF/HF= 2  for the frequency band ratios in the following.
We have also explored different aperture scalings as variations of
 {\it Large aperture-b} while performing this frequency weighting
 optimization, but we did not find strong improvement beyond the nominal
 scaling we show here.
We note that here we have assumed a simple foreground model with a
power-law scaling for both synchrotron and gray-body dust.
More complicated foreground models might move the optimum; this is a topic
for further study.

\begin{figure}[htbp]
 \begin{center}
  \includegraphics[width=0.6\textwidth]{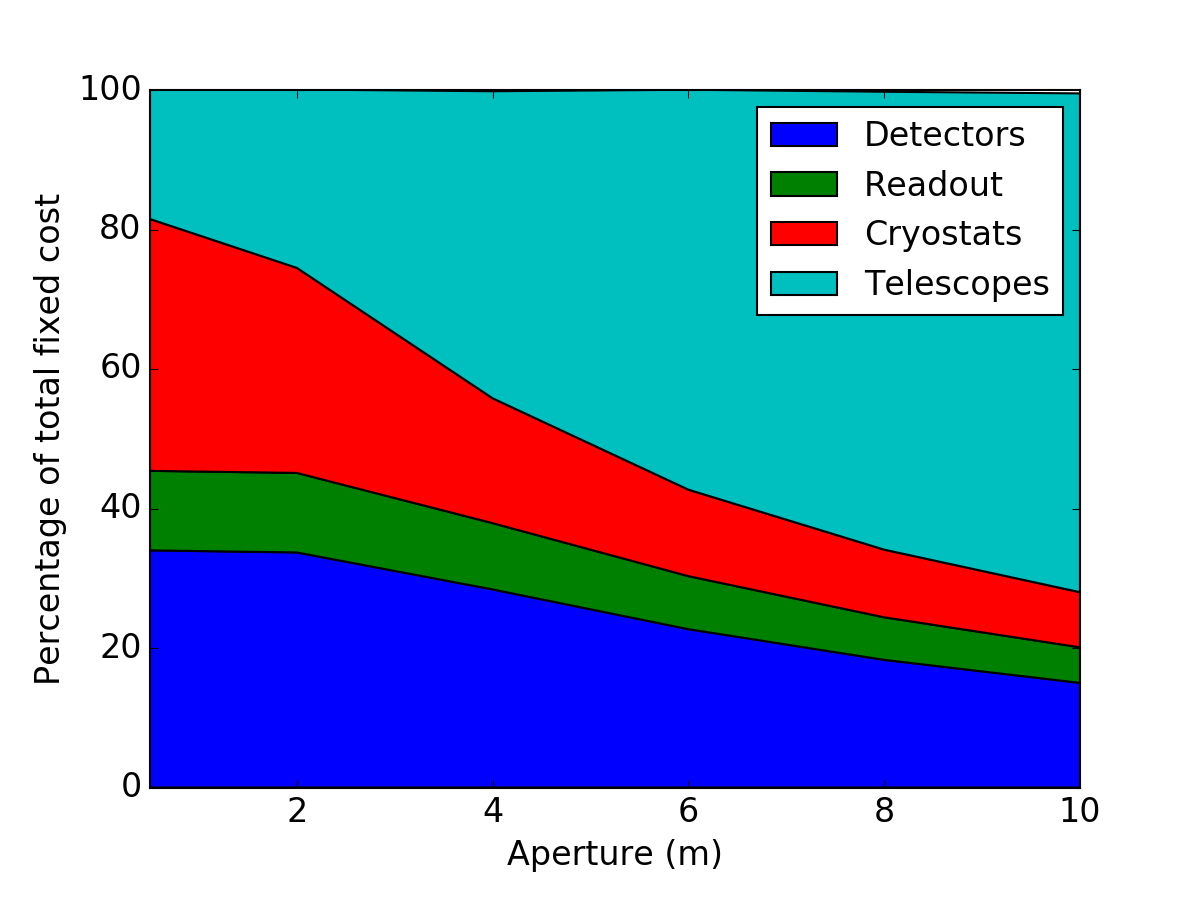}
  \caption{\label{fig:cost_distribution} 
  The cost distribution over telescope, cryostat, detector, and readout
  for the fixed aperture configurations ({\it Large aperture-a}) as a
  function of the telescope aperture size.
  }
 \end{center}
\end{figure}

Once the frequency weighting is fixed, the cost distribution among each
of the subsystems is uniquely determined in our model.
Figure~\ref{fig:cost_distribution} shows the distribution.
As can be seen in the figure, the telescope cost dominates 
at the limit of large $D_{\rm tel}$.

\subsubsection{Error on $r$ and $N_{\rm eff}$ vs Aperture}
\begin{figure}[htbp]
 \begin{center}
  \includegraphics[width=0.45\textwidth]{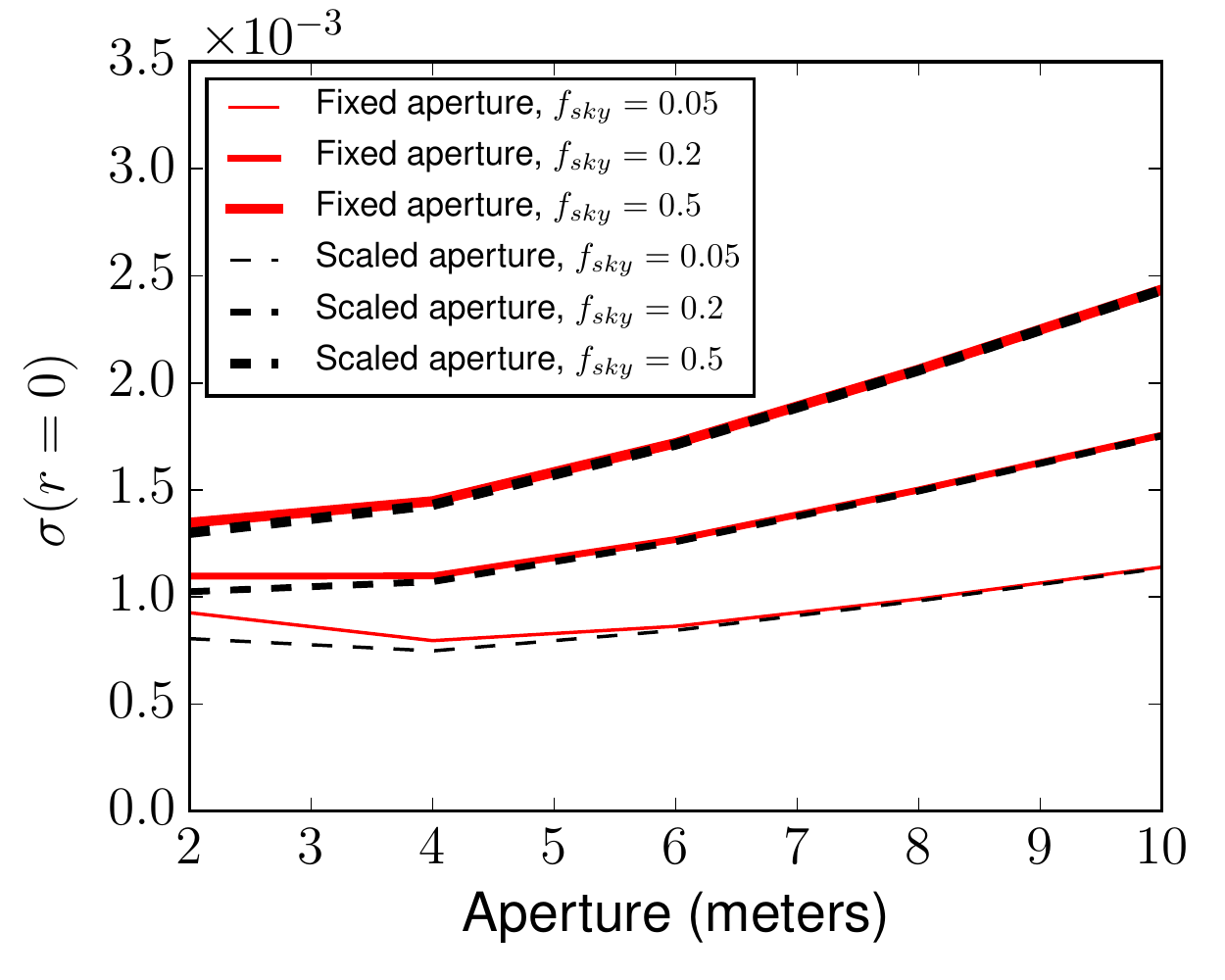}
  \hspace{0.05\textwidth}
  \includegraphics[width=0.45\textwidth]{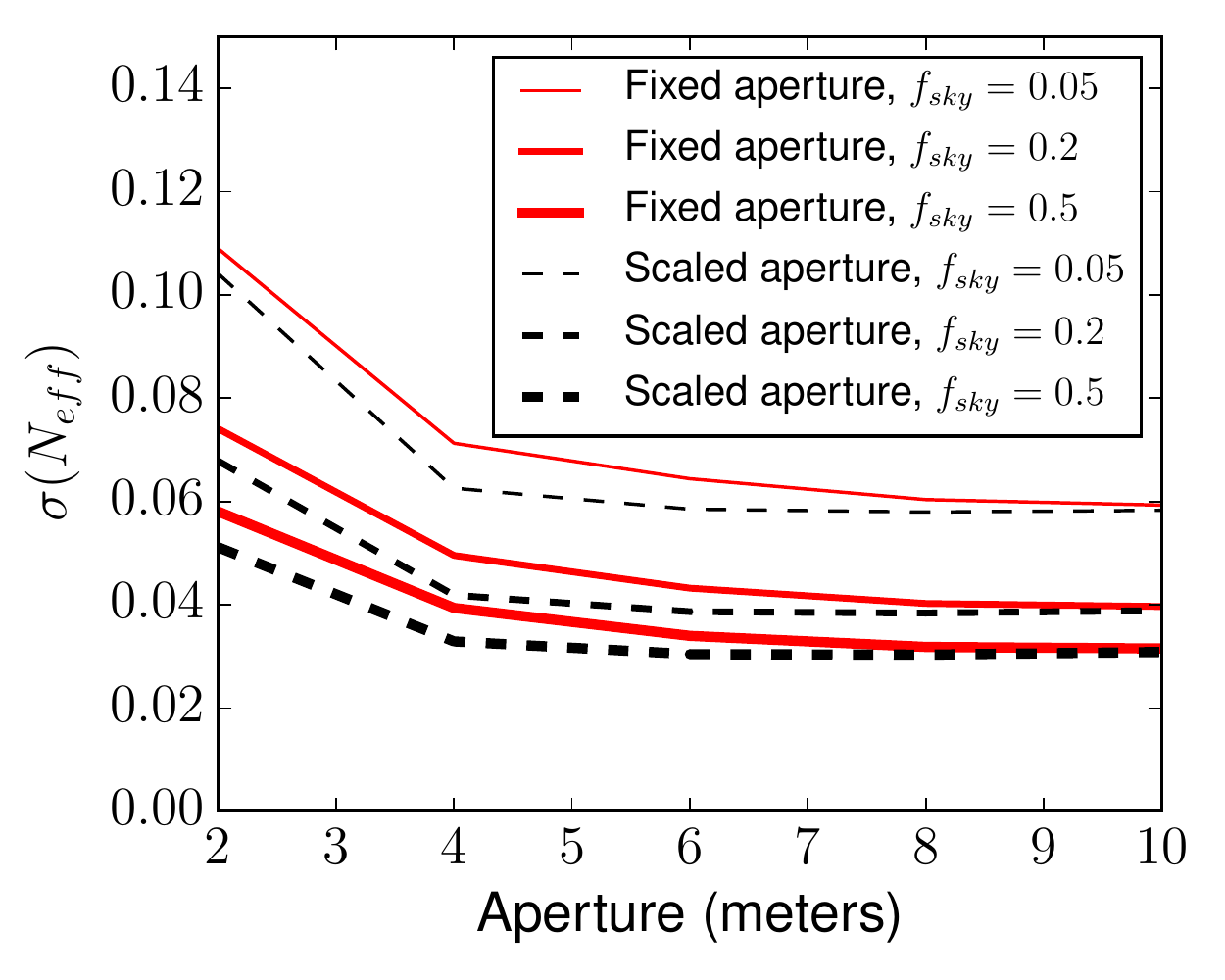}
  \caption{\label{fig:lo_aperture_scan} 
  Left: the error of $r$ as a function of the telescope aperture
  $D_{\rm tel}$ for a fixed total cost of 50\,PCU and a 2.5-year survey.
  Right: the error of $N_{\rm eff}$ as a function of the telescope aperture 
  $D_{\rm tel}$ for a fixed total cost of 50\,PCU and a 2.5-year survey.
  Both fixed aperture ({\it Larger aperture-a}) and scaled aperture
  ({\it Large aperture-b}) configuration types are shown.
  }
 \end{center}
\end{figure}

We now study how
performance varies with aperture and cost for
the large aperture arrays. 
Figure~\ref{fig:lo_aperture_scan} shows the error on $r$ and
$N_\mathrm{eff}$ as a function of the telescope
aperture size, $D_{\rm tel}$, for a large aperture array of telescopes with a
fixed total cost of 50\,PCU.  The errors are for a 2.5 year survey covering
areas ranging from  0.05 to 0.5 $ f_{\rm sky}$.  As can be seen from these plots, 
a smaller, deeper survey area is optimal for measuring $r$, and the optimum aperture is
$D_{\rm tel} \sim 4$ -- $6\,\mathrm{m}$.  This is primarily driven by the de-lensing
capability; while better resolution helps, larger
aperture size results in fewer detectors, leading to inferior sensitivity.
The optimum for $N_\mathrm{eff}$ is broad, 
$D_{\rm tel} \gtrsim 4\,\mathrm{m}$, yielding similarly good sensitivity.
The larger survey sky area leads to better sensitivity when measuring
$N_{\rm eff}$.
For both $r$ and $N_{\rm eff}$, there is only a minor difference between the
fixed  and scaled aperture sizes.
Since the optimum is broad for $N_{\rm eff}$,
$D_{\rm tel} \sim 6\,\mathrm{m}$ yields a balanced optimum for both $r$
and $N_{\rm eff}$.

\subsubsection{Error on Neutrino Mass and kSZ vs Aperture}
\begin{figure}[htbp]
 \begin{center}
  \includegraphics[width=0.45\textwidth]{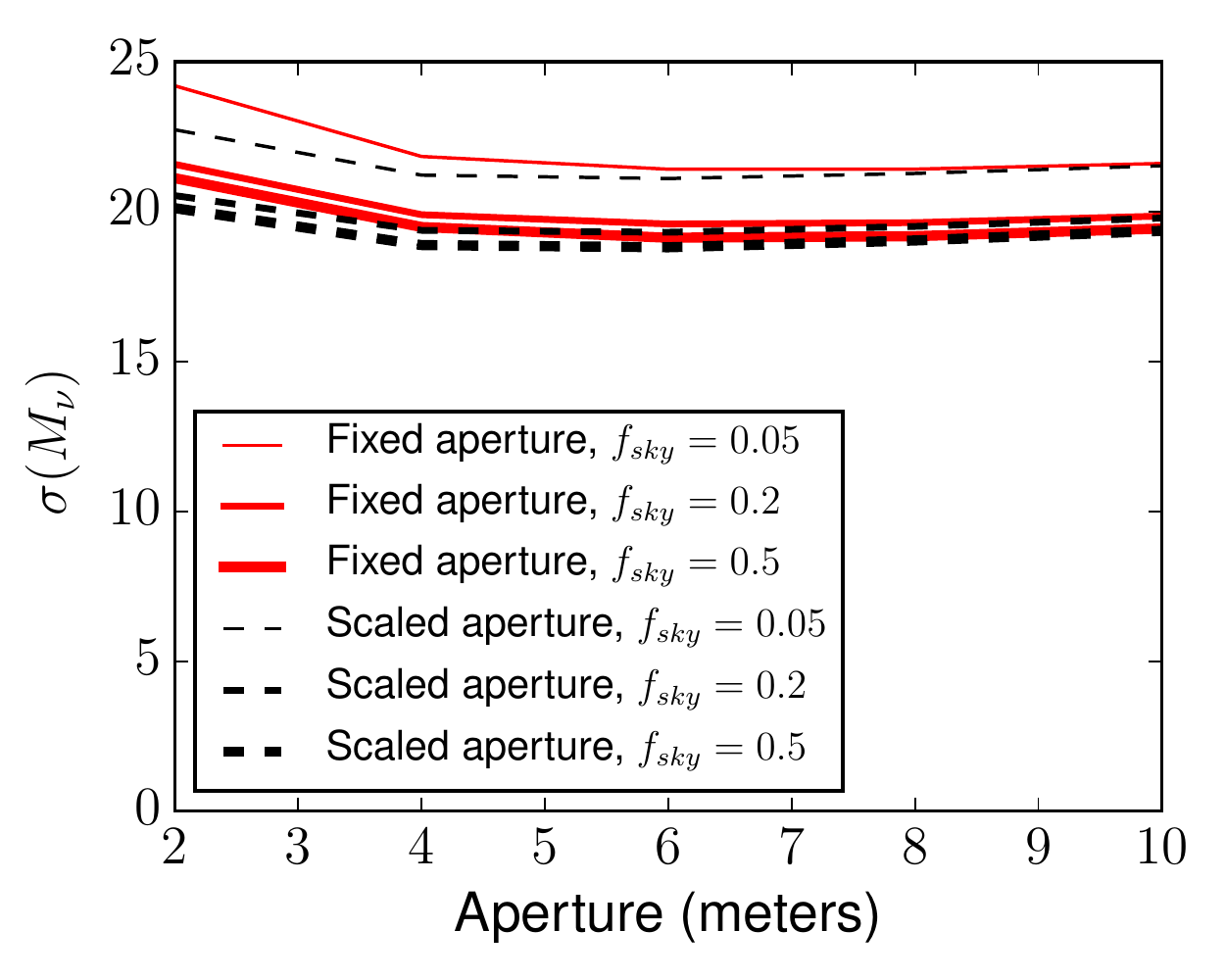}
  \hspace{0.05\textwidth}
  \includegraphics[width=0.45\textwidth]{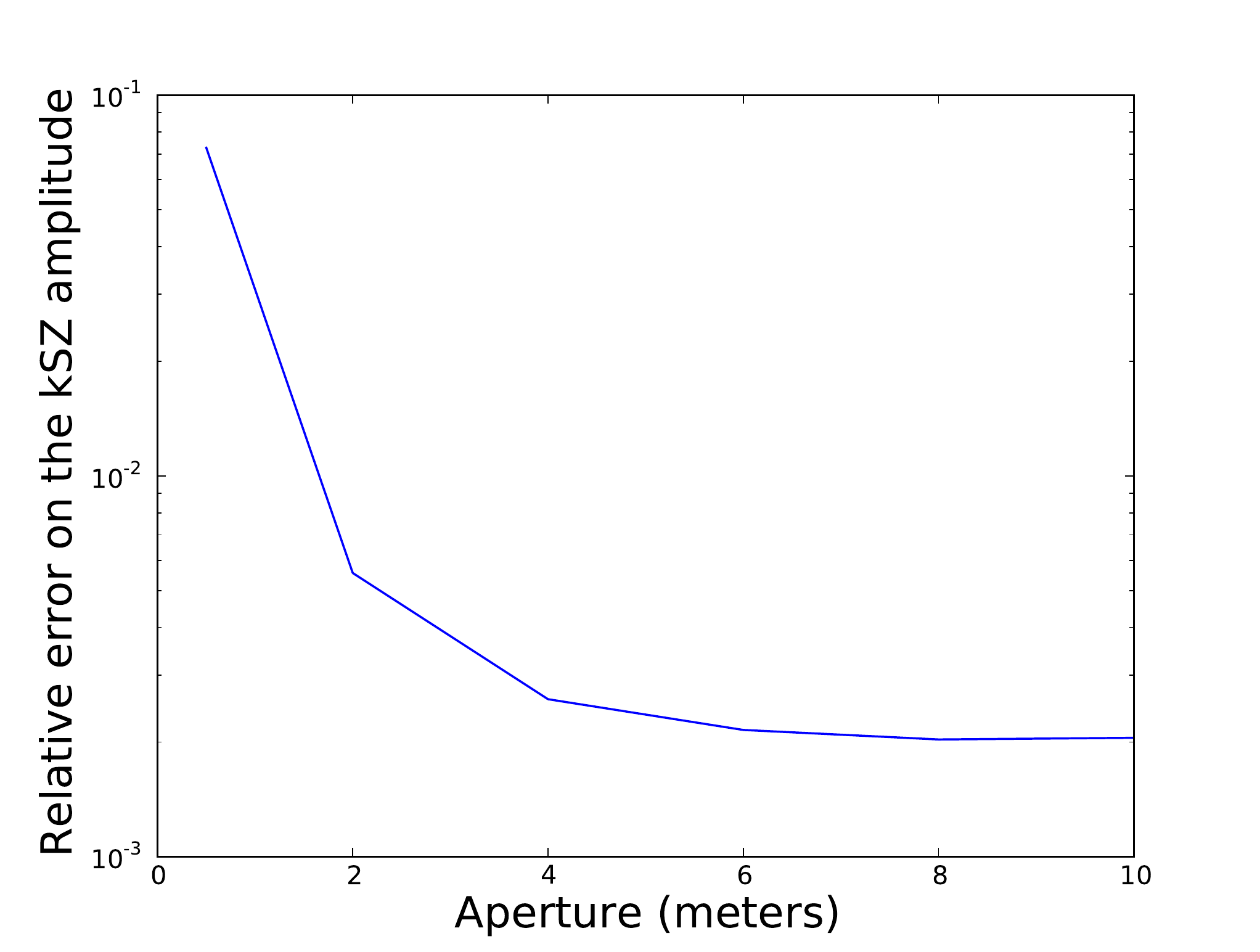}
  \caption{\label{fig:lo_aperture_scan_mnu_ksz} 
  Left: the error of $\sum m_\nu$ as a function of the telescope aperture
  $D_{\rm tel}$ for a fixed total cost of 50\,PCU with $f_{\rm sky}=0.05$,
  0.2, and 0.5.
  Right: the relative error (the inverse of the signal-to-noise ratio)
  for kSZ effect as a function of the
  telescope aperture 
  $D_{\rm tel}$ for a fixed total cost of 50\,PCU with $f_{\rm sky}=0.5$.
  For kSZ, we only use 150-GHz channels.
  }
 \end{center}
\end{figure}

Figure~\ref{fig:lo_aperture_scan_mnu_ksz} (left) shows the error of
 $\sum m_\nu$ as a function
 of the telescope aperture size. 
The trend is very similar to the case for
 $N_{\rm eff}$;
there is a broad optimum for $D_{\rm tel} \gtrsim 4\,\mathrm{m}$.
On the other hand, kSZ prefers slightly larger telescopes,
with a shallow optimum around $D_{\rm tel} \sim 8\,\mathrm{m}$.
However, the $D_{\rm tel} \sim 6\,\mathrm{m}$, which is favored by the optimization
 for $r$ and $N_{\rm eff}$, is not significantly worse than the optimum.

The pairwise kSZ
calculation is not based on \textsc{CMB4cast} and is calculated
separately using only the 150\,GHz channels.
Figure~\ref{fig:lo_aperture_scan_mnu_ksz} (right)
shows the
relative error on the kSZ amplitude from low-redshift tracers, which is
assumed to be the DESI spectroscopic galaxy catalog, comprising
$\sim$20 million objects over 14,000 sq. deg.
If the optical depth is known a priori (from other observables), this
 corresponds to the error on the growth factor of perturbations.
 Conversely, this measurement can be converted into a measurement of the
 gas distribution around the tracer galaxies, yielding information about
 galaxy formation and feedback processes as well as helping calibration
 of baryon effects in weak lensing surveys (since the gas is
 approximately 20\% of the total mass).
 
For this preliminary forecast, we assume that a foreground-cleaned map
with resolution corresponding to the 150GHz channel is available.
Empirically, we assume that component separation increases the effective
noise by a factor 1.4, which is similar to what is found with the Planck SMICA
map.
Although the gains in S/N appear to saturate at ~4--5m apertures in this
fixed cost model, the relative size of contributions from the ``1-halo
term'' (i.e., from gas bound to the galaxy itself) and ``2-halo term''
(i.e., gas in other galaxies and in the intracluster medium) vary, making
the gains in parameters improve with resolution above the 4m aperture.
A self-consistent treatment of high-$\ell$ component separation and
forecasts of constraints on physical parameters are important and the
subject of current work~\cite{Simone1,Simone2}.

\subsubsection{Limit of Diminishing Return vs. Total Cost}
\begin{figure}[htbp]
 \begin{center}
  \includegraphics[width=0.45\textwidth]{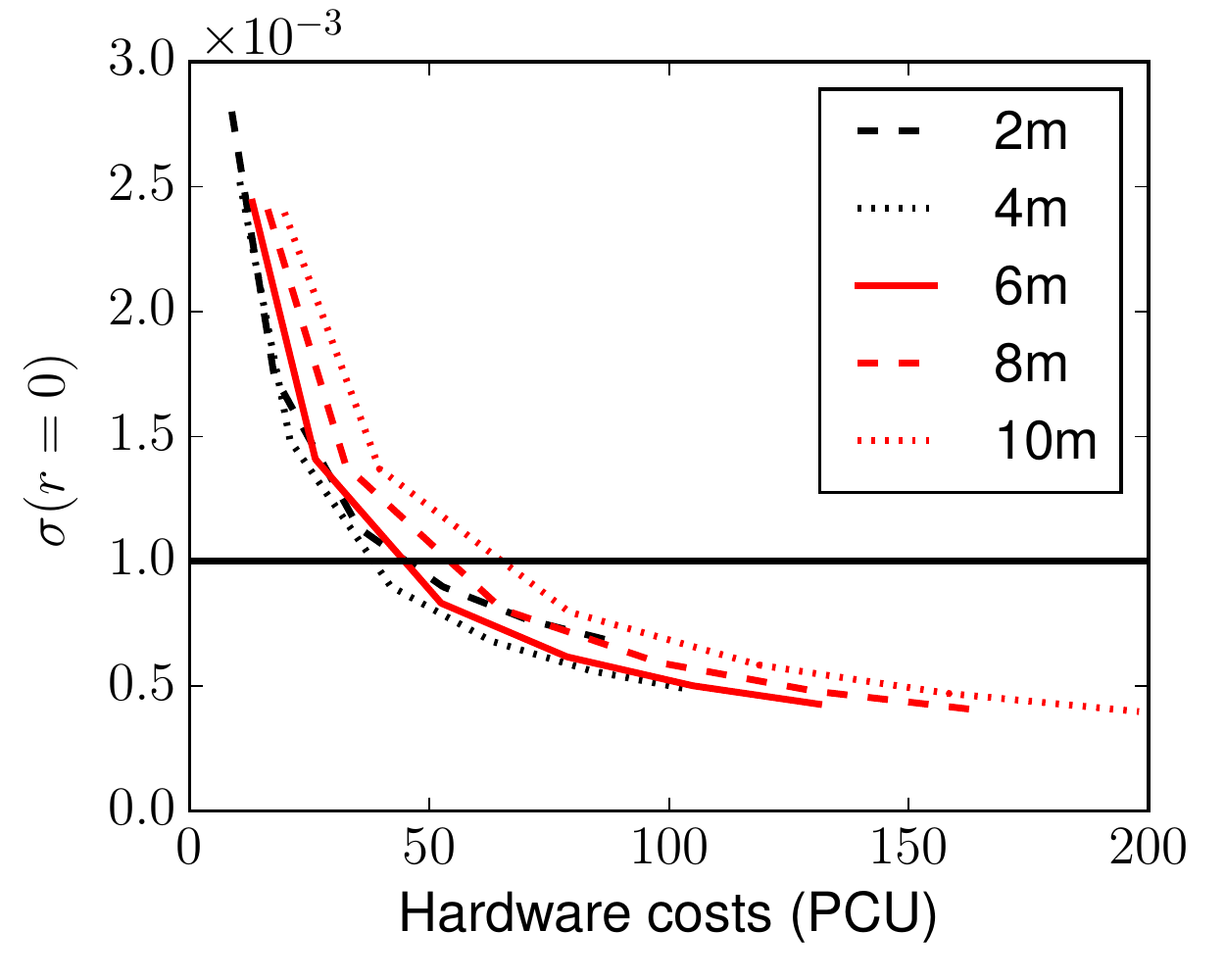}
  \hspace{0.05\textwidth}
  \includegraphics[width=0.45\textwidth]{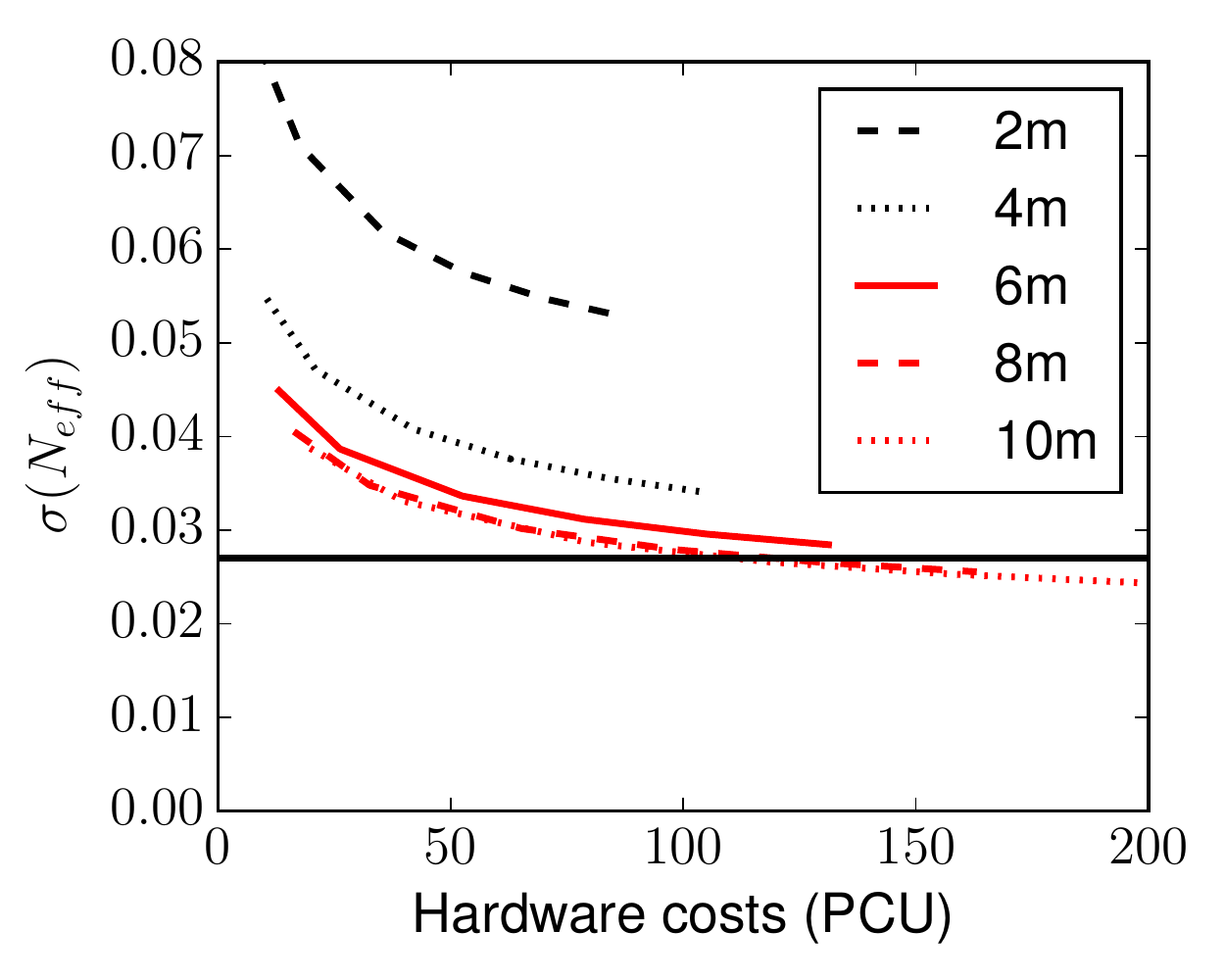}
  \caption{\label{fig:diminishing_return_r_neff} 
  Constraint on $r$ with $f_{\rm sky}=0.05$ (left) and $N_{\rm eff}$ with
  $f_{\rm sky}=0.5$ (right) for different apertures, as a
  function of the total cost of the project.
  Both are for the large aperture telescope array with fixed aperture sizes
  ({\it Large aperture-a}).
  For both, the improvement saturates approximately at a total hardware
  cost of 50\,PCU.
  The improvement of $r$ is not linear with the total cost, or with the total
  number of detectors, because the de-lensing noise levels do not improve
  as fast as the map depth.
  }
 \end{center}
\end{figure}

In addition to studying the optimal telescope aperture for a fixed total cost,
we look at the errors as a function of total cost to determine the limit of diminishing 
scientific return.  This is shown in Figure~\ref{fig:diminishing_return_r_neff} 
where we plot the errors on $r$ and $N_{\rm eff}$ for arrays of fixed
aperture size ({\it Large aperture-a}) with varying 
total cost to explore the point where the error saturates.

These plots  show that the limit of diminishing returns is reached at a total hardware
cost of approximately 50\,PCU and an error of
 $\sigma (r) \approx 0.75 \times 10^{-3}$
 for an array of $D_{\rm tel} \sim 4 - 6\,\mathrm{m}$ telescopes.
Doubling the cost to 100\,PCU reduces the error by 30\% to
 $\sigma (r) \approx 0.5 \times 10^{-3}$.
The errors on $N_{\rm eff}$ are saturated at 50\,PCU with $\sigma (N_{\rm eff}) \approx 0.03$ for telescopes
larger than 6\,m in aperture.  Improvement by increasing the total
instrument cost beyond 50\,PCU is even slower than that for $\sigma(r)$.

\subsubsection{Cost Model Variations}
\begin{figure}[htbp]
 \begin{center}
  \includegraphics[width=0.32\textwidth]{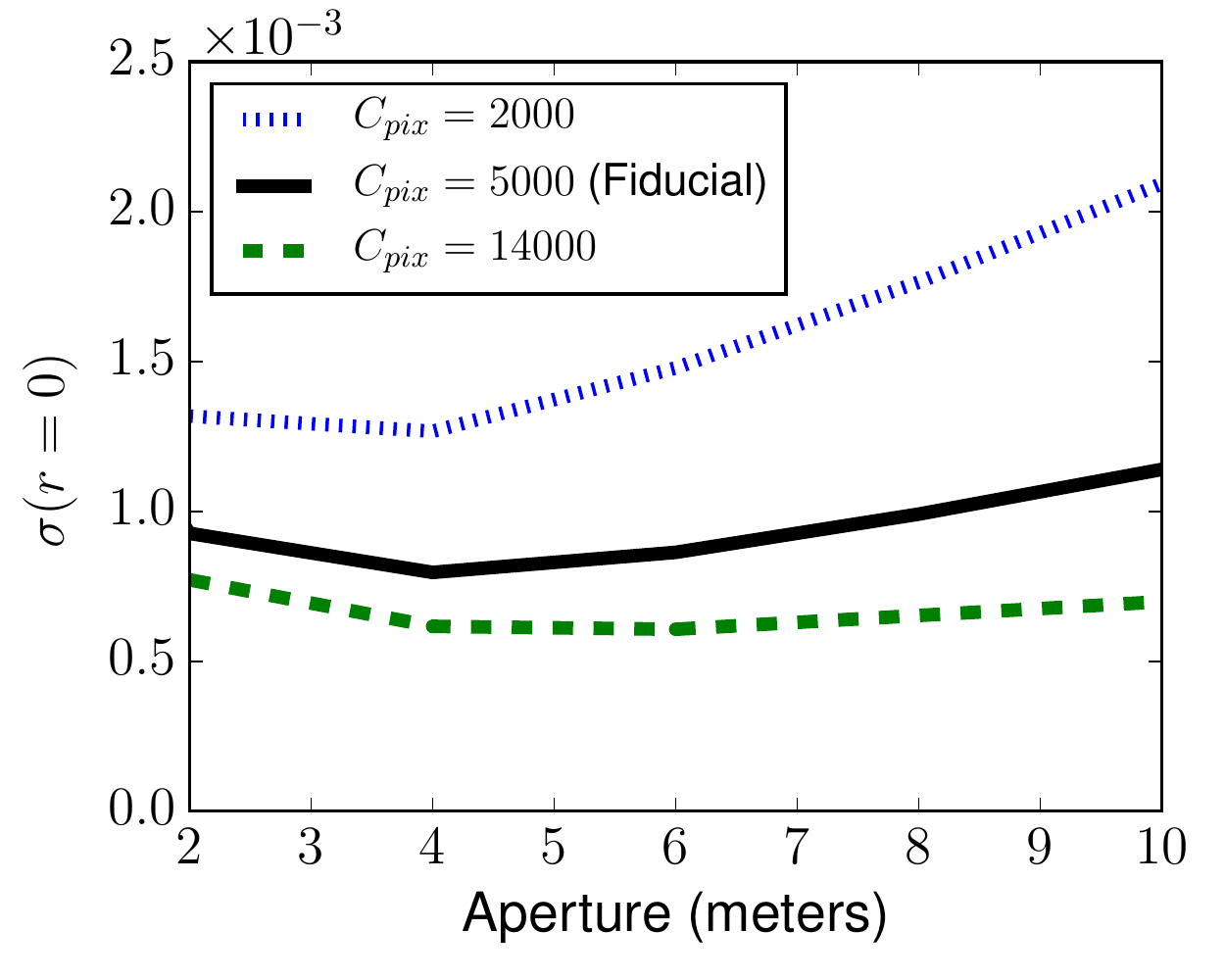}
  \hspace{0.005\textwidth}
  \includegraphics[width=0.32\textwidth]{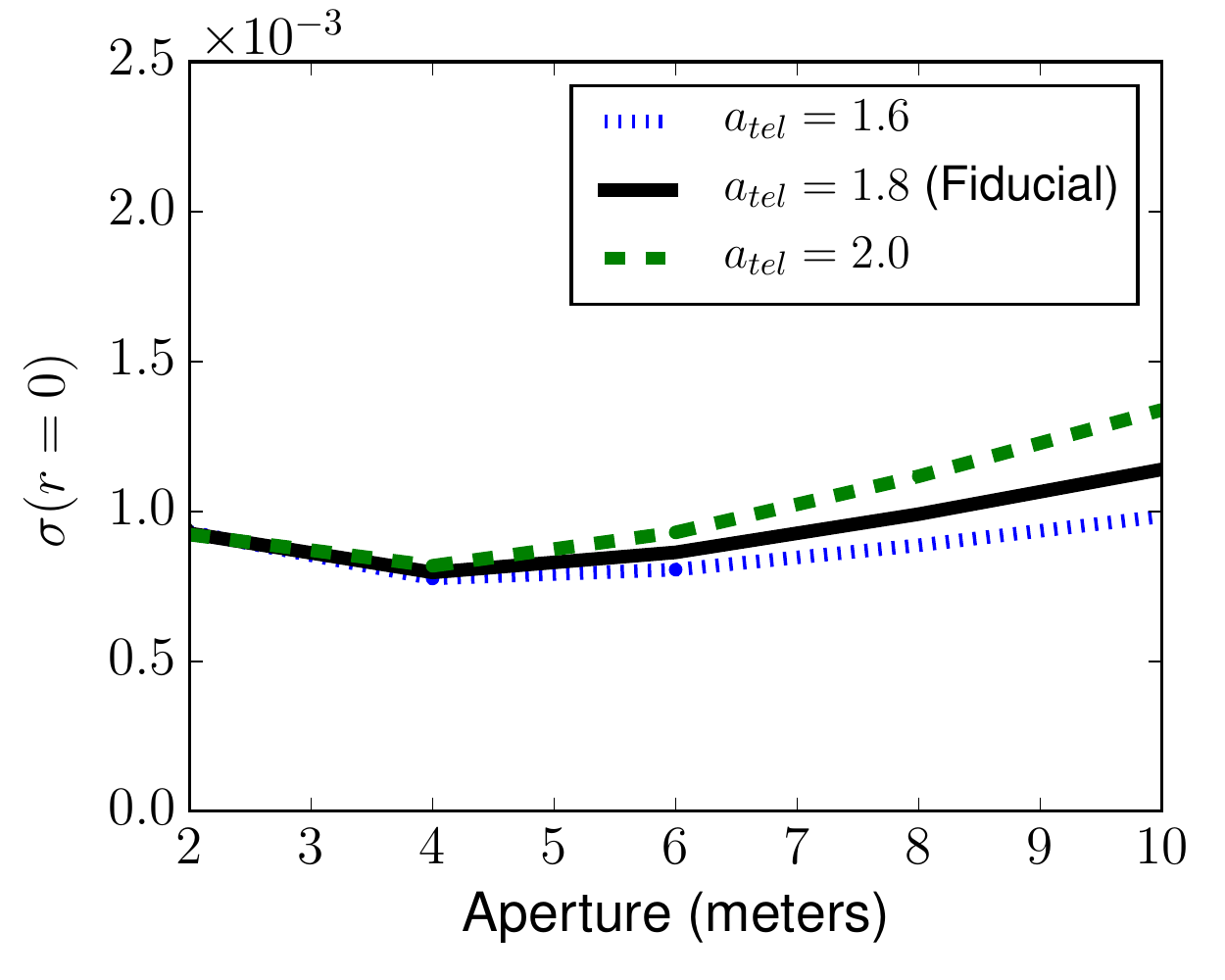}
  \hspace{0.005\textwidth}
  \includegraphics[width=0.32\textwidth]{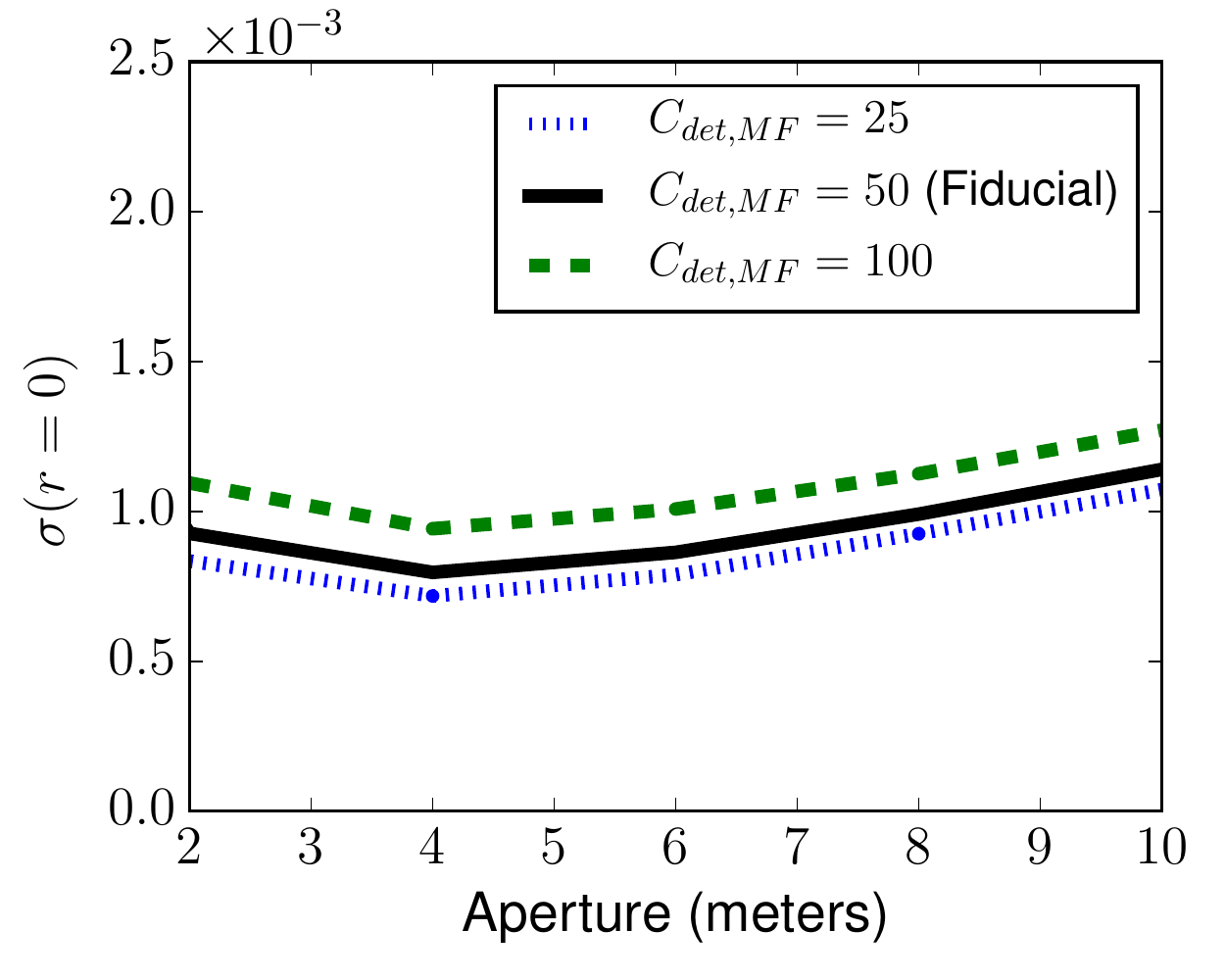}
  \caption{\label{fig:cost_model_variation} 
  The performance dependence on the cost assumptions.
  All panels show the error on $r$ as a function of
  the telescope aperture
  $D_{\rm tel}$ for a fixed total cost of 50 PCU and $f_{\rm sky}=0.05$
  Left: dependence on the telescope throughput for $C_{\rm pix}$ of 2000,
  5000 (fiducial), and 14000 are shown. Center: dependence on the telescope cost scaling with aperture size for $a_{\rm tel}$ of 1.6, 1.8 (fiducial), and 2.0.
  Right: dependence on the cost of the detector+readout for per-channel cost of the 150\,GHz detector
  of \$25 PCU, \$50 (fiducial), and \$100. The costs for the other frequencies are scaled accordingly.
  }
 \end{center}
\end{figure}

As discussed above, our cost model has 
uncertainties.  While we do not intend to present a
finalized cost model here,
we explore some variations of the cost model to show examples of
possible impact.
Figure~\ref{fig:cost_model_variation} shows the impact of different
telescope throughput ($C_{\rm pix}$) and the detector costs on the
results for the fixed aperture configuration ({\it Large aperture-a}).
Note that varying $C_{\rm pix}$ is equivalent to varying the telescope cost
scale, $C_{\rm tel}$, by the same factor.

As shown in the figure, a smaller telescope throughput, or a higher
telescope cost, results in the optimum moving towards smaller aperture and
a larger error on $r$.  It is also worth noting that the difference
between $C_{\rm pix}$ of 5000 and 14000 is relatively modest.
This is primarily because of two reasons: 1)
the telescope cost is already less than half of the total experimental
cost (see Fig.~\ref{fig:cost_distribution}), and thus reducing the telescope
cost by a factor of three (or more)
results in less than a factor of two increase in the detector count;
and 2) the constraint on $r$ is already reaching saturation and
the improvement is slower than the increase of the detector count,
as shown in Fig.~\ref{fig:diminishing_return_r_neff} (left).
The dependence on the detector cost is modest because
the detector cost does not dominate the total experimental cost.

\subsection{Hybrid Telescope Array Configurations}
\label{sec:hybridresults}
We now discuss hybrid configurations with a mix of apertures including
small apertures of 0.5\,m.
We study two types of hybrid configurations: 
{\it Hybrid-a} in which the large
  telescopes have the same aperture size, $D_{\rm tel}$
 (Fig.~\ref{fig:configuration_types}c),
and {\it Hybrid-b}
in which the large telescopes have scaled aperture sizes
(Fig.~\ref{fig:configuration_types}d).
The total cost of 50\,PCU is split into the large and small aperture
instruments.
We use a 50\%/50\% split as the nominal configuration,
which is near the optimum as we will show.
We assume that all the large aperture instruments have
$\ell^L_{\mathrm{knee}}$ of 500 and the small aperture instruments have
$\ell^S_{\mathrm{knee}}$ of 40.
While our choice of $\ell^L_{\mathrm{knee}}$ is conservative
and the actual instrument is likely to achieve a lower value,
this serves as a good example of a configuration in which
the small and large aperture instruments play distinct roles
scientifically due to their different $\ell$ coverage.

For the hybrid configuration, we mainly explore the error on $r$,
which strongly depends on the instrumental sensitivity at low $\ell$.
 Only the large-aperture telescopes in the hybrid configurations
contribute
to the other cosmological observables such as $N_{\rm eff}$,
 $\sum m_\nu$, and kSZ.  Performance for these observables can simply be
 extrapolated from the large-aperture configurations discussed
 above.

\subsubsection{Frequency Combination}
\begin{figure}[htbp]
 \begin{center}
  \includegraphics[width=0.45\textwidth]{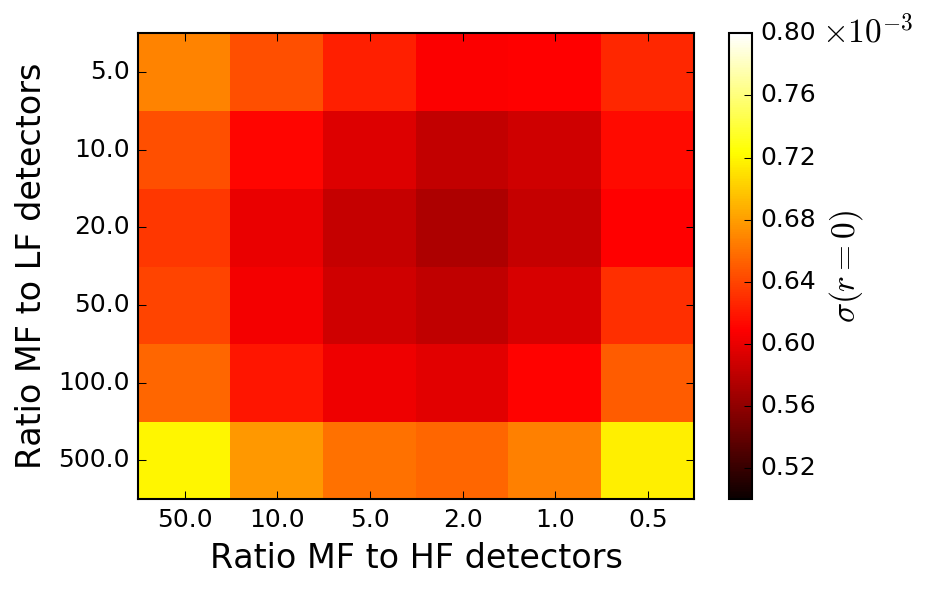}
  \includegraphics[width=0.45\textwidth]{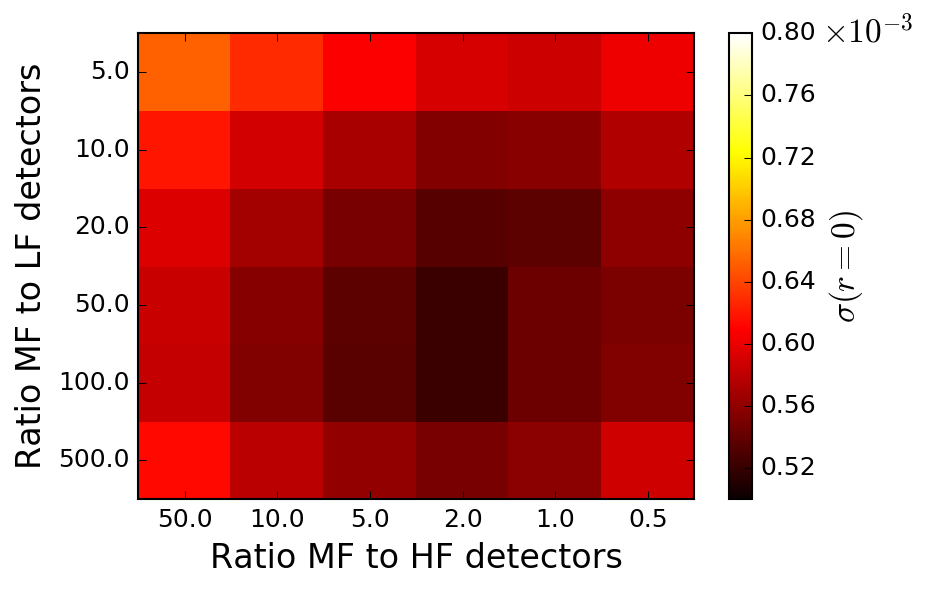}
  \\
  \vspace{0.5cm}
  \includegraphics[width=0.45\textwidth]{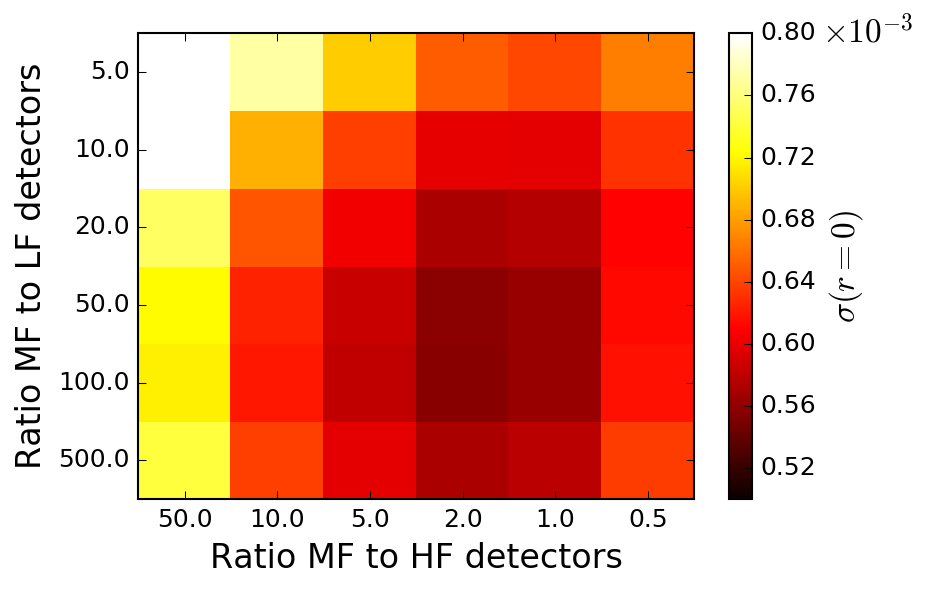}
  \includegraphics[width=0.45\textwidth]{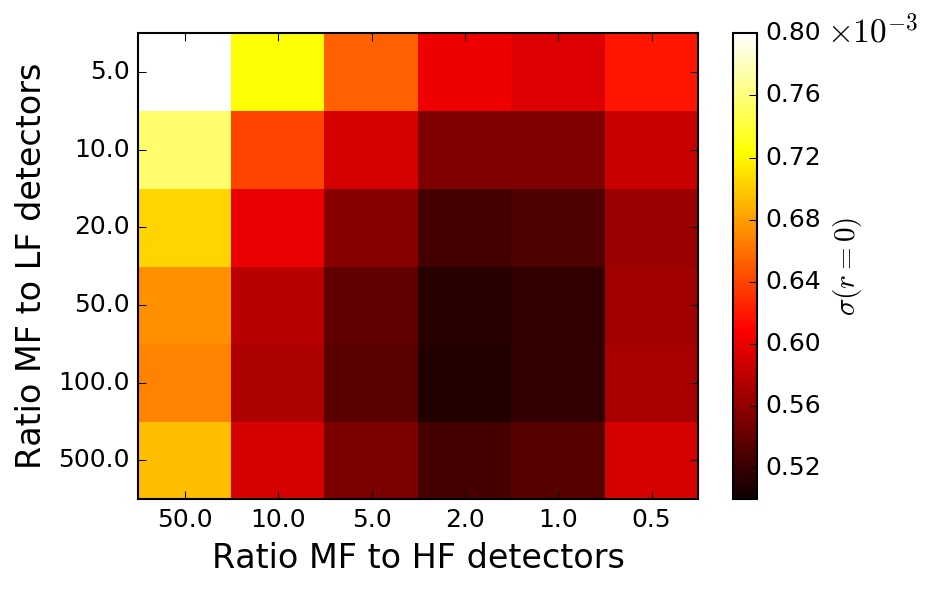}
  \caption{\label{fig:hybrid_frequency_optimizations} 
  Optimization of $\sigma(r)$ over frequency weighting for the hybrid
  configurations.
  In the top (bottom) panels, the ratio MF/LF and MF/HF are varied for 
  large-aperture (small-aperture)
  telescopes while keeping them at the nominal values, MF/LF=20 and
  MF/HF=2, for the small-aperture (large-aperture) telescopes.
  A sky coverage of 5\%  is assumed.
  The left panels show the case where the large-telescope apertures
  are fixed to 6\,m ({\it Hybrid-a} with
  $D_{\rm tel}=6$\,m).
  The right panels show the case
  where the large-telescope apertures are scaled with frequency to
  12\,m, 6\,m, and 3\,m for LF,
  MF, and HF, respectively ({\it Hybrid-b} with $D_{\rm tel}=6$\,m).
  For both, we again find that MF/LF=20 and MF/HF=2 are near the
  optimum.
  }
 \end{center}
\end{figure}

Following the same procedure employed for the large-aperture configurations,
we first optimize the weighting between the LF, MF, and HF detectors.
Figure~\ref{fig:hybrid_frequency_optimizations} shows the expected error
on $r$ as a function of the ratio of MF/LF and MF/HF,
with $D_{\rm tel}=6\,\mathrm{m}$.
We set the nominal ratio to be MF/LF=20 and MF/HF=2
and vary them separately for the large-aperture and small-aperture components
of the instrument while keeping the other at the nominal ratio.
As shown in the figure, the nominal ratio of MF/LF=20 and MF/HF=2
is sufficiently near the optimum.  Thus, in the
following, we use these ratios.

\begin{figure}[htbp]
 \begin{center}
  \includegraphics[width=0.6\textwidth]{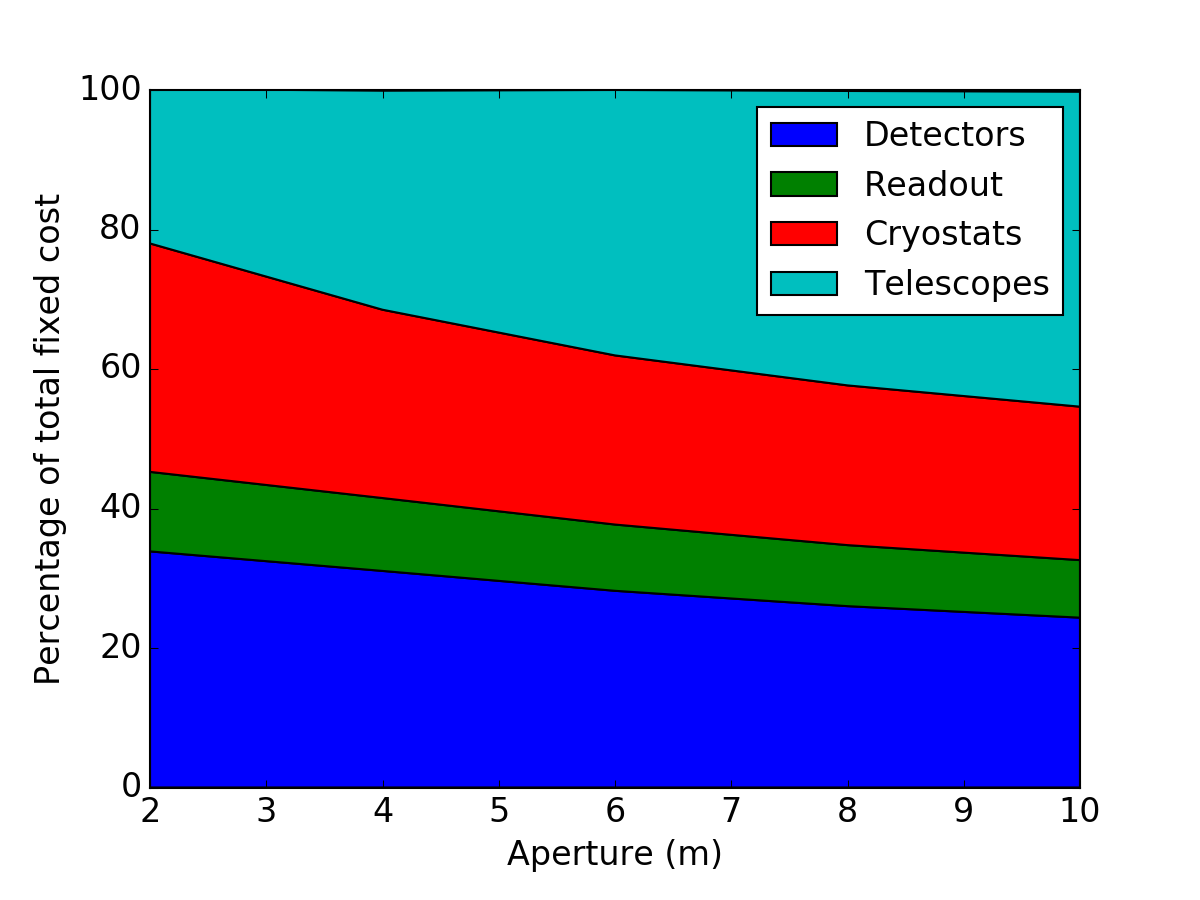}
  \caption{\label{fig:hybrid_cost_distribution} 
  The cost distribution over telescope, cryostat, detector, and readout
  for the hybrid configuration with fixed large-telescope aperture sizes
  ({\it Hybrid-a}).
  The cost distribution is shown as a 
  function of the large-telescope aperture size $D_{\rm tel}$.
  }
 \end{center}
\end{figure}

Once the frequency weighting is fixed, the cost distribution among each
of the subsystems is uniquely determined in our model.
Figure~\ref{fig:hybrid_cost_distribution} shows the distribution.
As expected, the fraction of the telescope cost is reduced
compared to the cost distribution of the large-aperture-only configurations
 (Fig.~\ref{fig:cost_distribution}).

\subsubsection{Fraction of Large vs. Small}
\begin{figure}[htbp]
 \begin{center}
  \includegraphics[width=0.5\textwidth]{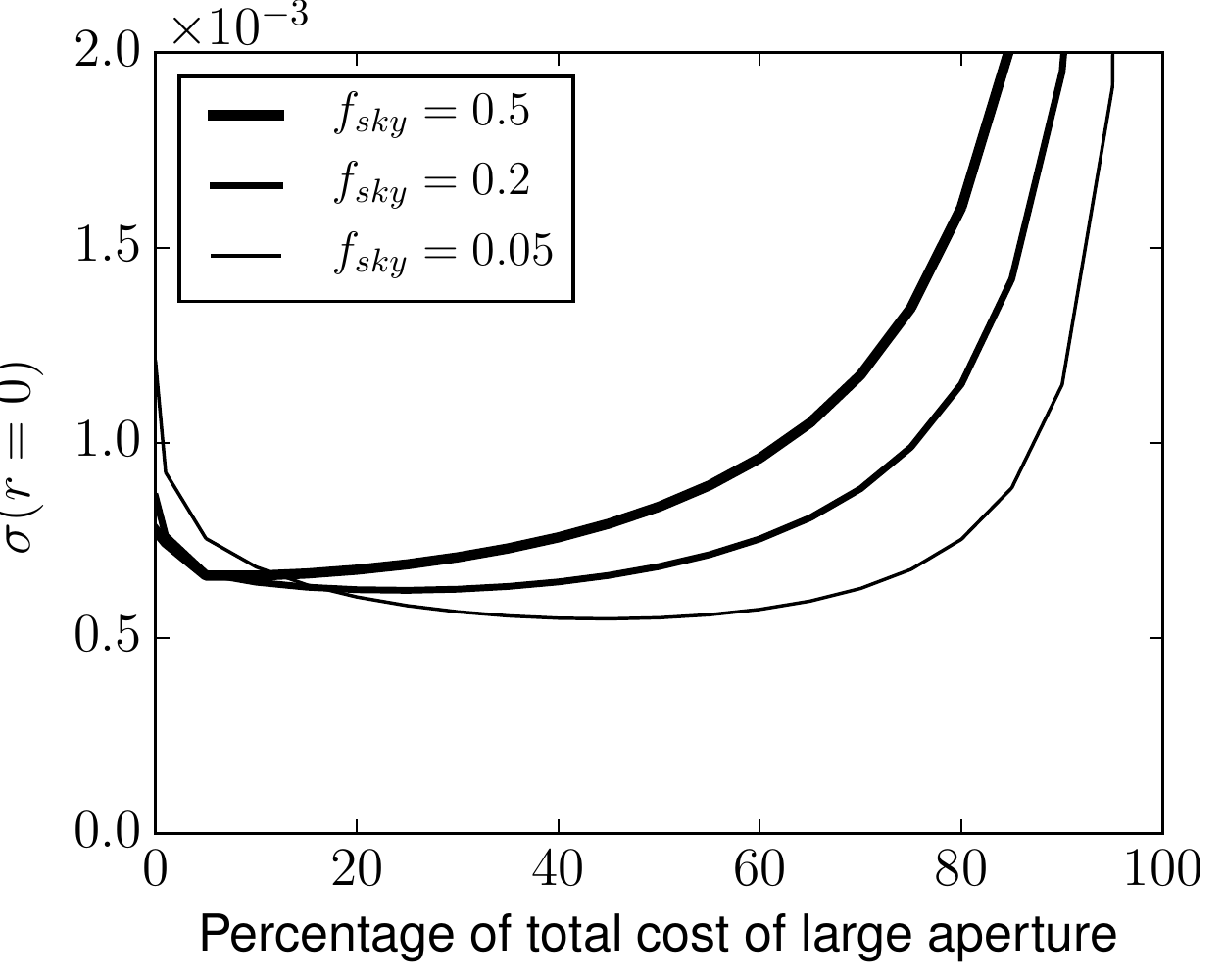}
  \caption{\label{fig:hybrid_fraction_r} 
  Error on $r$ as a function of the fraction of the cost allocated to the
  large aperture part of the instrument.
  The assumed configuration is {\it Hybrid-a}
  with $D_{\rm tel}=6\,\mathrm{m}$ and a total cost of 50\,PCU.
  The smaller $f_{\rm sky}$
  requires more de-lensing, favoring a larger fraction of large-aperture
  instruments.}
 \end{center}
\end{figure}

Figure~\ref{fig:hybrid_fraction_r} shows the constraint on $r$ as a
function of the fraction of cost spent for the large aperture telescopes.
The dependence is relatively shallow, and there is a broad optimum around 
the 50\%/50\% split between large and small aperture instruments.  A trend
can be seen in which a small value of
$f_{\rm sky}$ favors a larger fraction for the large aperture instrument
due to the de-lensing requirements.
In the following, we assume a 50\%/50\% cost distribution between the
large and small aperture instruments.

\subsubsection{Constraint on $r$ and Dependence on Aperture Size}
\begin{figure}[htbp]
 \begin{center}
  \includegraphics[width=0.6\textwidth]{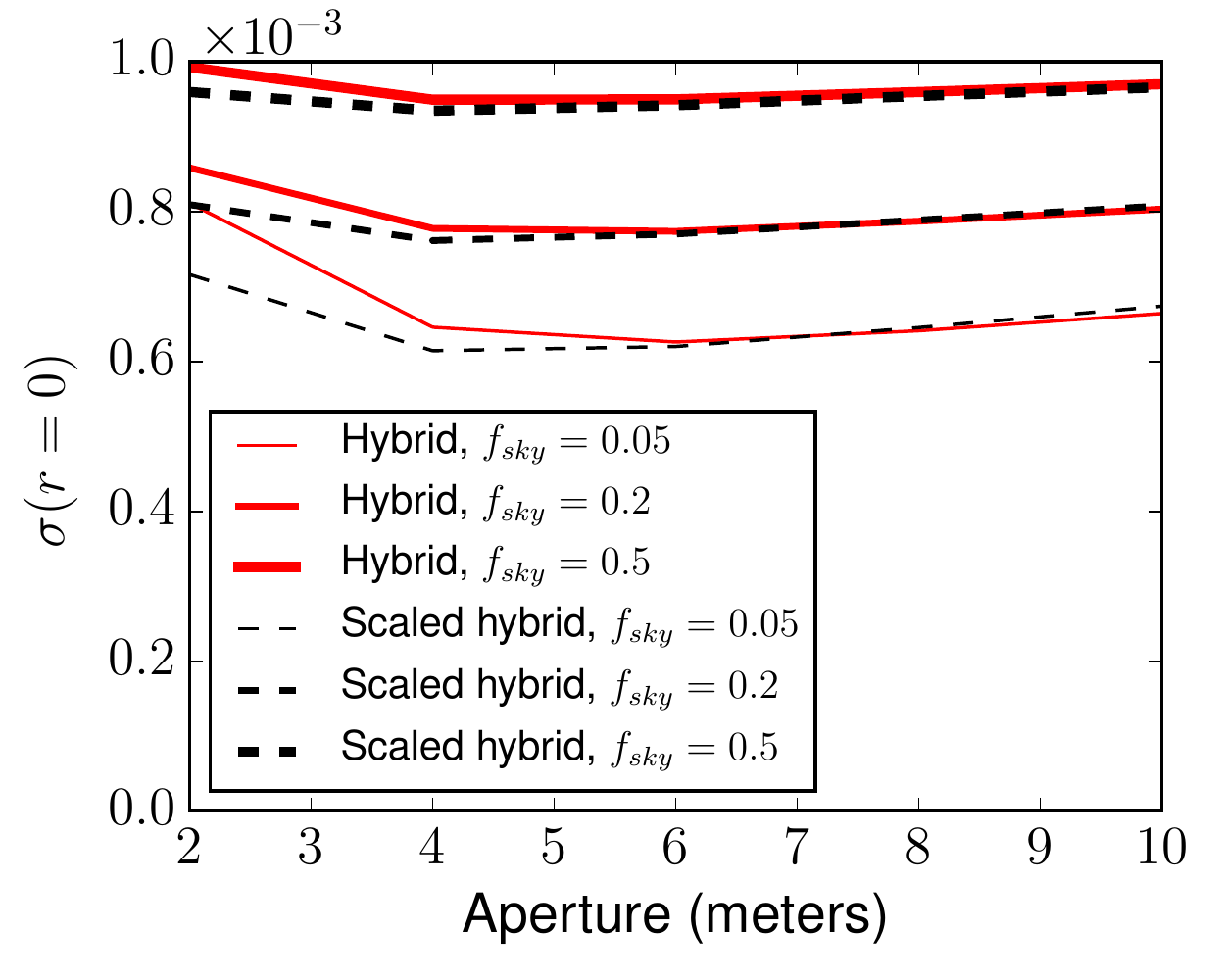}
  \caption{\label{fig:hybrid_aperture_scan} 
  The error of $r$ as a function of the telescope aperture
  $D_{\rm tel}$ for a fixed total cost of 50\,PCU
  for ``hybrid'' (fixed large-telescope aperture; {\it Hybrid-a})
  and ``scaled hybrid'' (scaled large-telescope aperture;
  {\it Hybrid-b}).
  The total cost of 50\,PCU is equally split between the large and small
  aperture instrument.
  }
 \end{center}
\end{figure}

Figure~\ref{fig:hybrid_aperture_scan} shows the error on $r$
as a function of the diameter of the
large-aperture instrument for a fixed total cost of 50\,PCU.
As can be seen, the optimum for $r$ is 
broad, around $D_{\rm tel} \sim 4$ -- $8\,\mathrm{m}$.
The trend differs from the case of large aperture only configurations
(Fig.~\ref{fig:lo_aperture_scan}) in that the performance does not
degrade for large $D_{\rm tel}$.  This can be understood as follows.
The sensitivity on $r$ requires both low-$\ell$ 
sensitivity to the primordial gravity wave signature  at $\ell\sim 100$ 
and de-lensing capability in the high-$\ell$ region.
The de-lensing capability stays roughly constant when $D_{\rm tel}$ increases
above 4\,m due to cancellation between two factors:
 sensitivity degradation due to the smaller number of
detectors as the telescope cost increases with aperture, 
and resolution improvement due to better angular resolution with increasing aperture.
The low-$\ell$ sensitivity is a function of the detector count,
and thus it degrades as $D_{\rm tel}$ increases for large-aperture-only
configurations.  On the other hand, for hybrid configurations, 
low-$\ell$ sensitivity is provided only  by the small-aperture instrument, which
does not depend on $D_{\rm tel}$.  As a result, the
dependence on $D_{\rm tel}$ is very shallow for hybrid configurations
so long as $D_{\rm tel} \gtrsim 4\,\mathrm{m}$.

\subsubsection{Cost and Instrumental Model Dependence}
\label{sec:with_and_without_lf_throughput_constraint}
\begin{figure}[htbp]
 \begin{center}

  \includegraphics[width=0.45\textwidth]{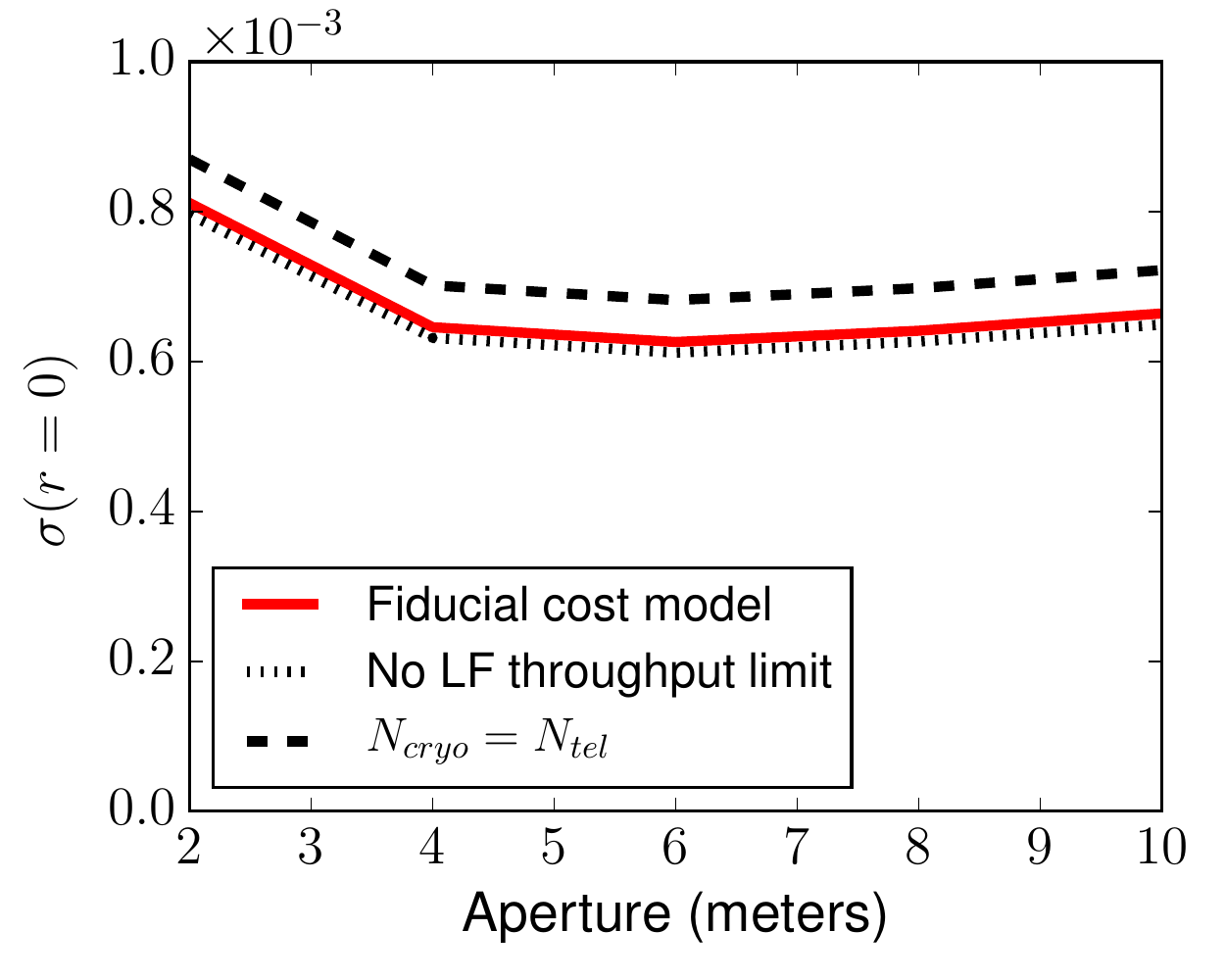}
  \hspace{0.05\textwidth}
  \includegraphics[width=0.45\textwidth]{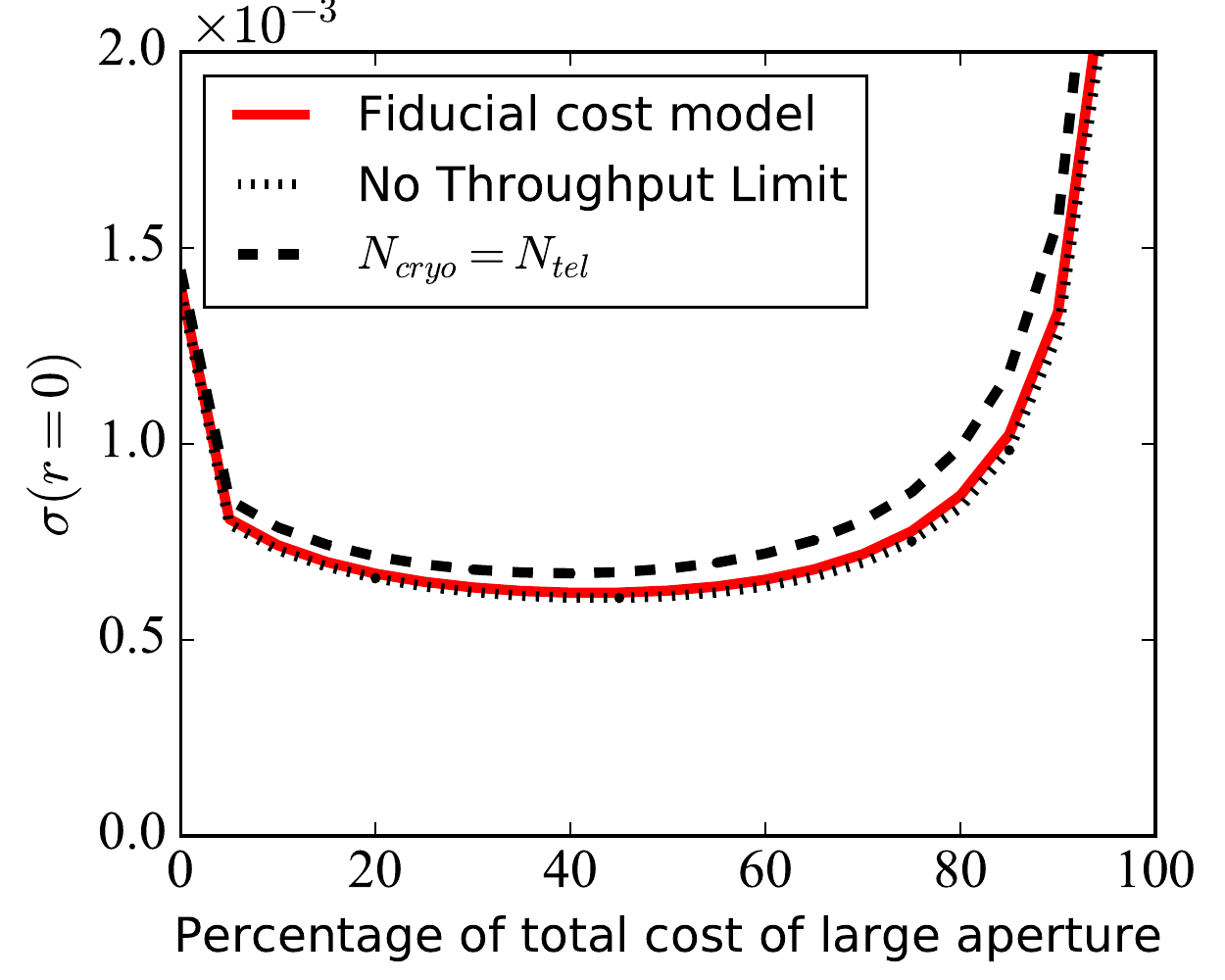}
  
  \caption{\label{fig:smallaperture_variation} 
  The fiducial ``hybrid'' (fixed large-telescope aperture; {\it Hybrid-a})
   is compared with variations where the constraint on the throughput of the 
   small-aperture instrument is removed (``No Throughput Limit'')
    as well as a variation where the limit is imposed together with
    an additional constraint that the number of telescopes and cryostats
	must be the same (``$N_{cryo}=N_{tel}$'').
  Left: The error of $r$ as a function of the telescope aperture
  $D_{\rm tel}$. The total cost of 50\,PCU is equally split between the 
  large and small aperture instrument. A sky coverage of 5\% is assumed.
  Right: The error of $r$ as a function of the fraction of the cost allocated to the
    large aperture part of the instrument, for a fixed sky coverage of 5\%.
  }
 \end{center}
\end{figure}

Here, we focus on the additional throughput constraint
imposed specifically on the small-aperture instrument
discussed in Sec.~\ref{sec:telescope_assumption} and its implication for cost modeling of the cryostat discussed in Section \ref{sec:cryostat_cost}. Figure \ref{fig:smallaperture_variation} compares the forecast results for variations on these
constraints on the small-aperture instrument in the cost model. 
Removing the constraint on the small-aperture throughput allows as many
wafers as the $N_{pix}$ scaling for the telescope throughput and $N_{pix}$ limit
of the cryostat allows, which reduces the cost per mapping speed of the small aperture instrument. As Figure \ref{fig:smallaperture_variation} shows, this has a negligible impact on the overall cost and forecast results of the hybrid array.
With the 7-wafer small-aperture throughput limit in place, we also studied the effect
of imposing an additional constraint that each small-aperture telescope requires an additional cryostat. This increases the cost per mapping speed of the small-aperture instrument and leads to the cryostat costs becoming a significant portion of the overall
small-aperture instrument costs. The effect of this constraint on the overall results
is also shown in Figure \ref{fig:smallaperture_variation}.

\subsubsection{Comparison with Larger Aperture Configurations}
\label{sec:lknee_vary}
\begin{figure}[htbp]
\begin{center}
 \includegraphics[width=0.65\textwidth]{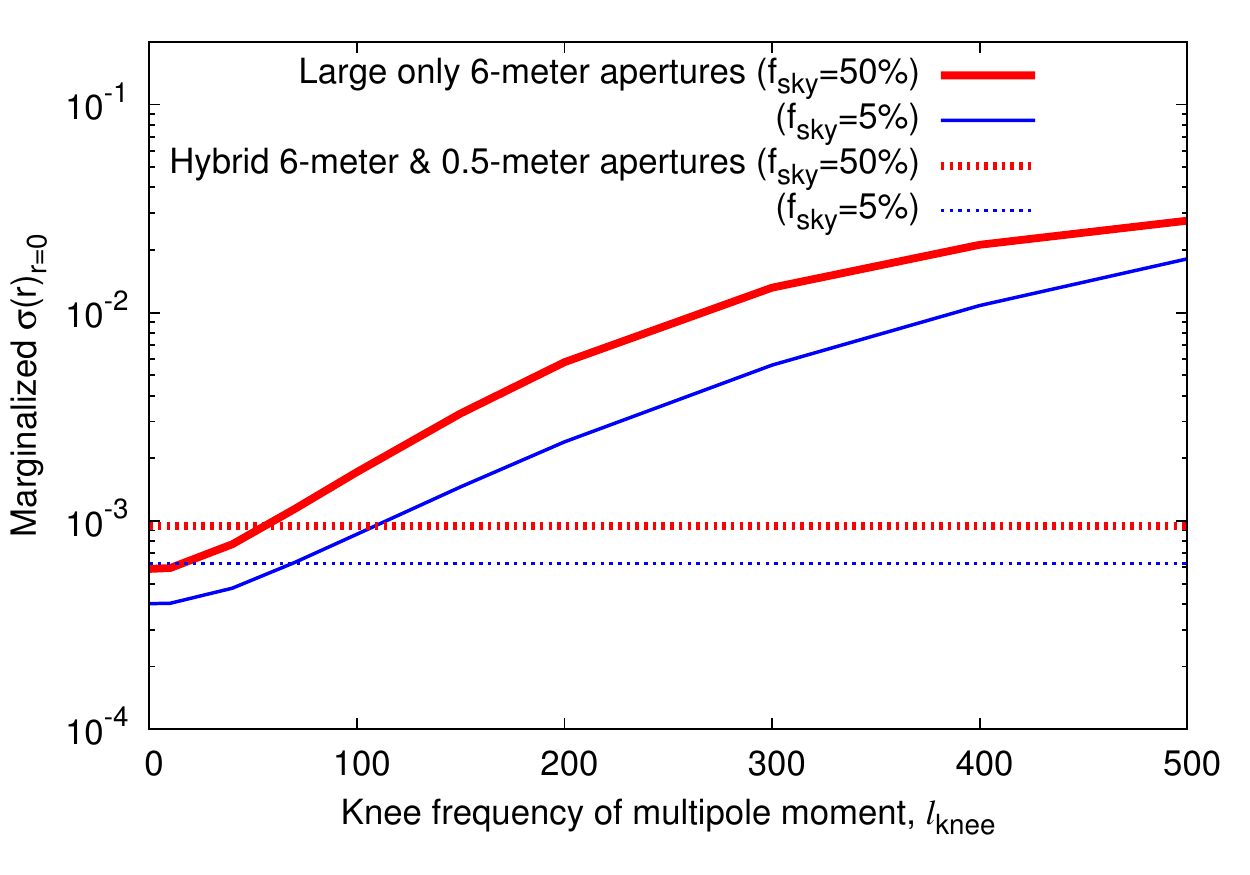}
 \caption{\label{fig:lknee_dependence} 
 Constraint on $r$ as a function of $\ell_{\rm knee}$
 for the large-aperture configuration ({\it Large aperture-a}) with
 $D_{\rm tel} = 6\,\mathrm{m}$ compared to the
 constraint for the hybrid configuration ({\it Hybrid-a}) 
 with fixed $\ell_{\rm knee}\,(\ell_{\rm knee}^{\rm \,0.5m}=40,
 \ell_{\rm knee}^{\rm \,6m}=500)$ are shown for comparison.
 As can be seen, the performance of the two configurations is approximately
 equal when $\ell_{\rm knee}$ of the large aperture configuration is around 80.
 }
\end{center}  
\end{figure}

Figure~\ref{fig:lknee_dependence} compares the constraint on $r$ for
the large aperture telescope configuration,
  {\it Large aperture-a} with $D_{\rm tel}=6\,\mathrm{m}$,
and the hybrid telescope configuration,
 {\it Hybrid-a} with $D_{\rm tel}=6\,\mathrm{m}$.
The results are shown for two choices of survey area: 5\% and 50\%.
In this comparison, we vary $\ell_\mathrm{knee}$ of the large-aperture
configuration.

As shown in Fig.~\ref{fig:lknee_dependence}, the performance of the two types of
configurations for $r$ are
approximately equal for $\ell_\mathrm{knee} \simeq 80$, and the
large-aperture configuration will perform better on large scale structure metrics.
Thus, from a purely statistical point of view, large-aperture configurations
are advantageous if the large aperture telescope can achieve
$\ell_\mathrm{knee} < 80$.
However, we note that a detection of the primordial gravitational wave signature
requires exquisite control of systematic errors, and 
redundancy is important for cross checks.  In this sense, the ability to make measurements
over a larger $\ell$ range, in particular toward the lower
 $\ell$ range of $\sim 40$, may be important.  In this respect, achievement of $\ell_{\rm knee}=80$ 
 may not be sufficient to fully justify the choice of the large-aperture
 configuration.

\subsection{Survey Strategy}
\label{sec:survey}
In this section, we explore the dependence of the cosmological constraints on the
survey strategy.
We consider two scenarios.  The first is the single survey strategy,
where we study the performance as a function of the sky coverage fraction
$f_\mathrm{sky}$.  The second is the so-called ``deep + wide'' survey
strategy, in which 
the survey consists of two sub-surveys,  covering a
deep/small-area and a shallow/wide-area; for this strategy,
we vary the fraction of time spent on each sub-survey.

We include two instrument configurations in this study: a large aperture
configuration ({\it Large aperture-a}) with $D_{\rm tel}=6\,\mathrm{m}$ and
$\ell_{\rm knee}=100$; and
a hybrid configuration ({\it Hybrid-a}) with $D_{\rm tel}=6\,\mathrm{m}$,
  $\ell^{0.5\,\mathrm{m}}_{\rm knee}=40$,
and $\ell^{4\,\mathrm{m}}_{\rm knee}=500$.
Each configuration has a fixed cost of 50\,PCU.
For the single survey, we assume an observing duration of 2.5 years,
while for the deep + wide survey, we assume a total of 5 years of observations
divided between the two sub-surveys.

\subsubsection{Single Survey: Dependence on $f_\mathrm{sky}$}
\begin{figure}[htbp]
\begin{center}
 \includegraphics[width=0.46\textwidth]{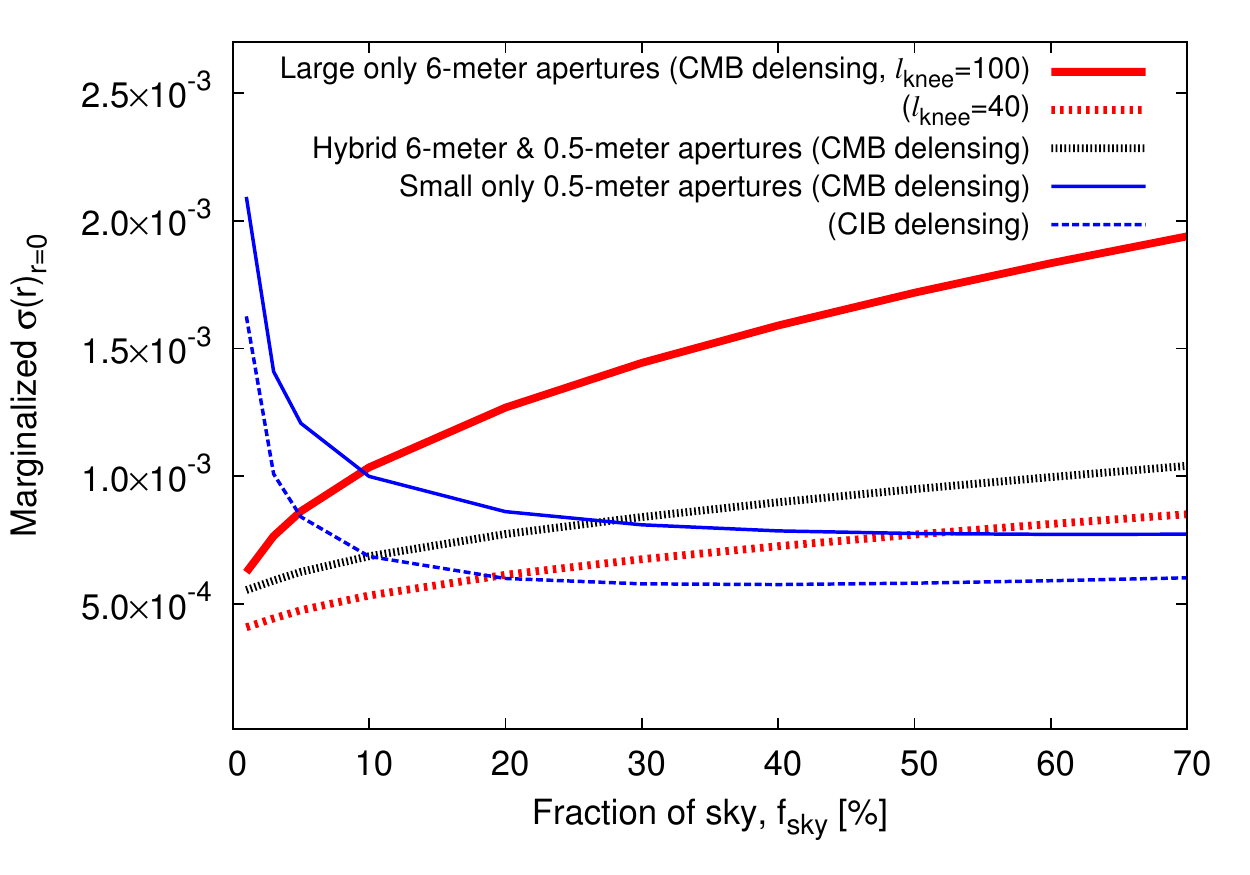}
 \hspace{0.05\textwidth}
 \includegraphics[width=0.46\textwidth]{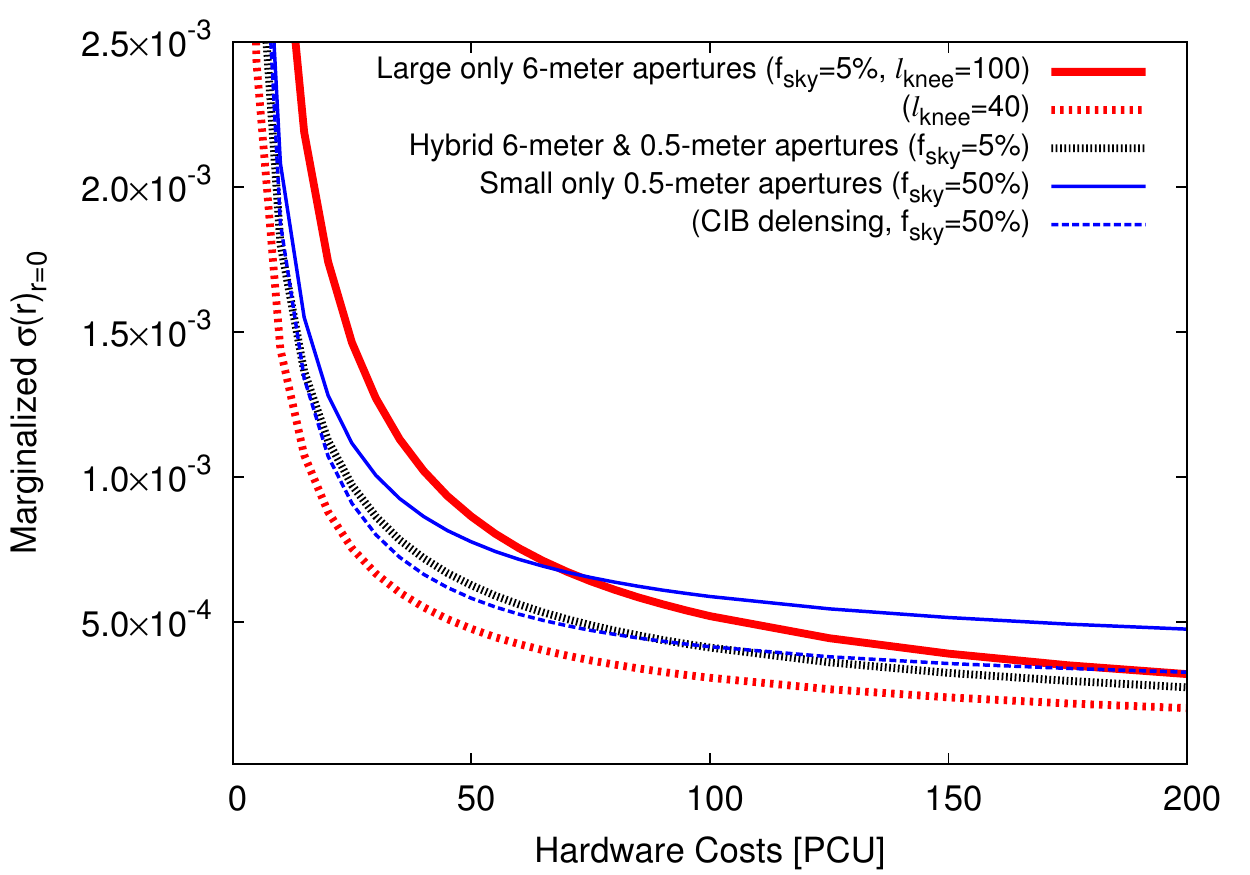}
 \caption{\label{fig:fsky_dependence_comb} 
 Left: Constraints on $r$ as a function of survey area $f_{\rm sky}$
 for a survey duration of 2.5 years.
 The following configurations are compared:
 1) large aperture configuration with $D_{\rm tel}=6\,\mathrm{m}$ and $\ell_{\rm knee}=100$;
 2) the same large aperture only configuration but with $\ell_{\rm knee}=40$;
 3) hybrid configuration with $D_{\rm tel}=6\,\mathrm{m}$, $\ell^{0.5\,\mathrm{m}}_{\rm knee}=40$, and $\ell^{6\,\mathrm{m}}_{\rm knee}=500$;
 4) a small-aperture-only configuration with $D_{\rm  tel}=0.5\,\mathrm{m}$ and $\ell_{\rm knee}=40$ with CMB self-de-lensing; and
 5) the same small-aperture-only configuration but with CIB de-lensing.
 The first three configurations, both with competitive de-lensing
 capabilities, favor small $f_{\rm sky}$.
 The latter two, on the other hand, favor large $f_{\rm sky}$
 in order to reduce the lensing sample variance.
 Right: $\sigma(r)$ as a function of the total experimental cost for the
 five cases enumerated.  The survey area $f_{\rm sky}=0.05$
 is chosen for large-aperture and hybrid configurations,
 while $f_{\rm sky}=0.5$ is used for the small-aperture-only
 configuration.
 }
 \end{center}  
\end{figure}

In the following, we show how the errors on the parameters $r$,
$N_\mathrm{eff}$, and $\sum m_\nu$ depend on the survey area
$f_\mathrm{sky}$.
Figure~\ref{fig:fsky_dependence_comb} (left) shows the error on $r$
as a function of $f_\mathrm{sky}$.
Both the large-aperture configuration and the hybrid configuration with
$D_{\rm tel}=6\,\mathrm{m}$ favor small $f_{\rm sky}$,
since they can de-lense and eliminate sample variance due to lensing.  It is worth
noting, however, that this trend is dependent on the experimental
sensitivity.  At the limit of very good sensitivity, where the
lensing noise completely dominates, the residual de-lensing noise does not
scale as favorably as the usual instrumental noise, and $\sigma(r)$
becomes more or less flat as a function of $f_{\rm sky}$.  
In Fig.~\ref{fig:fsky_dependence_comb} (left) for comparison,
we also show the case with only small-aperture telescopes
with $D_{\rm tel}=0.5\,\mathrm{m}$ and $\ell_{\rm knee}=40$
for both CMB self-de-lensing and CIB de-lensing.  In this case, the
de-lensing does not keep up with the instrumental noise at low
 $\ell$, favoring large $f_{\rm sky}$ to reduce the lensing sample variance.
Thus, these are the two limits where relatively large $f_{\rm sky}$ is
preferred.

The large-aperture-only configuration falls between the two;
it has competitive de-lensing capability and yet is not fully de-lensing
limited.  Thus, it prefers small $f_{\rm sky}$.  The hybrid
configuration is in between the cases with large aperture only and 
small aperture only, and thus prefers small $f_{\rm sky}$ but not as
strongly as the large-aperture-only case.

These dependences, as well as the $\sigma(r)$ itself, are up to the
map depth.
Figure~\ref{fig:fsky_dependence_comb} (right) illustrates such
dependences.  The  large-aperture-only configuration,
which heavily relies on de-lensing, eventually becomes the best, since
its improvement as a function of map-depth is the most steep among the
configurations compared here.
 
\begin{figure}[htbp]
\begin{center}
 \includegraphics[width=0.48\textwidth]{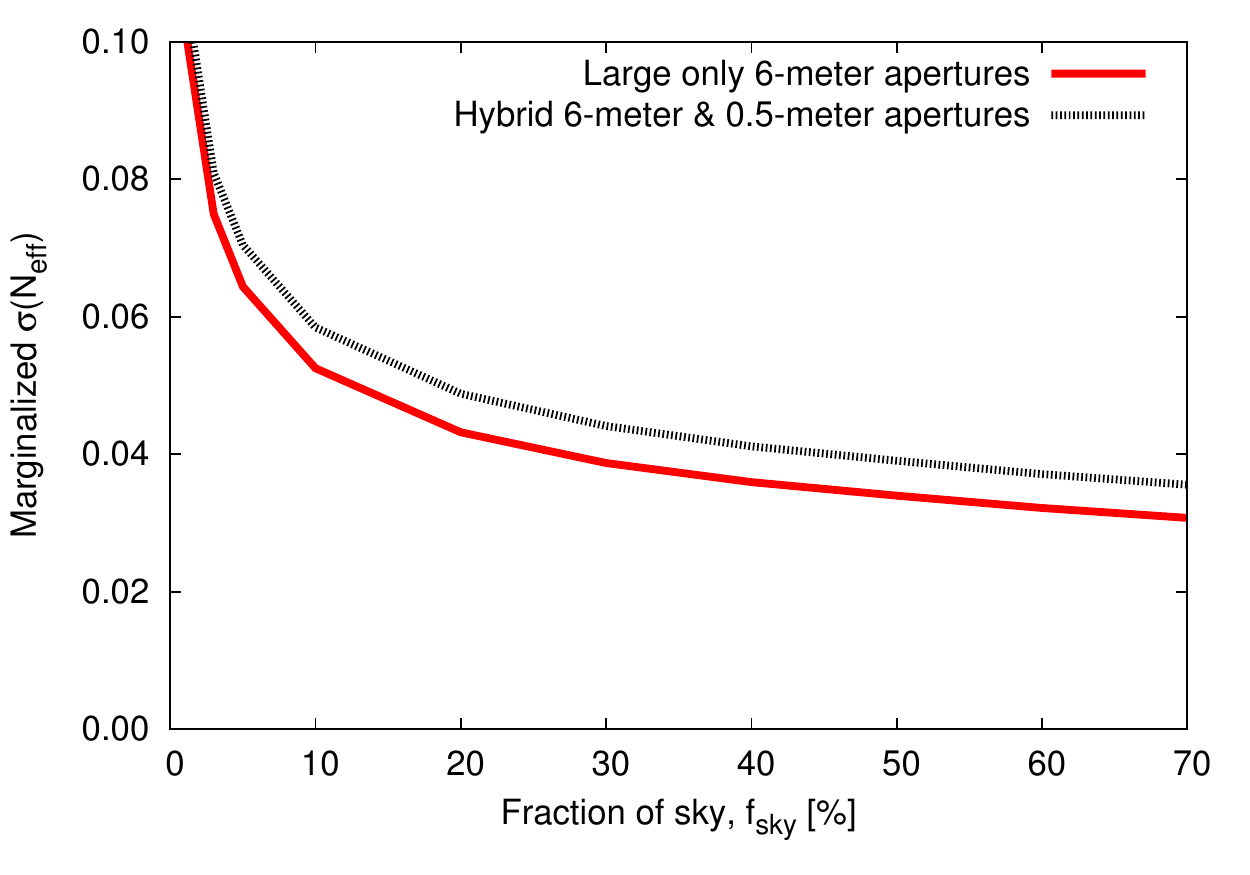}
 \hspace{0.02\textwidth}
 \includegraphics[width=0.48\textwidth]{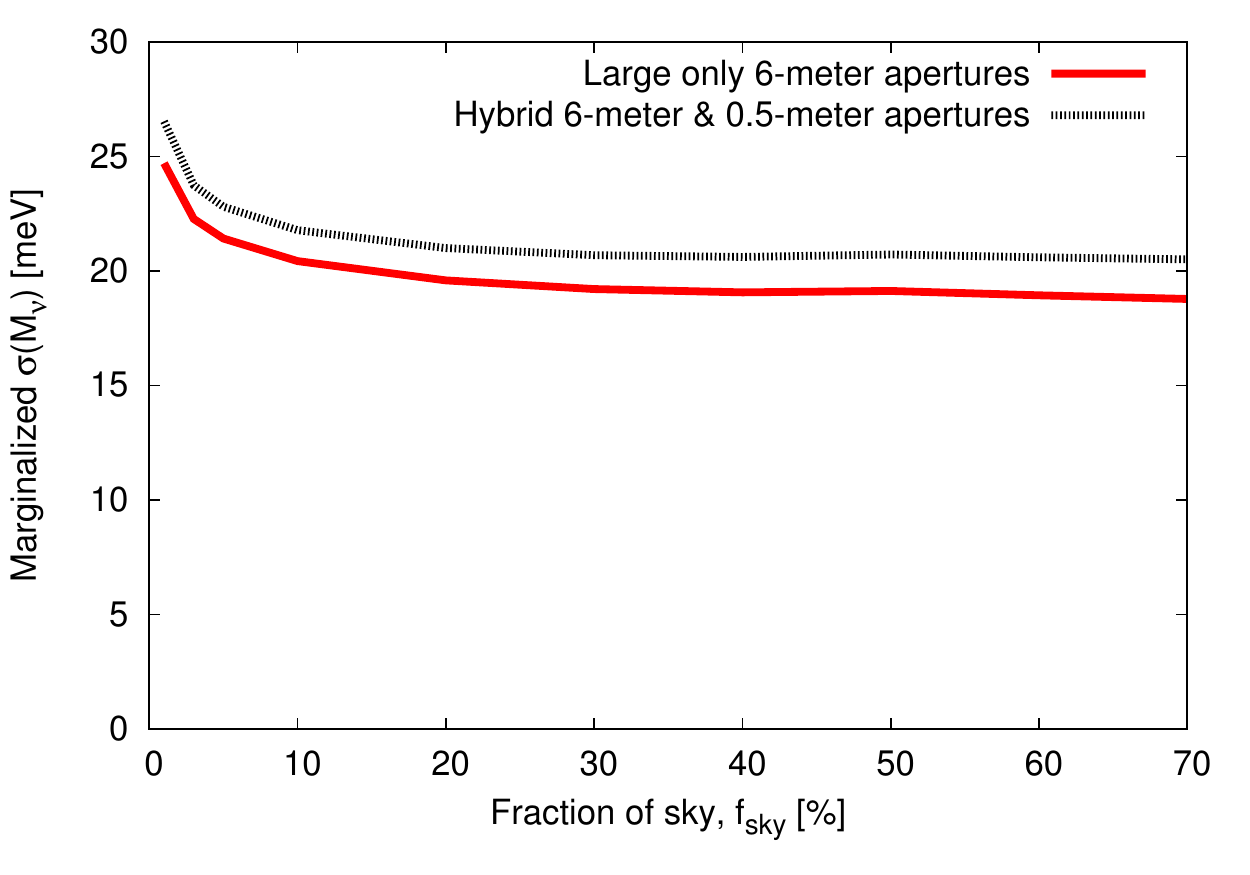}
 \caption{\label{fig:fsky_dependence_neff_mnu} 
 Constraints on $N_{\rm eff}$~(left) and
 $\sum m_\nu$~(right) as a function of $f_{\rm sky}$
 for a large-aperture-telescope configuration with
 $D_{\rm tel}=6\,\mathrm{m}$ and 
 a hybrid configuration with $D_{\rm tel}=6\,\mathrm{m}$ and $D_{\rm tel}=0.5\,\mathrm{m}$.
 }
\end{center}  
\end{figure}

Figure~\ref{fig:fsky_dependence_neff_mnu} shows the error on
$N_\mathrm{eff}$ and $\sum m_\nu$ as a function of
$f_\mathrm{sky}$ for the large-aperture telescope configuration with
 $D_{\rm tel}=6\,\mathrm{m}$.  A
larger survey area is favored for these cosmological parameters,
although the dependence is shallow, in particular for the neutrino mass.

\subsubsection{Deep + Wide Survey: Dependence on the Time Split}
In this scenario, we assume two sub-surveys with survey areas of
 5\% (deep/small area survey) and 50\% (shallow/wide area survey).
The survey strategy is parameterized by the fraction of time spent on
the wide-area survey: $R_t$.

\begin{figure}[htbp]
  \begin{center}
 \includegraphics[width=0.31\textwidth]{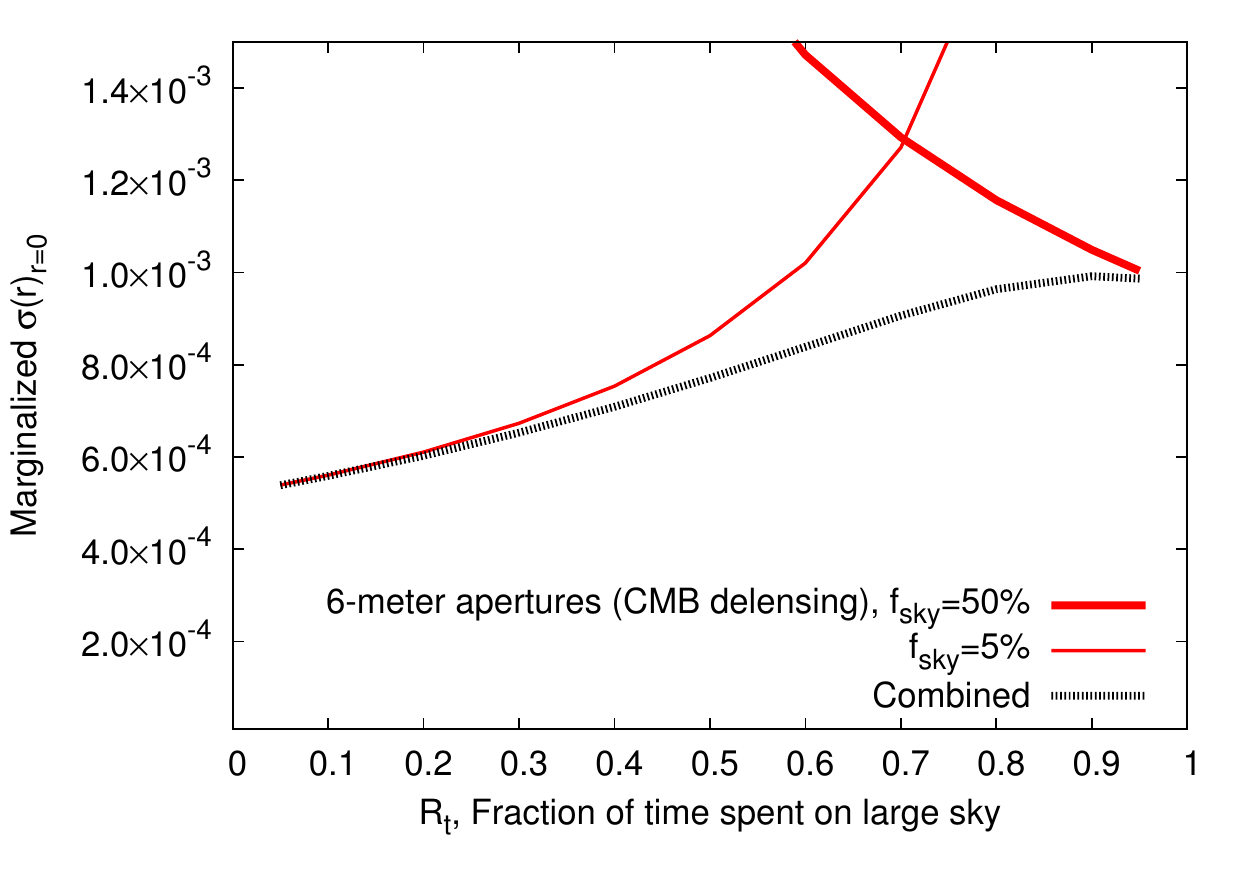}
 \hspace{0.02\textwidth}
 \includegraphics[width=0.31\textwidth]{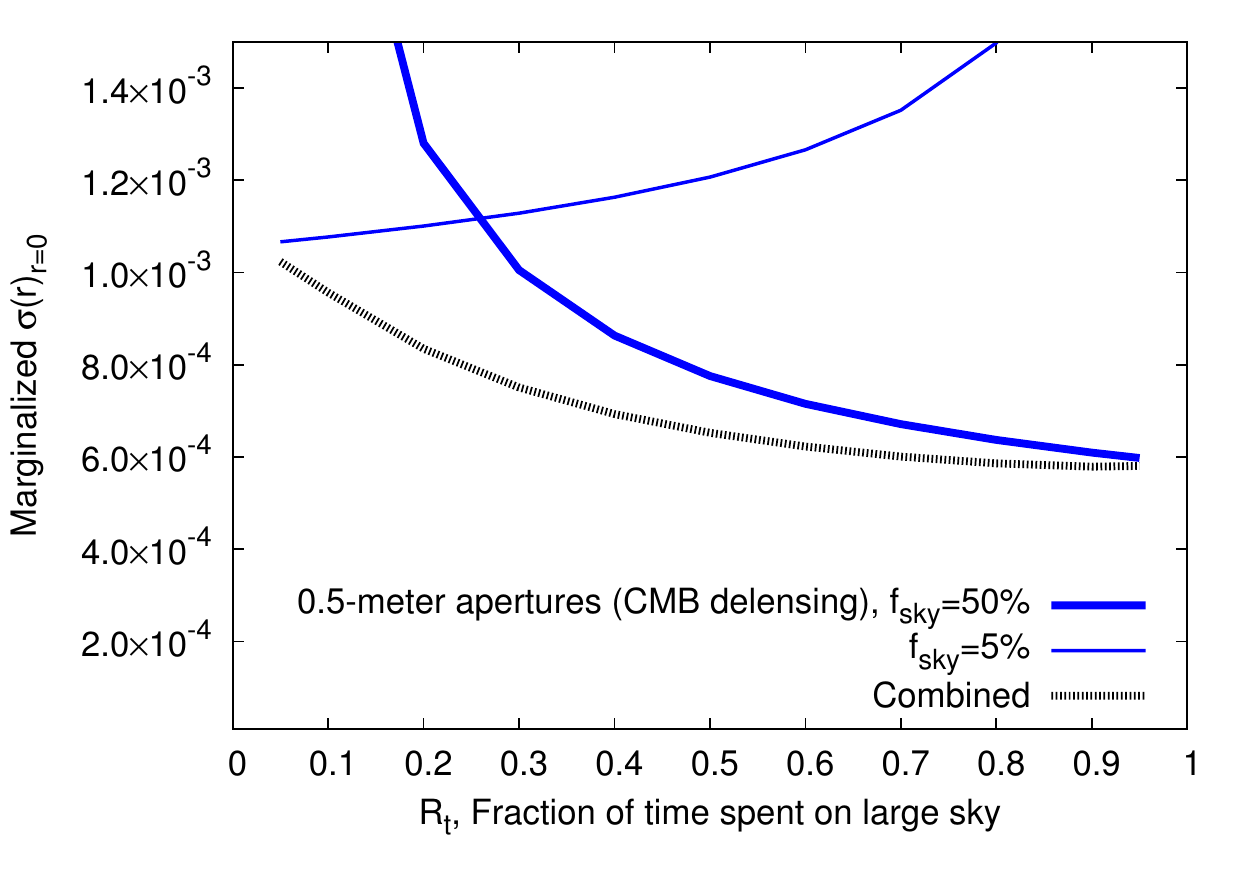}
 \hspace{0.02\textwidth}
 \includegraphics[width=0.31\textwidth]{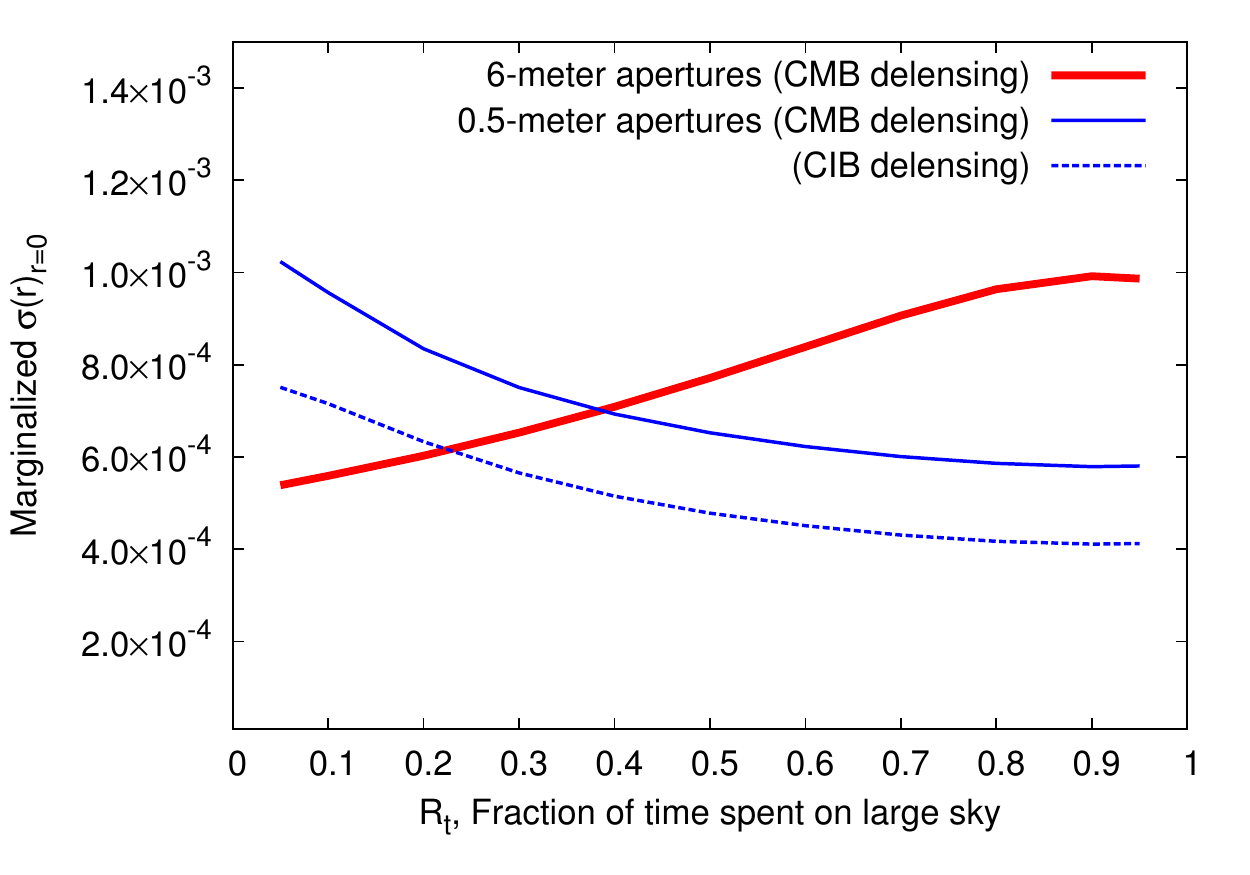}
   \caption{\label{fig:survey_rt_vs_constraints} 
   Constraints on $r$ as a function of the fractional observing time for
   the wide-area sub-survey ($f_{\rm sky}=50\%$) over the 5-year total observing time
   combining the wide-area sub-survey and the deep-area sub-survey ($f_{\rm sky}=5\%$).
   The left panel shows the case with a large aperture configuration
   with $D_{\rm tel}=6\,\mathrm{m}$ and $\ell_{\rm knee}=100$.
   The center panel shows the case for a small-aperture only
   configuration
   with $D_{\rm tel}=0.5\,\mathrm{m}$ and $\ell_{\rm knee}=40$
   with CMB self-de-lensing.
   For each, the total survey combined constraint is calculated
   as the inverse-variance weighted average of the two sub-surveys,
   neglecting the small overlap of the sub-survey areas;
   this is shown as the dotted line.
   The right panel shows the comparison of these total survey
   constraints for the two configurations, where both CMB and CIB 
   de-lensing are shown for the small-aperture configuration. }
  \end{center}
 \end{figure}

Figure~\ref{fig:survey_rt_vs_constraints} shows the constraints on $r$
as a function of $R_t$.
Here, we approximate the combined constraining power of the two surveys by simply
combining the constraints from the two surveys and
neglecting the small overlap between them.
The trend is consistent with expectations:
the configuration with $D_{\rm tel}=6\,\mathrm{m}$
favors a larger fraction for the small/deep sub-survey because of its strong
de-lensing capability, while the small-aperture only configuration favors a
larger fraction for the wide/shallow sub-survey in order to reduce the lensing sample
variance.

\begin{figure}[htbp]
  \begin{center}
 \includegraphics[width=0.48\textwidth]{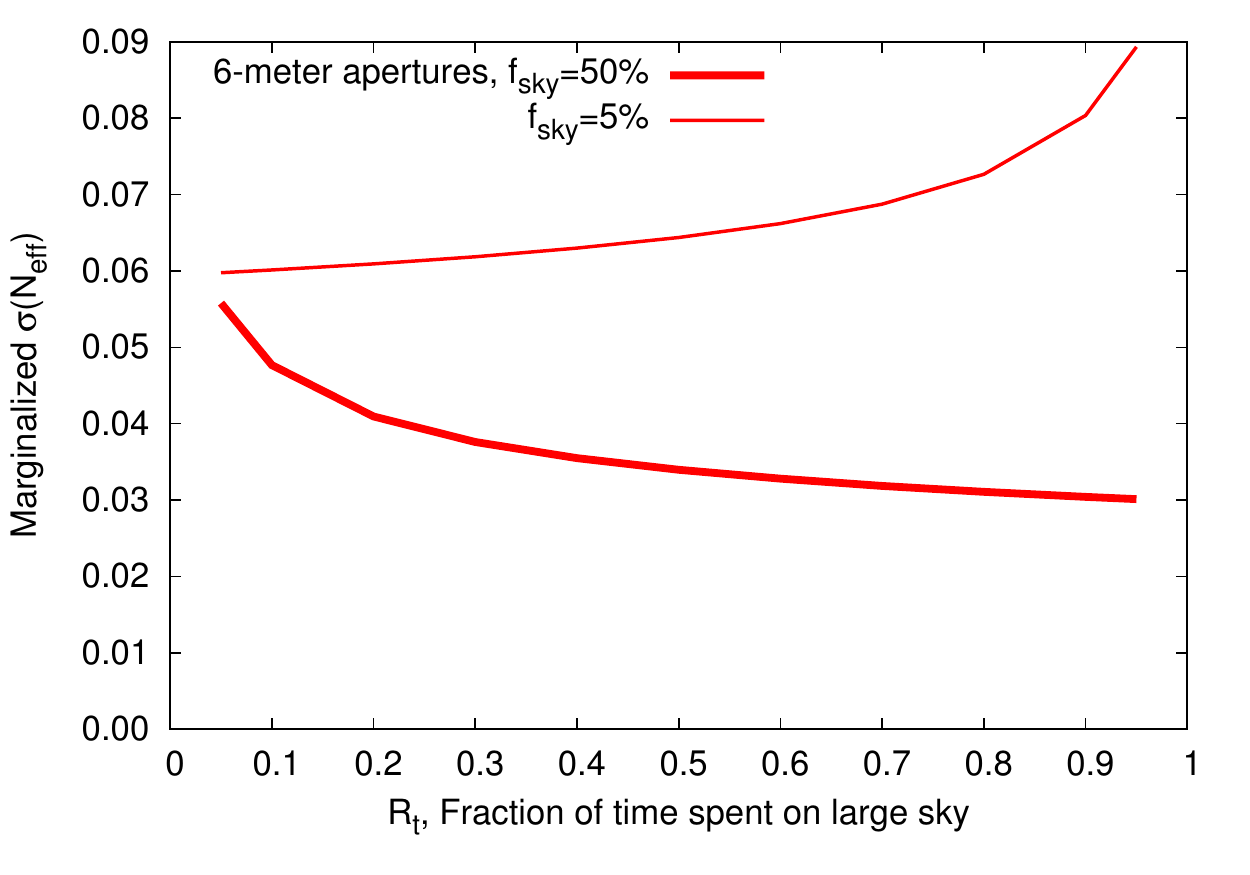}
 \hspace{0.02\textwidth}
   \includegraphics[width=0.48\textwidth]{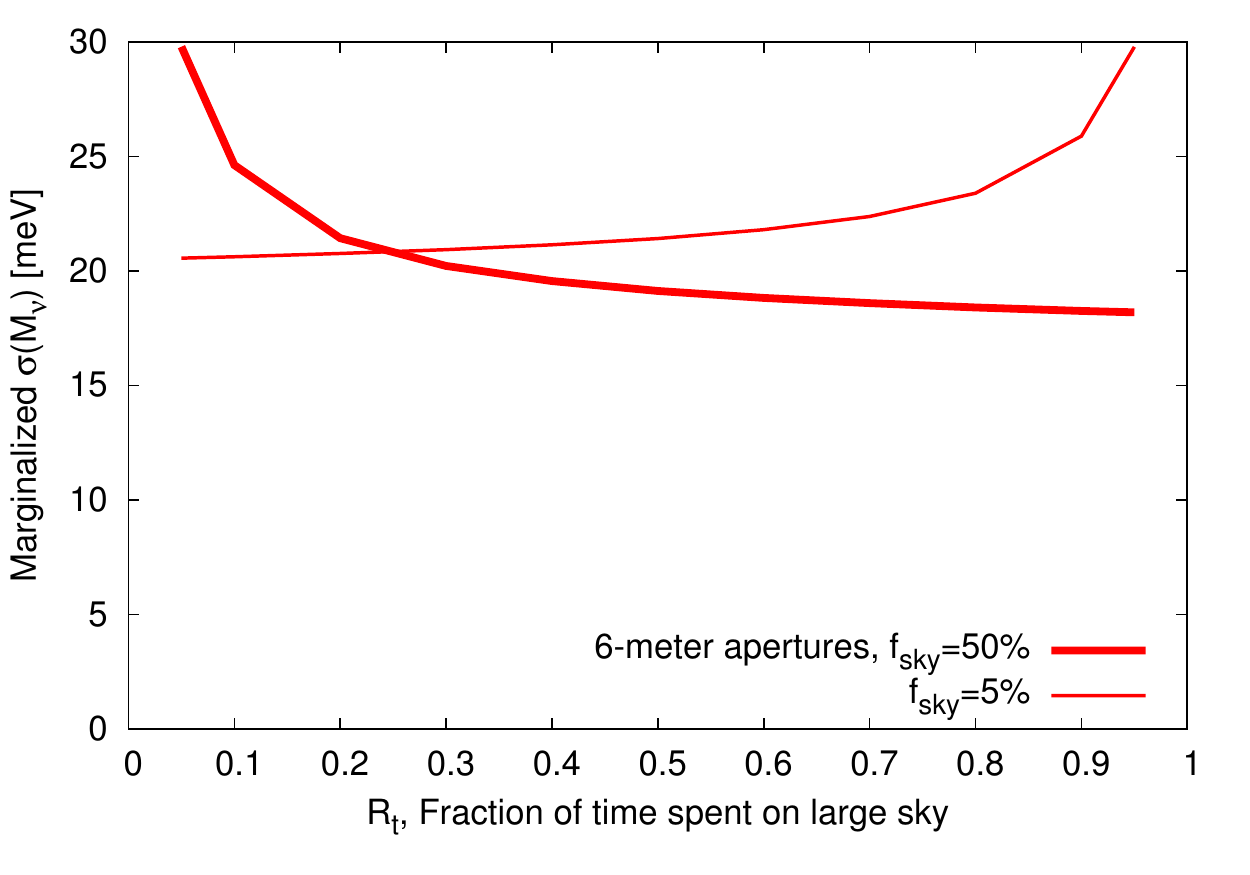}
   \caption{\label{fig:survey_rt_vs_constraints2} 
   Constraint on $N_{\rm eff}$~(left) and $\sum m_{\nu}$~(right) as a function of
   the fraction of time spent on the wide-area sub-survey, $R_t$.
   A large-aperture configuration with $D_{\rm tel}=6\,\mathrm{m}$ is assumed.}
  \end{center}
 \end{figure}

Figure~\ref{fig:survey_rt_vs_constraints2} shows the constraints on
$N_{\rm eff}$ and $\sum m_\nu$ as a function of $R_t$.
A large aperture configuration with $D_{\rm tel}=6\,\mathrm{m}$ is assumed. 
Here we show only the individual constraints from each sub-survey; the
measurement of $N_{\rm eff}$ is
dominated by the wide/shallow sub-survey.
While the trend is similar for $\sum m_\nu$,
the contribution of the deep/small sub-survey is closer to that of 
the wide/shallow sub-survey.

\section{Strawperson Configurations}
\label{sec:strawperson}
In the previous section, we presented optimizations in an abstract manner
(e.g., the number of telescopes is not constrained to be an integer).
In this section, we will discuss two
examples in detail to make the optimization process more concrete
and to shed light on additional practical issues that arise when
designing a real experiment.
In the following, we present the configurations for a total cost of
50\,PCU.
While we present the forecasted error on $r$ and
$N_\mathrm{eff}$ for this configuration,
we emphasize that they depend on the underlying assumptions.
For example, more
complicated foregrounds than our assumptions would inevitably degrade the
errors on the cosmological parameters.
The detector efficiency and yield we assume
(Sec.~\ref{sec:realistic_sensit}) may
be optimistic;
lower values of the efficiency and yield would
lead to requiring more detectors.
Thus, although the 50-PCU configurations correspond to $\sim 200{,}000$
detectors as we show below, possible variations in these assumptions
would lead to a larger number of detectors required for the same level
of errors in $r$ and $N_\mathrm{eff}$.
We note that the total cost of the instrument is linear at this
scale to a good approximation -- that is, configurations with twice as many
detectors and telescopes would cost 100\,PCU.

\subsection{Strawperson 1: 6\,m Large Aperture Configuration}
As shown in Sec.~\ref{sec:results}, we find $D_{\rm tel}=6\,\mathrm{m}$
to be approximately optimal for a large-aperture telescope array.  
We also find that there is no significant difference in performance for an array with
fixed size aperture vs. scaled aperture as a function of frequency.
Thus, we select a strawperson design based on a homogeneous configuration with 6\,m fixed-aperture
telescopes and a total cost of 50\,PCU.

Table~\ref{tab:strawperson_6m_homogeneous} shows the performance:
the total array sensitivity and the beam size vs. frequency of this configuration.
The sensitivity in the table includes the degradation factors discussed in
Sec.~\ref{sec:realistic_sensit}.
For this configuration, the raw output of the optimization
gives a fractional number of LF telescopes.  
Thus, for this strawperson, we choose to combine LF
and MF detectors within the focal plane.  The strawperson has 4 combined
LF/MF telescopes, each
supporting  2,300 LF detectors and 30,300 MF detectors, and 2 HF telescopes
with 30,200 detectors each, for a total of 190,400 detectors.
In a five-year survey
equally split between small and wide fields, this strawperson design
achieves
$\sigma(r) = 7.7\times 10^{-4}$ and $\sigma(N_{\rm eff})=0.034$,
where, for the latter, we assume the wide survey dominates the sensitivity.
As described in the previous section, $\ell_{\rm knee} = 100$ is
assumed here.

While we could in principle also consider combining LF or MF with HF detectors,
 we choose not to do so because 1) the HF telescope will
require better surface accuracy and thus not combining with LF
and MF detectors may be more cost effective, and 2) we know from the
optimization results that we could reduce cost by decreasing the
aperture diameter of the HF telescopes without degrading the
experimental sensitivity.  
This is  a topic for further
discussion as we advance in the design process.
Table~\ref{tab:strawperson_6m_homogeneous_tel} shows the details
of the 6-m strawperson array.  
The results of the optimization give approximately one cryostat per
telescope, and thus we assume one cryostat per telescope in this
strawperson.
Table~\ref{tab:strawperson_6m_homogeneous_cost} shows the cost
allocation for the various sub-components.

There are a few aspects  that are worth noting for
future refinements of the modeling and optimization process.
In particular, since we built the cost and instrumental models around the MF
instrument, some of the approximations may break down for the HF and LF instruments.
For example, one caveat for the HF is the packing efficiency of the pixels.
In the strawperson design, the HF instrument only requires 12
detector wafers, due to its high pixel density.  Each wafer in this design
has almost 5000 pixels.
While it is optically possible to pack the HF pixels this tightly
(Table~\ref{tab:wafer_cost_to_detector_cost}),
this high density may be challenging from a fabrication and integration
perspective.
While the solution to this problem may require a lower HF pixel density, 
this will not necessarily result in a
significant cost change, because the HF cryostats and telescopes in this specific
configuration have margin in capacity, and it is likely that a greater number
of wafers can be accommodated in the same number of the cryostats and
telescopes, if necessary. 

The LF instrument, on the other hand, has a large number
of wafers -- that is, almost as many as for the MF instrument (due to the larger
pixel size, which offsets the smaller number of detectors).
In our example, there may be too many LF + MF wafers 
allocated per telescope and cryostat (44 total), and,
as a result, the cost may be underestimated.
We note, however, that our optimization versus frequency had a very shallow
optimum in the ratio of MF/LF detectors.  The nominal value of MF/LF=20
could be increased to 50 or even 100 without 
a significant  degradation in sensitivity
 (Fig.~\ref{fig:frequency_optimization}).
 This choice, however, is also related to the need to characterize
foregrounds, which may be more complicated than our model
and clearly requires further study.

\begin{table}[htbp]
 \begin{center}
  \caption{The number of detectors and wafers,
  array sensitivity, and angular resolutions of the optimum homogeneous
 configuration with a 6-m telescope aperture diameter for all the
 frequencies.
  The target total cost is 50\,PCU.
  The number of detectors ($N_{\rm det}$)
  and the number of wafers ($N_{\rm wafer}$)
  are counted for each frequency groups,
  i.e., the number of detectors for each of 95 and 150\,GHz bands is
  60,500.
  The map depth is given for 2.5\,years of observation for each of the
  5\% and 50\% of $f_{\rm sky}$.
  The array NET is with the degradation factor $\beta$
  and the yield factor $Y$ applied,
  and the map depth assumes the efficiency factor $\varepsilon$
  (see Sec.~\ref{sec:realistic_sensit}).
  \label{tab:strawperson_6m_homogeneous}
 }
   \begin{tabular}{cccccccc}
    \hline \hline    
    Frequency & Frequency & \multirow{2}*{$N_{\rm det}$} & \multirow{2}*{$N_{\rm wafer}$} & Array NET &
    Beam FWHM & \multicolumn{2}{c}{Map Depth ($\mu \mathrm{K}$-arcmin)} \\
    Group &  (GHz) & & & ($\mu K_\mathrm{CMB} \sqrt{\rm s}$)  &  (arcmin)
			&	$f_{\rm sky}=0.05$ & $f_{\rm sky}=0.5$ \\
    \hline
    \multirow{3}*{LF} &  21 & \multirow{3}*{  9,000} & \multirow{3}*{ 76} & 
                       8.4 & 10.4 &  7.3 & 23.1 \\
        &  29 & & &    6.1 &  7.5 &  5.3 & 16.7 \\
        &  40 & & &    6.1 &  5.5 &  5.3 & 16.8 \\
    \hline
    \multirow{2}*{MF} &  95 & \multirow{2}*{121,000} & \multirow{2}*{100} & 
                       1.5 &  2.3 &  1.3 &  4.1 \\
        & 150 & & &    1.7 &  1.5 &  1.5 &  4.6 \\
    \hline
    \multirow{2}*{HF} & 220 & \multirow{2}*{ 60,400} & \multirow{2}*{ 12} & 
                       6.5 &  1.0 &  5.7 & 17.9 \\
        & 270 & & &   10.8 &  0.8 &  9.4 & 29.7 \\
    \hline
    Total & & 190,400 & 188 & & \\
    \hline
   \end{tabular}
 \end{center}
\end{table}
\begin{table}[htbp]
 \begin{center}
  \caption{\label{tab:strawperson_6m_homogeneous_tel}
  The parameters of the telescopes for the optimal homogeneous large-aperture
  configuration with a 6\,m telescope aperture diameter.
  The target total cost is 50\,PCU.
  The same telescopes accommodate both LF and MF, while HF has dedicated
  telescopes.}
  \begin{tabular}{cccc}
   \hline \hline
   & $N_{\rm det}$ per telescope & $N_{\rm wafer}$ per telescope & Number of
   telescopes \\
   \hline
    LF &  2,300 & 19.0 & \multirow{2}*{4}  \\
    MF & 30,300 & 25.0 &   \\
    HF & 30,200 &  6.0 & 2  \\
   \hline
  \end{tabular}
 \end{center}
\end{table}
\begin{table}[htbp]
 \begin{center}
  \caption{\label{tab:strawperson_6m_homogeneous_cost}
  The cost allocation over various subcomponents for the optimal
  homogeneous large aperture
  configuration with a 6\,m telescope aperture diameter.
  The cost is in Parametric Cost Units (PCU).
  The cost split of the cryostat and telescope between LF and MF is 
  not well defined since LF and MF share the same telescope and
  cryostat.  Here, we take a split of 3.6 vs. 0.4 just for the sake of 
completing the per-frequency column (right most).
  }
  \begin{tabular}{cccccccccc}
   \hline \hline
   & \multicolumn{2}{c}{Detector} & \multicolumn{2}{c}{Readout} &
   \multicolumn{2}{c}{Cryostat} & \multicolumn{2}{c}{Telescope} & Total \\
   & $N_{\rm wafer}$ & Cost & Channels & Cost &
   Count & Cost & Count & Cost & Cost \\
   \hline
    LF &  76 &  4.6 &   9,000 &  0.2 &  0.4 &  0.4 &  0.4 &   2.0 &   7.2 \\
    MF & 100 &  6.0 & 121,000 &  2.4 &  3.6 &  3.6 &  3.6 &  18.1 &  30.1 \\
    HF &  12 &  0.8 &  60,400 &  1.2 &  2.0 &  2.0 &  2.0 &  10.1 &  14.1 \\
    \hline
    Total & & 11.4 & &  3.8 & &  6.0 & &  30.2 &  51.4 \\
   \hline
  \end{tabular}
 \end{center}
\end{table}

\subsection{Strawperson 2: 6\,m / 0.5\,m Hybrid Configuration}
\label{sec:strawperson_2}
We find $D_{\rm tel}=6\,\mathrm{m}$
is also a good choice for the large-aperture telescope in the hybrid configuration
and find no significant difference
between the fixed and scaled aperture sizes vs. frequency.
A 50/50 cost split between the large-aperture and
small-aperture portions was found to be close to optimal for $r$.
Thus, we examine a hybrid configuration with 6-m large-aperture 
telescopes and 0.5-m small-aperture telescopes, with the
total cost of 50\,PCU  split equally between them.  
The large aperture portion of this configuration is simply half of
that shown in Tables~\ref{tab:strawperson_6m_homogeneous}-\ref{tab:strawperson_6m_homogeneous_cost},
and the sensitivity in Table~\ref{tab:strawperson_6m_homogeneous}
is $\sqrt{2}$ worse.  The small-aperture portion consists of seventeen 0.5-m telescopes;
Table~\ref{tab:strawperson_05m_hybrid} shows the performance of the
small-aperture telescopes, and
Table~\ref{tab:strawperson_05m_hybrid_tel} and
\ref{tab:strawperson_05m_hybrid_cost}
show the parameters of the  small telescopes
and the cost allocation over the subcomponents, respectively.
This strawperson design has a total of 207,300 detectors and in a five-year survey
equally split between small and wide fields achieves
$\sigma(r) = 5.2 \times 10^{-4}$ and $\sigma(N_{\rm eff})=0.039$,
where, for the latter, we assume the wide survey of the large-aperture
telescopes dominates the sensitivity.
As described in the previous section, $\ell_{\rm knee} = 40$ (500) is
assumed for the small-aperture (large-aperture) portion of the configuration.

Similarly to the homogeneous large-aperture configuration, it is worth
discussing the
LF part of this configuration.  In this configuration,
the small-aperture LF instrument comprises one large cryostat
that accommodates six small-aperture telescopes.  This cryostat has a capacity
slightly larger than that for large-aperture telescopes.
Fabrication of such a large cryostat may cost more than our assumption
and may take more pulse-tube cryocoolers or sub-Kelvin refrigerators
than we assumed, resulting in an increased cost.  It may turn out
that splitting this instrument into two cryostats is financially more
beneficial.
We also note that our optimization 
indicates that a smaller number of
LF channels will not lead to significant sensitivity
 degradation (Fig.~\ref{fig:hybrid_frequency_optimizations}).
Yet another point to note is that the 0.5-m aperture size is too small for
these frequencies,
and the instrument barely achieves the required resolution.  
On the other hand, the atmospheric fluctuation is smaller at these
frequencies, because the emission from oxygen dominates
as opposed to the water vapor.  Thus, it is plausible that a larger-aperture instrument could still achieve good low-frequency noise
performance.  
A dedicated design study of the LF instrument is clearly an area that needs
further study.

\begin{table}[htbp]
 \begin{center}
  \caption{The number of detectors and wafers,
  array sensitivity, and angular resolutions of the small-aperture (0.5\,m)
  portion of the optimum hybrid
  configuration.
  The total cost of the small aperture portion is 25\,PCU.
  The number of detectors ($N_{\rm det}$)
  and the number of wafers ($N_{\rm wafer}$)
  are counted for each frequency group.
  The array NET is with the degradation factor $\beta$
  and the yield factor $Y$ applied,
  and the map depth assumes the efficiency factor $\varepsilon$
  (see Sec.~\ref{sec:realistic_sensit}).
  \label{tab:strawperson_05m_hybrid}
 }
   \begin{tabular}{cccccccc}
    \hline \hline 
    Frequency & Frequency & \multirow{2}*{$N_{\rm det}$} & \multirow{2}*{$N_{\rm wafer}$} & Array NET &
    Beam FWHM & \multicolumn{2}{c}{Map Depth ($\mu \mathrm{K}$-arcmin)} \\
    Group &  (GHz) & & & ($\mu K_\mathrm{CMB} \sqrt{\rm s}$)  &  (arcmin)
			&	$f_{\rm sky}=0.05$ & $f_{\rm sky}=0.5$ \\
    \hline
    \multirow{3}*{LF} &  21 & \multirow{3}*{  5,300} & \multirow{3}*{ 45} & 
                       9.2 & 125.0 &  8.0 & 25.2 \\
        &  29 & & &    6.4 & 90.5 &  5.5 & 17.5 \\
        &  40 & & &    6.7 & 65.6 &  5.8 & 18.3 \\
    \hline
    \multirow{2}*{MF} &  95 & \multirow{2}*{ 71,200} & \multirow{2}*{ 59} & 
                       1.6 & 27.6 &  1.4 &  4.4 \\
        & 150 & & &    1.8 & 17.5 &  1.5 &  4.8 \\
    \hline
    \multirow{2}*{HF} & 220 & \multirow{2}*{ 35,600} & \multirow{2}*{  7} & 
                       6.8 & 11.9 &  5.9 & 18.7 \\
        & 270 & & &   11.6 &  9.7 & 10.0 & 31.8 \\
    \hline
    Total & & 112,100 & 111 & & \\
    \hline
   \end{tabular}
 \end{center}
\end{table}
\begin{table}[htbp]
 \begin{center}
  \caption{\label{tab:strawperson_05m_hybrid_tel}
  The parameters of the telescopes for the small-aperture (0.5\,m)
  portion of the optimum hybrid
  configuration.
  The total cost is 25\,PCU.}
  \begin{tabular}{cccc}
   \hline \hline
   & $N_{\rm det}$ per telescope & $N_{\rm wafer}$ per telescope & Number of
   telescopes \\
   \hline
    LF &    900 &  7.5 & 6 \\
    MF & 11,900 &  7.4 & 8 \\
    HF & 11,900 &  2.3 & 3 \\
   \hline
  \end{tabular}
 \end{center}
\end{table}
\begin{table}[htbp]
 \begin{center}
  \caption{\label{tab:strawperson_05m_hybrid_cost}
  The cost allocation over various subcomponents for  the small-aperture (0.5\,m)
  portion of the optimum hybrid
  configuration.
  The cost is in the Parametric Cost Unit (PCU).  }
  \begin{tabular}{cccccccccc}
   \hline \hline
   & \multicolumn{2}{c}{Detector} & \multicolumn{2}{c}{Readout} &
   \multicolumn{2}{c}{Cryostat} & \multicolumn{2}{c}{Telescope} & Total \\
   & $N_{\rm wafer}$ & Cost & Channels & Cost &
   Count & Cost & Count & Cost \\
   \hline
    LF &  45 &  2.7 &   5,300 &  0.1 &  1.0 &  1.0 &  6.0 &   1.6 &   5.4 \\
    MF &  59 &  3.6 &  71,200 &  1.4 &  8.0 &  8.0 &  8.0 &   2.1 &  15.1 \\
    HF &   7 &  0.4 &  35,600 &  0.7 &  3.0 &  3.0 &  3.0 &   0.8 &   4.9 \\
    \hline
    Total & &  6.7 & &  2.2 & & 12.0 & &   4.4 &  25.3 \\
    \hline
  \end{tabular}
 \end{center}
\end{table}

\section{Conclusions}
\label{sec:conclusion}
We have established a framework to optimize the science
output of CMB-S4 based on a simple cost and instrumental model
combined with the Fisher matrix forecasting code \textsc{CMB4cast}. 
We have carried out a variety of optimization exercises and
identified instrument configurations that are close to
optimal for a given fixed cost. 

We have examined four science goals: primordial gravitational waves,
or its amplitude $r$;
the number of relativistic species in the early universe $N_\mathrm{eff}$;
sum of the neutrino mass $M_\nu \equiv \sum m_\nu$;
and the kSZ effect.
The cost is modeled in the unit of ``Parametric Cost Model Unit (PCU).''
One PCU approximately corresponds to one million dollars of raw hardware
cost.  Inclusion of labor cost for integration and evaluation of
components would increase the cost (roughly) by a factor of three;
this is clearly an area that requires more sophisticated
estimate by further research.

We examined mainly two types of configurations.  One is with large-aperture telescopes only.  In this case, the large-aperture telescopes
are assumed to measure the entire angular scales required for the
science goals: from degree to arcminute scales.  Another type is hybrid
configurations, which combine the large aperture telescopes 
and small-aperture telescopes with a fixed aperture size of 0.5\,m.
In order to compare the two types in equal footing, the total cost is
kept the same.
For the large-aperture telescopes of both types, we have examined both
the case where the aperture size is fixed regardless frequency
and the case where the aperture size is scaled depending on the
frequency.  We did not find significant differences between the two cases.
In comparing the two types, the assumption on the low frequency noise
excess is important.  The characteristic angular scale of the noise
excess, $\ell_{\rm knee}$, is assumed to be 100 for the large-aperture-only configurations.  In the hybrid configurations,
we assume $\ell_{\rm knee}$ of 40 (500) for their small-aperture
(large-aperture) telescopes.

We optimized instrumental parameters to minimize uncertainties in
cosmological parameters for a fixed cost of 50\,PCU.
One parameter studied is the frequency distribution.  The ratio between low frequency
(LF),
mid-frequency (MF), and high-frequency (HF) shows a broad optimum.
The ratio between LF and MF is optimum with their ratio of 20--100,
and that between HF and MF is optimum at the ratio of 1--5.
This is the case in all configurations we examined,
and for both of the cosmological parameters $r$ and $N_{\rm eff}$.

For the configurations with large-aperture telescopes only,
the optimum of the aperture size is driven by $r$, resulting in a broad
optimal around 4--6\,m.  Smaller aperture size than this result in
inferior
de-lensing performance, while the larger aperture size leads to smaller
number of detectors and inferior noise performance.
The performance saturates at around 4--6\,m
for other science goals of $N_{\rm eff}$, $M_\nu$, and kSZ,
although they do not degrade at a larger aperture size since the
decrease of the number of detectors approximately balances with the
improvement in the angular resolution.

For the hybrid configurations, the optimum is broader even for $r$,
since the instrumental noise level remains constant for degree-angular
scales regardless the aperture size of the large-aperture telescopes.
The uncertainty on $r$ saturates at around an aperture size of 4\,m
and does not degrade up to 10\,m, which is the maximum size that we examined.
Comparison between the large-aperture and hybrid types depends on the
$\ell_{\rm knee}$.  From a purely statistical point of view, the
uncertainty on $r$ approximately equals when 
$\ell_{\rm knee}\simeq 80$ for the large-aperture type.
 
The optimum sky coverage depends on the configuration and science
target.
Large sky coverage ($f_{\rm sky} \gtrsim 0.3$) is preferred for
 $N_{\rm eff}$ and $M_\nu$.
On the other hand, small sky coverage, $f_{\rm sky} \lesssim 0.05$ is
preferred for $r$ but only with CMB de-lensing.
In practice, the CMB-S4 is likely to combine deep/small survey and
wide/shallow survey.  We confirmed that  the uncertainties in
cosmological parameters have shallow dependence on the ratio of the two
surveys.

Finally, we have presented a couple of strawperson configurations for
CMB-S4, one for each of the large-aperture and hybrid configurations.
The distribution of the detectors among frequencies, sensitivity,
and the cost distribution are presented.  It was also noted that the
instrumental model is prone to break down at the corners of parameter
space, in particular at the low-frequency and small-aperture end. This
is one of the areas that needs an improvement.

Our study serves as a stepping stone toward the conceptual design of the
CMB-S4.
There are several areas that deserve further improvement.
In our forecasting framework, we adopt simple foregrounds model.
We assume simple two-component foregrounds with spatially varying
power-law spectral indices.
We estimate de-lensing performance statistically;
non-idealities such as
anisotropic mode coverage, boundary effect, and possible 
foreground residual may degrade the performance.
Instrumental systematics, which were not accounted for in our analyses, 
may also influence the conceptual design.
Further study of the instrumental and cost models
will be one of the main areas of study by the whole CMB-S4 community.
We hope that our framework will be useful for estimating the influence
of these improvement in future, and we will incorporate them in order to
 further improve the optimization.

\section*{Acknowledgement}
We thank Charles~Lawrence, Pat~McDonald,
Michael~Niemack, and Edward~Wollack for useful discussions.
We acknowledge inputs from the CMB-S4 community at the
collaboration workshops at LBNL in March 2016 and at U. Chicago in
September 2016.
We thank Mark~Devlin, Brian~Keating, Stephen~Padin, and Lyman~Page for
providing telescope cost information.
D.B. acknowledges support from the NSF Astronomy and Astrophysics Postdoctoral Fellowship under grant number NSF AST 1501422 and additional support from the UC Berkeley Space Sciences Lab Charles H. Townes Fellowship and the Lawrence Berkeley National Laboratory Chamberlain Fellowship.
Work at LBNL
is supported in part by the U.S. Department of Energy, Office of
Science,  Office of High Energy Physics, under contract
No. DE-AC02-05CH11231.
Part of the work was performed using the computational resources at the
National Energy Research Scientific Computing Center, a DOE Office of
Science User Facility supported by the Office of Science of the
U.S. Department of Energy under Contract No. DE-AC02-05CH11231.
The Flatiron Institute is supported by the Simons Foundation.
The original data of the global distribution of mean clear-sky PWV were provided by Jonathan Y. Suen, and we thank him.

\bibliography{berkeley_cmbs4_optimization}

\providecommand{\href}[2]{#2}\begingroup\raggedright\begin{thebibliography}{10}

\bibitem{2015APh....63...55A}
K.~N. {Abazajian}, K.~{Arnold}, J.~{Austermann}, B.~A. {Benson}, C.~{Bischoff},
  J.~{Bock} et~al., \emph{{Inflation physics from the cosmic microwave
  background and large scale structure}},
  \href{http://dx.doi.org/10.1016/j.astropartphys.2014.05.013}{\emph{Astroparticle
  Physics} {\bf 63} (Mar., 2015) 55--65},
  [\href{https://arxiv.org/abs/1309.5381}{{\tt 1309.5381}}].

\bibitem{2015APh....63...66A}
K.~N. {Abazajian}, K.~{Arnold}, J.~{Austermann}, B.~A. {Benson}, C.~{Bischoff},
  J.~{Bock} et~al., \emph{{Neutrino physics from the cosmic microwave
  background and large scale structure}},
  \href{http://dx.doi.org/10.1016/j.astropartphys.2014.05.014}{\emph{Astroparticle
  Physics} {\bf 63} (Mar., 2015) 66--80},
  [\href{https://arxiv.org/abs/1309.5383}{{\tt 1309.5383}}].

\bibitem{2016arXiv161002743A}
K.~N. {Abazajian}, P.~{Adshead}, Z.~{Ahmed}, S.~W. {Allen}, D.~{Alonso}, K.~S.
  {Arnold} et~al., \emph{{CMB-S4 Science Book, First Edition}}, {\emph{ArXiv
  e-prints} (Oct., 2016) }, [\href{https://arxiv.org/abs/1610.02743}{{\tt
  1610.02743}}].

\bibitem{Errard2016}
J.~{Errard}, S.~M. {Feeney}, H.~V. {Peiris} and A.~H. {Jaffe}, \emph{{Robust
  forecasts on fundamental physics from the foreground-obscured,
  gravitationally-lensed CMB polarization}},
  \href{http://dx.doi.org/10.1088/1475-7516/2016/03/052}{\emph{\jcap} {\bf 3}
  (Mar., 2016) 052}, [\href{https://arxiv.org/abs/1509.06770}{{\tt
  1509.06770}}].

\bibitem{1981PhRvD..23..347G}
A.~H. {Guth}, \emph{{Inflationary universe: A possible solution to the horizon
  and flatness problems}},
  \href{http://dx.doi.org/10.1103/PhysRevD.23.347}{\emph{\prd} {\bf 23} (Jan.,
  1981) 347--356}.

\bibitem{Linde:2005ht}
A.~D. Linde, \emph{{Particle physics and inflationary cosmology}},
  {\emph{Contemp. Concepts Phys.} {\bf 5} (1990) 1--362},
  [\href{https://arxiv.org/abs/hep-th/0503203}{{\tt hep-th/0503203}}].

\bibitem{Rubakov:1982df}
V.~A. Rubakov, M.~V. Sazhin and A.~V. Veryaskin, \emph{{Graviton Creation in
  the Inflationary Universe and the Grand Unification Scale}},
  \href{http://dx.doi.org/10.1016/0370-2693(82)90641-4}{\emph{Phys. Lett.} {\bf
  B115} (1982) 189--192}.

\bibitem{Fabbri:1983us}
R.~Fabbri and M.~d. Pollock, \emph{{The Effect of Primordially Produced
  Gravitons upon the Anisotropy of the Cosmological Microwave Background
  Radiation}},
  \href{http://dx.doi.org/10.1016/0370-2693(83)91322-9}{\emph{Phys. Lett.} {\bf
  B125} (1983) 445--448}.

\bibitem{Abbott:1984fp}
L.~F. Abbott and M.~B. Wise, \emph{{Constraints on Generalized Inflationary
  Cosmologies}},
  \href{http://dx.doi.org/10.1016/0550-3213(84)90329-8}{\emph{Nucl. Phys.} {\bf
  B244} (1984) 541--548}.

\bibitem{Seljak:1996ti}
U.~Seljak, \emph{{Measuring polarization in cosmic microwave background}},
  \href{http://dx.doi.org/10.1086/304123}{\emph{Astrophys. J.} {\bf 482} (1997)
  6}, [\href{https://arxiv.org/abs/astro-ph/9608131}{{\tt astro-ph/9608131}}].

\bibitem{Kamionkowski:1996zd}
M.~Kamionkowski, A.~Kosowsky and A.~Stebbins, \emph{{A Probe of primordial
  gravity waves and vorticity}},
  \href{http://dx.doi.org/10.1103/PhysRevLett.78.2058}{\emph{Phys. Rev. Lett.}
  {\bf 78} (1997) 2058--2061},
  [\href{https://arxiv.org/abs/astro-ph/9609132}{{\tt astro-ph/9609132}}].

\bibitem{Seljak:1996gy}
U.~Seljak and M.~Zaldarriaga, \emph{{Signature of gravity waves in polarization
  of the microwave background}},
  \href{http://dx.doi.org/10.1103/PhysRevLett.78.2054}{\emph{Phys. Rev. Lett.}
  {\bf 78} (1997) 2054--2057},
  [\href{https://arxiv.org/abs/astro-ph/9609169}{{\tt astro-ph/9609169}}].

\bibitem{Zaldarriaga:1998ar}
M.~Zaldarriaga and U.~Seljak, \emph{{Gravitational lensing effect on cosmic
  microwave background polarization}},
  \href{http://dx.doi.org/10.1103/PhysRevD.58.023003}{\emph{Phys. Rev.} {\bf
  D58} (1998) 023003}, [\href{https://arxiv.org/abs/astro-ph/9803150}{{\tt
  astro-ph/9803150}}].

\bibitem{Bashinsky:2003tk}
S.~Bashinsky and U.~Seljak, \emph{{Neutrino perturbations in CMB anisotropy and
  matter clustering}},
  \href{http://dx.doi.org/10.1103/PhysRevD.69.083002}{\emph{Phys. Rev.} {\bf
  D69} (2004) 083002}, [\href{https://arxiv.org/abs/astro-ph/0310198}{{\tt
  astro-ph/0310198}}].

\bibitem{Baumann:2015rya}
D.~Baumann, D.~Green, J.~Meyers and B.~Wallisch, \emph{{Phases of New Physics
  in the CMB}},
  \href{http://dx.doi.org/10.1088/1475-7516/2016/01/007}{\emph{JCAP} {\bf 1601}
  (2016) 007}, [\href{https://arxiv.org/abs/1508.06342}{{\tt 1508.06342}}].

\bibitem{Agashe:2014kda}
{\scshape Particle Data Group} collaboration, K.~A. Olive et~al., \emph{{Review
  of Particle Physics}},
  \href{http://dx.doi.org/10.1088/1674-1137/38/9/090001}{\emph{Chin. Phys.}
  {\bf C38} (2014) 090001}.

\bibitem{Lewis:2006fu}
A.~Lewis and A.~Challinor, \emph{{Weak gravitational lensing of the cmb}},
  \href{http://dx.doi.org/10.1016/j.physrep.2006.03.002}{\emph{Phys. Rept.}
  {\bf 429} (2006) 1--65}, [\href{https://arxiv.org/abs/astro-ph/0601594}{{\tt
  astro-ph/0601594}}].

\bibitem{Sunyaev:1980vz}
R.~A. Sunyaev and {\relax Ya}.~B. Zeldovich, \emph{{Microwave background
  radiation as a probe of the contemporary structure and history of the
  universe}},
  \href{http://dx.doi.org/10.1146/annurev.aa.18.090180.002541}{\emph{Ann. Rev.
  Astron. Astrophys.} {\bf 18} (1980) 537--560}.

\bibitem{Sunyaev:1972eq}
R.~A. Sunyaev and {\relax Ya}.~B. Zeldovich, \emph{{The Observations of relic
  radiation as a test of the nature of X-Ray radiation from the clusters of
  galaxies}}, {\emph{Comments Astrophys. Space Phys.} {\bf 4} (1972) 173--178}.

\bibitem{Crichton:2015joa}
D.~Crichton et~al., \emph{{Evidence for the thermal Sunyaev--Zel'dovich effect
  associated with quasar feedback}},
  \href{http://dx.doi.org/10.1093/mnras/stw344}{\emph{Mon. Not. Roy. Astron.
  Soc.} {\bf 458} (2016) 1478--1492},
  [\href{https://arxiv.org/abs/1510.05656}{{\tt 1510.05656}}].

\bibitem{Greco:2014vwa}
J.~P. Greco, J.~C. Hill, D.~N. Spergel and N.~Battaglia, \emph{{The Stacked
  Thermal Sunyaev--zel'dovich Signal of Locally Brightest Galaxies in Planck
  Full Mission Data: Evidence for Galaxy Feedback?}},
  \href{http://dx.doi.org/10.1088/0004-637X/808/2/151}{\emph{Astrophys. J.}
  {\bf 808} (2015) 151}, [\href{https://arxiv.org/abs/1409.6747}{{\tt
  1409.6747}}].

\bibitem{Ruan:2015vca}
J.~J. Ruan, M.~McQuinn and S.~F. Anderson, \emph{{Detection of Quasar Feedback
  from the Thermal Sunyaev-Zel'dovich Effect in Planck}},
  \href{http://dx.doi.org/10.1088/0004-637X/802/2/135}{\emph{Astrophys. J.}
  {\bf 802} (2015) 135}, [\href{https://arxiv.org/abs/1502.01723}{{\tt
  1502.01723}}].

\bibitem{Ade:2015fva}
{\scshape Planck} collaboration, P.~A.~R. Ade et~al., \emph{{Planck 2015
  results. XXIV. Cosmology from Sunyaev-Zeldovich cluster counts}},
  \href{https://arxiv.org/abs/1502.01597}{{\tt 1502.01597}}.

\bibitem{deHaan:2016qvy}
{\scshape SPT} collaboration, T.~de~Haan et~al., \emph{{Cosmological
  Constraints from Galaxy Clusters in the 2500 square-degree SPT-SZ Survey}},
  {\emph{Submitted to: Astrophys. J.} (2016) },
  [\href{https://arxiv.org/abs/1603.06522}{{\tt 1603.06522}}].

\bibitem{Sehgal:2010ca}
N.~Sehgal et~al., \emph{{The Atacama Cosmology Telescope: Cosmology from Galaxy
  Clusters Detected via the Sunyaev-Zel'dovich Effect}},
  \href{http://dx.doi.org/10.1088/0004-637X/732/1/44}{\emph{Astrophys. J.} {\bf
  732} (2011) 44}, [\href{https://arxiv.org/abs/1010.1025}{{\tt 1010.1025}}].

\bibitem{Ferraro:2010gh}
S.~Ferraro, F.~Schmidt and W.~Hu, \emph{{Cluster Abundance in f(R) Gravity
  Models}}, \href{http://dx.doi.org/10.1103/PhysRevD.83.063503}{\emph{Phys.
  Rev.} {\bf D83} (2011) 063503}, [\href{https://arxiv.org/abs/1011.0992}{{\tt
  1011.0992}}].

\bibitem{Schmidt:2009am}
F.~Schmidt, A.~Vikhlinin and W.~Hu, \emph{{Cluster Constraints on f(R)
  Gravity}}, \href{http://dx.doi.org/10.1103/PhysRevD.80.083505}{\emph{Phys.
  Rev.} {\bf D80} (2009) 083505}, [\href{https://arxiv.org/abs/0908.2457}{{\tt
  0908.2457}}].

\bibitem{Ostriker:1986fua}
J.~P. Ostriker and E.~T. Vishniac, \emph{{Generation of microwave background
  fluctuations from nonlinear perturbations at the ERA of galaxy formation}},
  \href{http://dx.doi.org/10.1086/184704}{\emph{Astrophys. J.} {\bf 306} (1986)
  L51}.

\bibitem{Ade:2015lza}
{\scshape Planck} collaboration, P.~A.~R. Ade et~al., \emph{{Planck
  intermediate results. XXXVII. Evidence of unbound gas from the kinetic
  Sunyaev-Zeldovich effect}},
  \href{http://dx.doi.org/10.1051/0004-6361/201526328}{\emph{Astron.
  Astrophys.} {\bf 586} (2016) A140},
  [\href{https://arxiv.org/abs/1504.03339}{{\tt 1504.03339}}].

\bibitem{Schaan:2015uaa}
{\scshape ACTPol} collaboration, E.~Schaan et~al., \emph{{Evidence for the
  kinematic Sunyaev-Zel'dovich effect with the Atacama Cosmology Telescope and
  velocity reconstruction from the Baryon Oscillation Spectroscopic Survey}},
  \href{http://dx.doi.org/10.1103/PhysRevD.93.082002}{\emph{Phys. Rev.} {\bf
  D93} (2016) 082002}, [\href{https://arxiv.org/abs/1510.06442}{{\tt
  1510.06442}}].

\bibitem{Kosowsky:2009nc}
A.~Kosowsky and S.~Bhattacharya, \emph{{A Future Test of Gravitation Using
  Galaxy Cluster Velocities}},
  \href{http://dx.doi.org/10.1103/PhysRevD.80.062003}{\emph{Phys. Rev.} {\bf
  D80} (2009) 062003}, [\href{https://arxiv.org/abs/0907.4202}{{\tt
  0907.4202}}].

\bibitem{Mueller:2014dba}
E.-M. Mueller, F.~de~Bernardis, R.~Bean and M.~D. Niemack, \emph{{Constraints
  on massive neutrinos from the pairwise kinematic Sunyaev-Zel'dovich effect}},
  \href{http://dx.doi.org/10.1103/PhysRevD.92.063501}{\emph{Phys. Rev.} {\bf
  D92} (2015) 063501}, [\href{https://arxiv.org/abs/1412.0592}{{\tt
  1412.0592}}].

\bibitem{Hu:2007bt}
W.~Hu, S.~DeDeo and C.~Vale, \emph{{Cluster Mass Estimators from CMB
  Temperature and Polarization Lensing}},
  \href{http://dx.doi.org/10.1088/1367-2630/9/12/441}{\emph{New J. Phys.} {\bf
  9} (2007) 441}, [\href{https://arxiv.org/abs/astro-ph/0701276}{{\tt
  astro-ph/0701276}}].

\bibitem{Seljak:1999zn}
U.~Seljak and M.~Zaldarriaga, \emph{{Lensing induced cluster signatures in
  cosmic microwave background}},
  \href{http://dx.doi.org/10.1086/309098}{\emph{Astrophys. J.} {\bf 538} (2000)
  57--64}, [\href{https://arxiv.org/abs/astro-ph/9907254}{{\tt
  astro-ph/9907254}}].

\bibitem{Calabrese:2014gwa}
E.~Calabrese et~al., \emph{{Precision Epoch of Reionization studies with
  next-generation CMB experiments}},
  \href{http://dx.doi.org/10.1088/1475-7516/2014/08/010}{\emph{JCAP} {\bf 1408}
  (2014) 010}, [\href{https://arxiv.org/abs/1406.4794}{{\tt 1406.4794}}].

\bibitem{smithferraro2016}
K.~M. Smith and S.~Ferraro, \emph{{Detecting patchy reionization in the CMB}},
  \href{https://arxiv.org/abs/1607.01769}{{\tt 1607.01769}}.

\bibitem{2016arXiv160502985P}
{Planck Collaboration}, N.~{Aghanim}, M.~{Ashdown}, J.~{Aumont},
  C.~{Baccigalupi}, M.~{Ballardini} et~al., \emph{{Planck intermediate results.
  XLVI. Reduction of large-scale systematic effects in HFI polarization maps
  and estimation of the reionization optical depth}}, {\emph{ArXiv e-prints}
  (May, 2016) }, [\href{https://arxiv.org/abs/1605.02985}{{\tt 1605.02985}}].

\bibitem{Brandt1994}
W.~N. {Brandt}, C.~R. {Lawrence}, A.~C.~S. {Readhead}, J.~N. {Pakianathan} and
  T.~M. {Fiola}, \emph{{Separation of foreground radiation from cosmic
  microwave background anisotropy using multifrequency measurements}},
  \href{http://dx.doi.org/10.1086/173867}{\emph{\apj} {\bf 424} (Mar., 1994)
  1--21}.

\bibitem{Eriksen2006}
H.~K. {Eriksen}, C.~{Dickinson}, C.~R. {Lawrence}, C.~{Baccigalupi}, A.~J.
  {Banday}, K.~M. {G{\'o}rski} et~al., \emph{{Cosmic Microwave Background
  Component Separation by Parameter Estimation}},
  \href{http://dx.doi.org/10.1086/500499}{\emph{\apj} {\bf 641} (Apr., 2006)
  665--682}, [\href{https://arxiv.org/abs/astro-ph/0508268}{{\tt
  astro-ph/0508268}}].

\bibitem{Stompor2009}
R.~{Stompor}, S.~{Leach}, F.~{Stivoli} and C.~{Baccigalupi}, \emph{{Maximum
  likelihood algorithm for parametric component separation in cosmic microwave
  background experiments}},
  \href{http://dx.doi.org/10.1111/j.1365-2966.2008.14023.x}{\emph{\mnras} {\bf
  392} (Jan., 2009) 216--232}, [\href{https://arxiv.org/abs/0804.2645}{{\tt
  0804.2645}}].

\bibitem{Errard2011}
J.~{Errard}, F.~{Stivoli} and R.~{Stompor}, \emph{{Publisher's Note: Framework
  for performance forecasting and optimization of CMB B-mode observations in
  the presence of astrophysical foregrounds [Phys. Rev. D 84, 063005 (2011)]}},
  \href{http://dx.doi.org/10.1103/PhysRevD.84.069907}{\emph{\prd} {\bf 84}
  (Sept., 2011) 069907}, [\href{https://arxiv.org/abs/1105.3859}{{\tt
  1105.3859}}].

\bibitem{Stompor2016}
R.~{Stompor}, J.~{Errard} and D.~{Poletti}, \emph{{Forecasting performance of
  CMB experiments in the presence of complex foreground contaminations}},
  \href{http://dx.doi.org/10.1103/PhysRevD.94.083526}{\emph{\prd} {\bf 94}
  (Oct., 2016) 083526}, [\href{https://arxiv.org/abs/1609.03807}{{\tt
  1609.03807}}].

\bibitem{smith2012}
K.~M. {Smith}, D.~{Hanson}, M.~{LoVerde}, C.~M. {Hirata} and O.~{Zahn},
  \emph{{Delensing CMB polarization with external datasets}},
  \href{http://dx.doi.org/10.1088/1475-7516/2012/06/014}{\emph{\jcap} {\bf 6}
  (June, 2012) 014}, [\href{https://arxiv.org/abs/1010.0048}{{\tt 1010.0048}}].

\bibitem{Hirata2003}
C.~M. {Hirata} and U.~{Seljak}, \emph{{Reconstruction of lensing from the
  cosmic microwave background polarization}},
  \href{http://dx.doi.org/10.1103/PhysRevD.68.083002}{\emph{\prd} {\bf 68}
  (Oct., 2003) 083002}, [\href{https://arxiv.org/abs/astro-ph/0306354}{{\tt
  astro-ph/0306354}}].

\bibitem{Smith2009}
K.~M. {Smith}, A.~{Cooray}, S.~{Das}, O.~{Dor{\'e}}, D.~{Hanson}, C.~{Hirata}
  et~al., \emph{{Gravitational Lensing}},  in \emph{American Institute of
  Physics Conference Series} (S.~{Dodelson}, D.~{Baumann}, A.~{Cooray},
  J.~{Dunkley}, A.~{Fraisse}, M.~G. {Jackson} et~al., eds.), vol.~1141 of
  \emph{American Institute of Physics Conference Series}, pp.~121--178, June,
  2009.
\newblock \href{https://arxiv.org/abs/0811.3916}{{\tt 0811.3916}}.
\newblock \href{http://dx.doi.org/10.1063/1.3160886}{DOI}.

\bibitem{Lesgourgues}
J.~{Lesgourgues} and S.~{Pastor}, \emph{{Massive neutrinos and cosmology}},
  \href{http://dx.doi.org/10.1016/j.physrep.2006.04.001}{\emph{\physrep} {\bf
  429} (July, 2006) 307--379},
  [\href{https://arxiv.org/abs/astro-ph/0603494}{{\tt astro-ph/0603494}}].

\bibitem{Knox1995}
L.~{Knox}, \emph{{Determination of inflationary observables by cosmic microwave
  background anisotropy experiments}},
  \href{http://dx.doi.org/10.1103/PhysRevD.52.4307}{\emph{\prd} {\bf 52} (Oct.,
  1995) 4307--4318}, [\href{https://arxiv.org/abs/astro-ph/9504054}{{\tt
  astro-ph/9504054}}].

\bibitem{Richards}
P.~Richards, \emph{Bolometers for infrared and millimeter waves},
  \href{http://dx.doi.org/10.1063/1.357128}{\emph{Journal of Applied Physics}
  {\bf 76} (1994) 1--24}.

\bibitem{Griffin:02}
M.~J. Griffin, J.~J. Bock and W.~K. Gear, \emph{Relative performance of filled
  and feedhorn-coupled focal-plane architectures},
  \href{http://dx.doi.org/10.1364/AO.41.006543}{\emph{Appl. Opt.} {\bf 41}
  (2002) 6543--6554}.

\bibitem{2016JLTP..184..824M}
T.~{Matsumura}, Y.~{Akiba}, K.~{Arnold}, J.~{Borrill}, R.~{Chendra},
  Y.~{Chinone} et~al., \emph{{LiteBIRD: Mission Overview and Focal Plane
  Layout}}, \href{http://dx.doi.org/10.1007/s10909-016-1542-8}{\emph{Journal of
  Low Temperature Physics} {\bf 184} (Aug., 2016) 824--831}.

\bibitem{2016arXiv160506569T}
R.~J. {Thornton}, P.~A.~R. {Ade}, S.~{Aiola}, F.~E. {Angile}, M.~{Amiri}, J.~A.
  {Beall} et~al., \emph{{The Atacama Cosmology Telescope: The
  polarization-sensitive ACTPol instrument}}, {\emph{ArXiv e-prints} (May,
  2016) }, [\href{https://arxiv.org/abs/1605.06569}{{\tt 1605.06569}}].

\bibitem{Suzuki_LTD15}
A.~Suzuki et~al., \emph{Multi-chroic dual-polarization bolometric detectors for
  studies of the cosmic microwave background}, {\emph{Journal of Low
  Temperature Physics} (2013 Submitted) }.

\bibitem{2012SPIE.8452E..3EM}
T.~{Matsumura}, P.~{Ade}, K.~{Arnold}, D.~{Barron}, J.~{Borrill}, S.~{Chapman}
  et~al., \emph{{POLARBEAR-2 optical and polarimeter designs}},  in
  \emph{Millimeter, Submillimeter, and Far-Infrared Detectors and
  Instrumentation for Astronomy VI}, vol.~8452 of \emph{\procspie}, p.~84523E,
  Sept., 2012.
\newblock \href{http://dx.doi.org/10.1117/12.926770}{DOI}.

\bibitem{2014SPIE.9153E..1PB}
B.~A. {Benson}, P.~A.~R. {Ade}, Z.~{Ahmed}, S.~W. {Allen}, K.~{Arnold}, J.~E.
  {Austermann} et~al., \emph{{SPT-3G: a next-generation cosmic microwave
  background polarization experiment on the South Pole telescope}},  in
  \emph{Millimeter, Submillimeter, and Far-Infrared Detectors and
  Instrumentation for Astronomy VII}, vol.~9153 of \emph{\procspie}, p.~91531P,
  July, 2014.
\newblock \href{https://arxiv.org/abs/1407.2973}{{\tt 1407.2973}}.
\newblock \href{http://dx.doi.org/10.1117/12.2057305}{DOI}.

\bibitem{2003ApJ...585..566P}
L.~{Page}, C.~{Jackson}, C.~{Barnes}, C.~{Bennett}, M.~{Halpern}, G.~{Hinshaw}
  et~al., \emph{{The Optical Design and Characterization of the Microwave
  Anisotropy Probe}}, \href{http://dx.doi.org/10.1086/346078}{\emph{\apj} {\bf
  585} (Mar., 2003) 566--586},
  [\href{https://arxiv.org/abs/astro-ph/0301160}{{\tt astro-ph/0301160}}].

\bibitem{2010A&A...520A...2T}
J.~A. {Tauber}, H.~U. {Norgaard-Nielsen}, P.~A.~R. {Ade}, J.~{Amiri Parian},
  T.~{Banos}, M.~{Bersanelli} et~al., \emph{{Planck pre-launch status: The
  optical system}},
  \href{http://dx.doi.org/10.1051/0004-6361/200912911}{\emph{\aap} {\bf 520}
  (Sept., 2010) A2}.

\bibitem{2013ApJ...768....9B}
C.~{Bischoff}, A.~{Brizius}, I.~{Buder}, Y.~{Chinone}, K.~{Cleary}, R.~N.
  {Dumoulin} et~al., \emph{{The Q/U Imaging ExperimenT Instrument}},
  \href{http://dx.doi.org/10.1088/0004-637X/768/1/9}{\emph{\apj} {\bf 768}
  (May, 2013) 9}, [\href{https://arxiv.org/abs/1207.5562}{{\tt 1207.5562}}].

\bibitem{2010arXiv1008.3915E}
T.~{Essinger-Hileman}, J.~W. {Appel}, J.~A. {Beall}, H.~M. {Cho}, J.~{Fowler},
  M.~{Halpern} et~al., \emph{{The Atacama B-Mode Search: CMB Polarimetry with
  Transition-Edge-Sensor Bolometers}}, {\emph{ArXiv e-prints} (Aug., 2010) },
  [\href{https://arxiv.org/abs/1008.3915}{{\tt 1008.3915}}].

\bibitem{2012SPIE.8444E..2YR}
J.~A. {Rubi{\~n}o-Mart{\'{\i}}n}, R.~{Rebolo}, M.~{Aguiar},
  R.~{G{\'e}nova-Santos}, F.~{G{\'o}mez-Re{\~n}asco}, J.~M. {Herreros} et~al.,
  \emph{{The QUIJOTE-CMB experiment: studying the polarisation of the galactic
  and cosmological microwave emissions}},  in \emph{Ground-based and Airborne
  Telescopes IV}, vol.~8444 of \emph{\procspie}, p.~84442Y, Sept., 2012.
\newblock \href{http://dx.doi.org/10.1117/12.926581}{DOI}.

\bibitem{2012SPIE.8452E..1MT}
O.~{Tajima}, J.~{Choi}, M.~{Hazumi}, H.~{Ishitsuka}, M.~{Kawai} and
  M.~{Yoshida}, \emph{{GroundBIRD: an experiment for CMB polarization
  measurements at a large angular scale from the ground}},  in
  \emph{Millimeter, Submillimeter, and Far-Infrared Detectors and
  Instrumentation for Astronomy VI}, vol.~8452 of \emph{\procspie}, p.~84521M,
  Sept., 2012.
\newblock \href{http://dx.doi.org/10.1117/12.925816}{DOI}.

\bibitem{doi:10.1117/12.2232008}
H.~Sugai, S.~Kashima, K.~Kimura, T.~Matsumura, M.~Inoue, M.~Ito et~al.,
  \emph{Optical designing of litebird},  2016.
\newblock 10.1117/12.2232008.

\bibitem{2014SPIE.9153E..1NA}
Z.~{Ahmed}, M.~{Amiri}, S.~J. {Benton}, J.~J. {Bock}, R.~{Bowens-Rubin},
  I.~{Buder} et~al., \emph{{BICEP3: a 95GHz refracting telescope for
  degree-scale CMB polarization}},  in \emph{Millimeter, Submillimeter, and
  Far-Infrared Detectors and Instrumentation for Astronomy VII}, vol.~9153 of
  \emph{\procspie}, p.~91531N, Aug., 2014.
\newblock \href{https://arxiv.org/abs/1407.5928}{{\tt 1407.5928}}.
\newblock \href{http://dx.doi.org/10.1117/12.2057224}{DOI}.

\bibitem{2010SPIE.7741E..1OR}
M.~C. {Runyan}, P.~A.~R. {Ade}, M.~{Amiri}, S.~{Benton}, R.~{Bihary}, J.~J.
  {Bock} et~al., \emph{{Design and performance of the SPIDER instrument}},  in
  \emph{Millimeter, Submillimeter, and Far-Infrared Detectors and
  Instrumentation for Astronomy V}, vol.~7741 of \emph{\procspie}, p.~77411O,
  July, 2010.
\newblock \href{https://arxiv.org/abs/1106.2173}{{\tt 1106.2173}}.
\newblock \href{http://dx.doi.org/10.1117/12.857715}{DOI}.

\bibitem{2016ApOpt..55.1686N}
M.~D. {Niemack}, \emph{{Designs for a large-aperture telescope to map the CMB
  10{$\times$} faster}},
  \href{http://dx.doi.org/10.1364/AO.55.001686}{\emph{\ao} {\bf 55} (Mar.,
  2016) 1686}, [\href{https://arxiv.org/abs/1511.04506}{{\tt 1511.04506}}].

\bibitem{6710211}
J.~Y. Suen, M.~T. Fang and P.~M. Lubin, \emph{Global distribution of water
  vapor and cloud cover --;sites for high-performance thz applications},
  \href{http://dx.doi.org/10.1109/TTHZ.2013.2294018}{\emph{IEEE Transactions on
  Terahertz Science and Technology} {\bf 4} (Jan, 2014) 86--100}.

\bibitem{bicep2_instrument2014}
P.~A.~R. Ade, R.~W. Aikin, M.~Amiri, D.~Barkats, S.~J. Benton, C.~A. Bischoff
  et~al., \emph{Bicep2. ii. experiment and three-year data set}, {\emph{The
  Astrophysical Journal} {\bf 792} (2014) 62}.

\bibitem{PB_ClBB_2014}
T.~P. C. P. A.~R. Ade, Y.~Akiba, A.~E. Anthony, K.~Arnold, M.~Atlas, D.~Barron
  et~al., \emph{A measurement of the cosmic microwave background b-mode
  polarization power spectrum at sub-degree scales with polarbear}, {\emph{The
  Astrophysical Journal} {\bf 794} (2014) 171}.

\bibitem{BICEPcost}
B.~Keating. Private Communication, 2016.

\bibitem{PBcost}
A.~Lee. Private Communication, 2016.

\bibitem{ACTcost}
M.~Devlin and L.~Page. Private Communication, 2016.

\bibitem{SPTcost}
S.~Padin. Private Communication, 2016.

\bibitem{2016PhRvL.116c1302B}
{BICEP2 Collaboration}, {Keck Array Collaboration}, P.~A.~R. {Ade}, Z.~{Ahmed},
  R.~W. {Aikin}, K.~D. {Alexander} et~al., \emph{{Improved Constraints on
  Cosmology and Foregrounds from BICEP2 and Keck Array Cosmic Microwave
  Background Data with Inclusion of 95 GHz Band}},
  \href{http://dx.doi.org/10.1103/PhysRevLett.116.031302}{\emph{Physical Review
  Letters} {\bf 116} (Jan., 2016) 031302},
  [\href{https://arxiv.org/abs/1510.09217}{{\tt 1510.09217}}].

\bibitem{Simone1}
S.~{Ferraro}. in preparation.

\bibitem{Simone2}
N.~Battaglia, E.~Schaan, S.~Ferraro and D.~Spergel. in preparation.

\bibitem{IrwinHilton}
K.~Irwin and G.~Hilton, \emph{Transition-edge sensors},  in \emph{Cryogenic
  Particle Detection} (C.~Enss, ed.), vol.~99 of \emph{Topics in Applied
  Physics}, pp.~63--150.
\newblock Springer Berlin Heidelberg, 2005.
\newblock \href{http://dx.doi.org/10.1007/10933596_3}{DOI}.

\bibitem{Mather:82}
J.~C. Mather, \emph{Bolometer noise: nonequilibrium theory},
  \href{http://dx.doi.org/10.1364/AO.21.001125}{\emph{Appl. Opt.} {\bf 21}
  (1982) 1125--1129}.

\end{thebibliography}\endgroup

\appendix

\section{Detector Sensitivity}\label{sec:detsens}
In this Appendix, we present our method for estimating the sensitivity in
Table~\ref{tab:NETSummary2}.
We summarize assumptions that were made to calculate the noise equivalent temperature (NET) of a detector.
The detectors are split into the frequency bands as shown in Table~\ref{tab:NETSummary2}.

\subsection{Optics}
We consider two configurations.
The first configuration is a small-aperture instrument with fully cryogenic optics (conf1 in Sec.~\ref{sec:detector_assumptions} ).
Table~\ref{tab:OptElement2} summarizes properties of optical elements of the configuration.
The second configuration is a large-aperture telescope with two room-temperature mirrors and cryogenic re-imaging optics (conf2 in
Sec.~\ref{sec:detector_assumptions}).
 Table~\ref{tab:OptElement} summarizes properties of optical elements in this configuration.
We use a standard method to calculate the optical loading and noise equivalent photons\cite{Richards,Griffin:02}. 

\begin{table}[htbp]
\centering
\caption{Temperature, efficiency, and emissivity of hypothetical
 small-aperture reflective telescope elements used for the sensitivity
 calculation.  Values used for the 150-GHz band is shown. Emissivity for each optical element decreases for lower frequency channels and increases for higher frequency channels\label{tab:OptElement2}}
	\begin{tabular}{c c c c}
	\hline
	\hline
	Optical element & Temperature & Efficiency & Emissivity\\
	\hline
	CMB & 2.725 & 1.000 & 1.000 \\
	Atmosphere & 277 & 0.968 & 0.032 \\
	Window & 250 & 0.980 & 0.020 \\
	50 Kelvin IR Filter & 50 & 0.950 & 0.050 \\	
	Half-Wave Plate & 50 & 0.970 & 0.030 \\
	Lyot Stop & 4 & $e^{-\frac{\pi^2}{2}\left(\frac{w_0}{F\lambda}\right)^2}$ & $1-e^{-\frac{\pi^2}{2}\left(\frac{w_0}{F\lambda}\right)^2}$ \\
	4 Kelvin Filter & 4 & 0.950 & 0.050 \\
	Primary Mirror & 4 & 0.993 & 0.007 \\
	Secondary Mirror & 4 & 0.993 & 0.007 \\
	1 Kelvin Filter & 1 & 0.950 & 0.050 \\
	Detecor & 0.1 & 0.700 & 0.300 \\
	\hline
	\hline
	\end{tabular}
\end{table} 

\begin{table}[htbp]
\centering
\caption{Temperature, efficiency, and emissivity of hypothetical
 large-aperture  telescope and receiver elements used for this
 sensitivity calculation.  Values used for the 150-GHz band is shown. Emissivity for each optical element decreases for lower frequency channels and increases for higher frequency channels\label{tab:OptElement}}
	\begin{tabular}{c c c c}
	\hline
	\hline
	Optical element & Temperature & Efficiency & Emissivity\\
	\hline
	CMB & 2.725 & 1.000 & 1.000 \\
	Atmosphere & 277 & 0.968 & 0.032 \\
	Primary Mirror & 277 & 0.993 & 0.007 \\
	Secondary Mirror & 277 & 0.993 & 0.007 \\
	Window & 250 & 0.980 & 0.020 \\
	Half-Wave Plate & 100 & 0.970 & 0.030 \\
	50 Kelvin IR Filter & 50 & 0.950 & 0.050 \\
	4 Kelvin Filter & 4 & 0.950 & 0.050 \\
	Field Lens & 4 & 0.970 & 0.030 \\
	Aperture Lens & 4 & 0.970 & 0.030 \\
	Lyot Stop & 4 & $e^{-\frac{\pi^2}{2}\left(\frac{w_0}{F\lambda}\right)^2}$ & $1-e^{-\frac{\pi^2}{2}\left(\frac{w_0}{F\lambda}\right)^2}$ \\
	Collimating Lens & 4 & 0.970 & 0.030 \\
	1 Kelvin Filter & 1 & 0.950 & 0.050 \\
	Detector & 0.1 & 0.700 & 0.300 \\
	\hline
	\hline
	\end{tabular}
\end{table}

\subsection{Pixel Size}
As outined in Table~\ref{tab:NETSummary2}, we assumed multi-chroic pixel. 
Each configuration's noise versus detector pixel size is a balance between noise per pixel versus the total number of pixels. 
In a limited field-of-view limit (limited focal plane area), smaller-sized pixels increase the number of pixels on a focal plane, but it reduces the signal-to-noise ratio per pixel. 
If the experiment is limited by the number of detectors (e.g., due to readout cost), it is beneficial to make the detector as big as possible for a given focal plane area. 
Ground experiments with hot optics and a bright sky prefer smaller pixels for a field-of-view limited case. 
In this report, we used $f\lambda = 1.5$ at the center frequency of a multi-chroic pixel,
where $f$ is f-number of optics at detector.
For example, for an f-number = 2 system, the low-frequency pixel size is 30\,mm, the mid-frequency pixel size is 7\,m,  and the high-frequency pixel size is 3.75\,mm.
We assumed that the diffraction aperture size of the pixel remains constant as a function of frequency; as a result, the aperture efficiency changes as a function of frequency.
We assumed a Gaussian beam to calculate the aperture efficiency.
The Gaussian beam waist for the Gaussian beam is assumed to be a factor of 2.6 smaller than the pixel size
\subsection{Sensor}
TES bolometer was assumed for the noise calculation.
Assumption for TES bolometer model is given in Table ~\ref{tab:NETAssumption}.

We followed Irwin and Hilton to calculate the thermal carrier noise of a TES bolometer \cite{IrwinHilton,Mather:82}. 

\begin{table}[htbp]
\centering
\caption{Summary of assumptions made for the sensitivity calculation. \label{tab:NETAssumption}}
	\begin{tabular}{c c}
	\hline
	\hline
	 Bath Temperature 		& 0.100 Kelvin \\
	 Transition Temperature 	& 0.170 Kelvin \\
	Bolometer Saturation power to Optical Power ratio 	& 2.5 \\
	Thermal Conductivity index for Bolometer Leg & 3\\
	Readout Chain Noise Equivalent Current & 7 $ \frac{\mathrm{pA}}{\sqrt{\mathrm{Hz}}}$ \\
	Gaussian Beam Waist to Pixel to Pixel Spacing 	& 2.6 \\
	Band Shape 		& Top Hat \\
	\hline
	\hline
	\end{tabular}
\end{table} 

\subsection{Base temperature}
The detector noise performance will depend on the focal plane temperature. 
We assumed a focal plane temperature of 100 mili-Kelvin, which is a typical base temperature of continuous dilusion refregirator system and adiabatic demagnetization refrigerator system.
The CMB instrument can be designed to achieve photon-noise-limited performance with a 250 milli-Kelvin focal plane temperature,
but the margin of error to achieve such performance is small. 
Reducing the focal plane temperature to 100 milli-Kelvin provides more margin of error in detector fabrication. 
Increasing the margin of error relaxes the requirement on detector fabrication, which will increase detector fabrication yield and throughput.
High yield and shorter production periods will result in reduction of detector fabrication cost, which should be compared against the cost and difficulty of achieving a 100 milli-Kelvin system.

\subsection{Readout}
For the readout system, we assumed a noise equivalent current
performance of $7~\mathrm{pA}/\sqrt{\mathrm{Hz}}$.
We multiply this by the voltage bias value optimized for bolometer performance to calculate the noise-equivalent power for readout noise.

\subsection{Total Noise}
We added photon noise, thermal carrier noise, and readout noise in quadrature.

\end{document}